\documentclass[final,5p,times,twocolumn]{elsarticle}
\usepackage{amssymb}
\usepackage{fixltx2e}
\usepackage[dvipsnames,table,xcdraw]{xcolor}
\usepackage{textcomp}
\usepackage[fleqn]{mathtools}
\definecolor{MyGreen-rgb}{rgb}{0,0.91,0.04001}
\definecolor{MyGreen-hsb}{hsb}{0.34065,1,0.91}
\definecolor{MyGreen-gray}{gray}{0.5383}
\colorlet{st}{MyGreen-rgb!75!black}
\colorlet{lv}{MyGreen-hsb!20!blue}
\colorlet{du}{MyGreen-gray!85!black}
\usepackage{amsmath}
\usepackage{endnotes}
\usepackage{stackengine}
\usepackage{babel}
\usepackage{xr-hyper}
\usepackage[colorlinks]{hyperref}
\usepackage[nameinlink,capitalise]{cleveref}
\usepackage[normalem]{ulem}
\usepackage{tikz}
\usepackage{floatrow}
\usepackage{makecell}
\usepackage{graphicx}
\usepackage{epstopdf} 
\usepackage{subcaption}
\usepackage{caption}
\floatstyle{plaintop}
\restylefloat{table}
\usepackage{multirow}
\usepackage{enumitem}
\usepackage[export]{adjustbox}
\usepackage{comment}

\makeatletter

\newcommand{\mathleft}{\@fleqntrue\@mathmargin0pt}
\newcommand{\mathnew}{\@fleqntrue\@mathmargin53pt}
\newcommand{\mathcenter}{\@fleqnfalse}
\newcommand\upstrut{\rule{0pt}{10pt}}
\newcommand\downstrut{\rule{0pt}{0.5pt}}
\newcommand\mystrut{\upstrut\downstrut}
\makeatother

\begin{document}

\begin{frontmatter}

\title{Improved Abdominal Multi-Organ Segmentation via 3D Boundary-Constrained Deep Neural Networks}

\author[a]{Samra Irshad\corref{aaa}}
\ead{sam.ershad@yahoo.com}
\author[b]{Douglas P.S. Gomes}
\author[c]{Seong Tae Kim}
\address[a]{Swinburne University of Technology, Hawthorn, Australia}
\address[b]{Victoria University, Melbourne, Australia}
\address[c]{Kyung Hee University, Yongin-si, Gyeonggi-do, South Korea}
\cortext[aaa]{Corresponding Author.}
\begin{abstract}
Background and Objective: Quantitative assessment of the abdominal region from clinically acquired CT scans requires the simultaneous segmentation of abdominal organs. Therefore, for the past two decades, automatic abdominal image segmentation has been the subject of intensive research to facilitate the health professionals easing the clinical workflow. Thanks to the availability of high-performance and powerful computational resources, deep learning-based methods have resulted in state-of-the-art performance for the segmentation of 3D abdominal CT scans. However, the complex characterization of organs with fuzzy and weak boundaries prevents the deep learning methods from accurately segmenting these anatomical organs. Specifically, the voxels on the boundary of organs are more vulnerable to misprediction due to the highly-varying intensity of inter-organ boundaries, and the misprediction of these voxels is detrimental to overall segmentation performance. This paper investigates the possibility of improving the abdominal image segmentation performance of the existing 3D encoder-decoder networks by leveraging organ-boundary prediction as a complementary task.\\

Method: To address the problem of abdominal multi-organ segmentation, we train the 3D encoder-decoder network to simultaneously segment the abdominal organs and their corresponding boundaries in CT scans via multi-task learning. The network is trained end-to-end using a loss function that combines two task-specific losses, i.e., complete organ segmentation loss and boundary prediction loss. We explore two different network topologies based on the extent of weights shared between the two tasks within a unified multi-task framework. In the first topology, the whole-organ prediction task and the boundary detection task share all the layers in the encoder-decoder network except for the last task-specific prediction layers. In contrast, the second topology employs a single shared encoder but two separate task-specific decoders. To evaluate the utilization of complementary boundary prediction task in improving the abdominal multi-organ segmentation, we use three state-of-the-art encoder-decoder networks: 3D UNet, 3D UNet$_{++}$, and 3D Attention-UNet. \\

Results: The effectiveness of utilizing the organs' boundary information for abdominal multi-organ segmentation is evaluated on two publically available abdominal CT datasets: Pancreas-CT and the BTCV dataset. The improvements shown in segmentation results (evaluated via Dice Score, Average Hausdorff Distance, Recall, and Precision) reveal the advantage of the multi-task training that forces the network to pay attention to ambiguous boundaries of organs. A maximum relative improvement of 3.5\% and 3.6\% is observed in Mean Dice Score for Pancreas-CT and BTCV datasets, respectively. All source codes are publically available on \url{https://github.com/samra-irshad/3d-boundary-constrained-networks}.
\end{abstract}


\end{frontmatter}


\section{Introduction}
\label{sec:intro}
Multi-organ segmentation on abdominal Computed Tomography (CT) scans is an essential prerequisite for computer-assisted surgery and organ transplantation \cite{okada2012multi}, \cite{gibson2018automatic}. Particularly, quantitative assessment of abdominal regions enables accurate organ dose calculation, required in numerous radiotherapy treatment options. Erroneous delineation of abdominal organs prevents harnessing the benefits of radiotherapeutic advancements. In clinical practice, physicians delineate abdominal organs using manual segmentation tools, which are time-consuming, observer-dependent, and error-prone. With the increased use of imaging facilities and production of a large number of abdominal CT scans, the utilization of automated, robust, and efficient organ-delineation tools has become compulsory \cite{gibson2018automatic}, \cite{hu2017automatic}, \cite{article1kim}. Automatic segmentation tools delineate the abdominal structures much faster and overcome the issues like variability in human expertise and inherent subjectivity.

Abdominal CT scans often present weak inter-organ boundaries characterized by regions of similar voxel intensities, which in turn results in low-contrast representations. Such appearances are usually caused by the representation of abdominal soft tissues in a narrow band of Hounsfield (HU) values. Another factor that enhances the already complex representation of abdominal organs is the existence of artifacts occurring due to blood flow, respiratory, and cardiac motion. Accurate delineation of abdominal organs with unclear boundaries and complex geometrical shapes is one of the ongoing challenges that hurdles the abdominal-related clinical diagnosis. 
\begin{figure}[!h]
\begin{subfigure}[b]{.3\textwidth}
\hspace{-1mm}
\includegraphics[width=\textwidth]{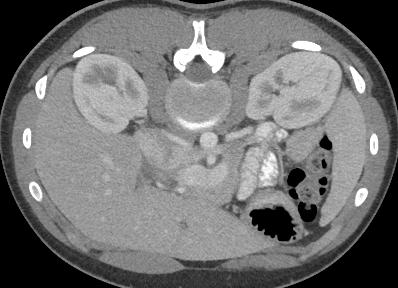} 
\caption{}
\label{fig:sub1}
\end{subfigure}
\begin{subfigure}[b]{.3\textwidth}
\includegraphics[width=\textwidth]{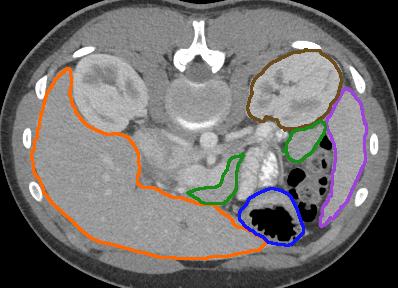} 
\caption{}
\label{fig:sub2}
\end{subfigure}%
\begin{subfigure}[b]{.3\textwidth}
\includegraphics[width=\textwidth]{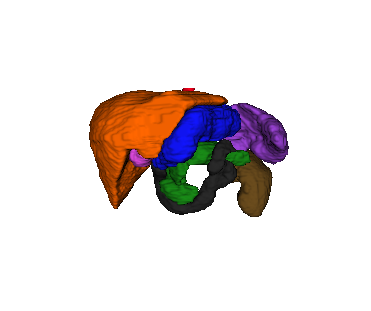} 
\caption{}
\label{fig:sub3}
\end{subfigure}%
\caption{Exemplary 2D abdominal CT image showing the visual characteristics of organs. (a) 2D abdominal image, (b) Abdominal organs annotated on CT image: pancreas ( \raisebox{0.3em}{\fcolorbox{RawSienna}{RawSienna}{\rule{0pt}{0pt}\rule{0pt}{0pt}}} ), spleen ( \raisebox{0.3em}{\fcolorbox{Purple}{Purple}{\rule{0pt}{0pt}\rule{0pt}{0pt}}} ), liver ( \raisebox{0.3em}{\fcolorbox{BurntOrange}{BurntOrange}{\rule{0pt}{0pt}\rule{0pt}{0pt}}} ), stomach ( \raisebox{0.3em}{\fcolorbox{lv}{lv}{\rule{0pt}{0pt}\rule{0pt}{0pt}}} ), gallbladder ( \raisebox{0.3em}{\fcolorbox{st}{st}{\rule{0pt}{0pt}\rule{0pt}{0pt}}} ), (c) 3D multi-organ voxel map.}\label{fig:ctscan2}
\end{figure}

Earlier methods proposed for the abdominal multi-organ segmentation mainly were based on multi-atlas \cite{xu2015efficient}, \cite{suzuki2012multi} or statistical models \cite{cerrolaza2015automatic}, \cite{okada2015abdominal}. Some methods also made use of handcrafted or learned features to segment abdominal organs \cite{campadelli2009automatic}, \cite{Selver2014SegmentationOA}. However, the recent Fully Convolutional Network (FCN) based approaches have presented better results due to the improved organ representation learning \cite{gibson2018automatic} \cite{heinrich2019obelisk}. Being able to preserve the image structure and provision of efficient learning as well as inference, FCN-based methods are currently considered state-of-the-art for abdominal multi-organ segmentation \cite{gibson2018automatic}, \cite{bobo2018fully}, \cite{sinha2020multi}, \cite{Lu2022}. Specifically, these networks follow the encoder-decoder architectural design \cite{ronneberger2015u}. In such networks, the shallow layers in the encoder aim to extract low-level features, and the deep layers encode high-level features. While the mirrored-decoder maps back the learned features to generate an output of the same size as input with skip connections assisting in retaining the crucial features extracted in the encoding path \cite{ronneberger2015u}.

Existing FCN-based methods for abdominal multi-organ segmentation employ either 2D or 3D convolutional architectures \cite{sinha2020multi}, \cite{bobo2018fully}. 2D methods process the CT scans in a slice-by-slice fashion and predict the organ labels on individual slices \cite{sinha2020multi}. Despite being memory- and parameter-efficient, 2D methods are unable to make full use of 3D contextual information \cite{gibson2018automatic}. 3D methods make use of rich volumetric context by processing the whole CT volume and generating voxel-maps in a single forward propagation pass, leading to better abdominal CT segmentation performance than 2D approaches \cite{Roth2018AnAO}, \cite{Zhou2018PerformanceEO}.

The existing 3D methods have primarily focused on designing better architectures for improved abdominal multi-organ representation learning \cite{bobo2018fully}, \cite{gibson2018automatic}. However, they treat all the anatomical parts within a single organ equally since they solely rely on voxel-level information and do not specifically focus on improving the segmentation of voxels in vulnerable regions/parts of organs. As an example, we highlight some of the important characteristics of abdominal organs in \autoref{fig:ctscan2}. From \Cref{fig:sub1,fig:sub2}, it can be noticed that the adjacent organs have weak contours which sometimes touch each other. As an example, observe the low-contrasted and touching boundaries between stomach ( \raisebox{0.3em}{\fcolorbox{lv}{lv}{\rule{0pt}{0pt}\rule{0pt}{0pt}}} ) and pancreas ( \raisebox{0.3em}{\fcolorbox{RawSienna}{RawSienna}{\rule{0pt}{0pt}\rule{0pt}{0pt}}} ). Moreover, 3D multi-organ visualization in \autoref{fig:sub3} shows that the adjacent positioning of organs in the abdominal cavity aggravates the complex spatial relationship among the organs. Simultaneously segmenting the abdominal organs with soft contours and complex spatial relationships is a challenging task.

The boundaries of anatomical regions in medical scans serve as an important cue for facilitating manual and automated delineation \cite{Guo2021}. Numerous existing deep learning-based studies leveraged learning of features corresponding to boundary of regions for improved medical image segmentation via multitask learning paradigm \cite{chen2016dcan}, \cite{8363791}, \cite{8906008}, \cite{9313420}, \cite{MEI2020101988}. In recent years, deep multitask learning paradigm has been widely used due to its potential to solve multiple tasks in one forward propagation and ability to learn better representations because of the multiple supervisory signals \cite{Amyar2020MultitaskDL}, \cite{Du2019CrossInfoNetMI}. In this paper, we propose to improve the segmentation of abdominal organs on CT scans by enhancing the segmentation of boundary of organs. Particularly, we train the 3D deep learning networks to simultaneously predict the boundary and the entire region of organs. The inclusion of boundary information is motivated by the fact that the voxels on the boundary of organs are more vulnerable to misprediction because of their ambiguous appearance and complex relationship with adjacent organs. Specifically, our work makes the following contributions:
\begin{enumerate}[label=(\roman*)]
\item We develop an end-to-end trainable 3D multi-task learning framework that simultaneously predicts the voxel-labels of abdominal organs and their corresponding boundaries. By integrating the boundary features, our proposed boundary-constrained 3D deep learning framework focuses on the accurate prediction of the edges of organs in addition to whole organs.
\item Instead of relying on a single network topology, we explore and compare two network topologies for conducting multi-task learning. In the first topology, the whole encoder-decoder network is shared with separate task-specific prediction layers at the end for predicting boundaries and entire organs' maps. In the second topology, an encoder is shared with separate task-specific decoders for decoding the features, jointly learned by the shared encoder to predict the boundary and organ probability maps. With an extensive comparison, we reveal that integration of boundary features invariably improves the multi-organ segmentation performance, independent of the multi-task network design.
\item We utilize three state-of-the-art 3D encoder-decoder architectures, i.e., UNet \cite{cciccek20163d}, UNet$_{++}$ \cite{zhou2019unet++}, and Attention-UNet \cite{schlemper2019attention} as baseline networks for evaluating the effect of incorporating boundary information. We modifiy each baseline architecture according to our proposed multi-task topologies. We demonstrate significant performance improvements with a negligible increase in trainable parameters.
\item We validate the performance of baseline and counterpart boundary-constrained models on two publically available datasets (Pancreas-CT \cite{Roth2016} and BTCV \cite{landmanbvc}) using Dice Score, Average Hausdorff Distance, Recall, and Precision. Furthermore, we conduct additional experiments to evaluate the improvement in the segmentation of regions around the boundaries. The results show that the boundary-constrained networks learn feature representations that focus on the accurate organs segmentation and the challenging parts around the border of the organs. 
\end{enumerate}
The rest of the article is organized as follows. In \autoref{sec:related}, we review the existing methods for abdominal multi-organ segmentation. \Cref{sec:method} describes our framework for incorporating the boundary information into the 3D fully convolutional networks, including the multi-task loss function and the details of boundary-constrained network topologies. Next, we describe the dataset specifications and implementation details in \autoref{sec:experiment}. We then present the experimental results, comparisons with existing single-task approaches, and in-depth performance analysis of boundary-constrained models in \autoref{sec:result}. Finally, we discuss the important highlights and some directions for future work in \autoref{sec:discuss} and present the conclusion in \autoref{sec:conc}. 
\begin{figure*}[!h]
\begin{subfigure}[b]{.45\textwidth}
\hspace{3mm}
\includegraphics[width=\textwidth]{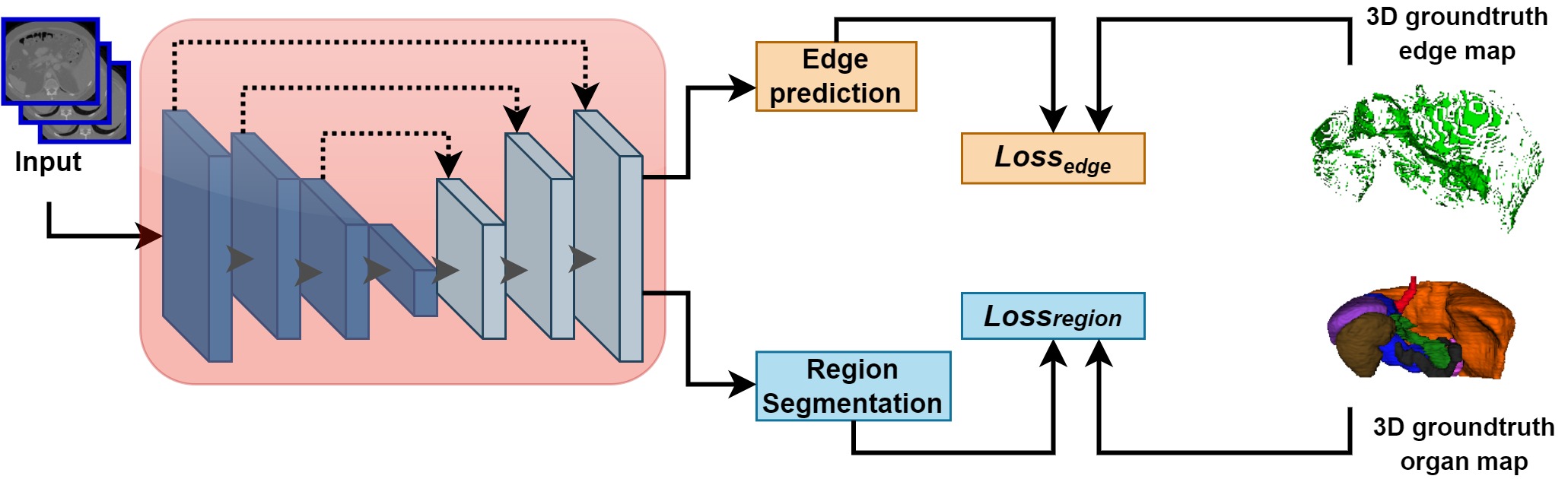} 
\caption{}
\label{fig:unet_mtl_gen1}
\end{subfigure}\hfill
\begin{subfigure}[b]{.42\textwidth}
\hspace{-5mm}
\includegraphics[width=\textwidth]{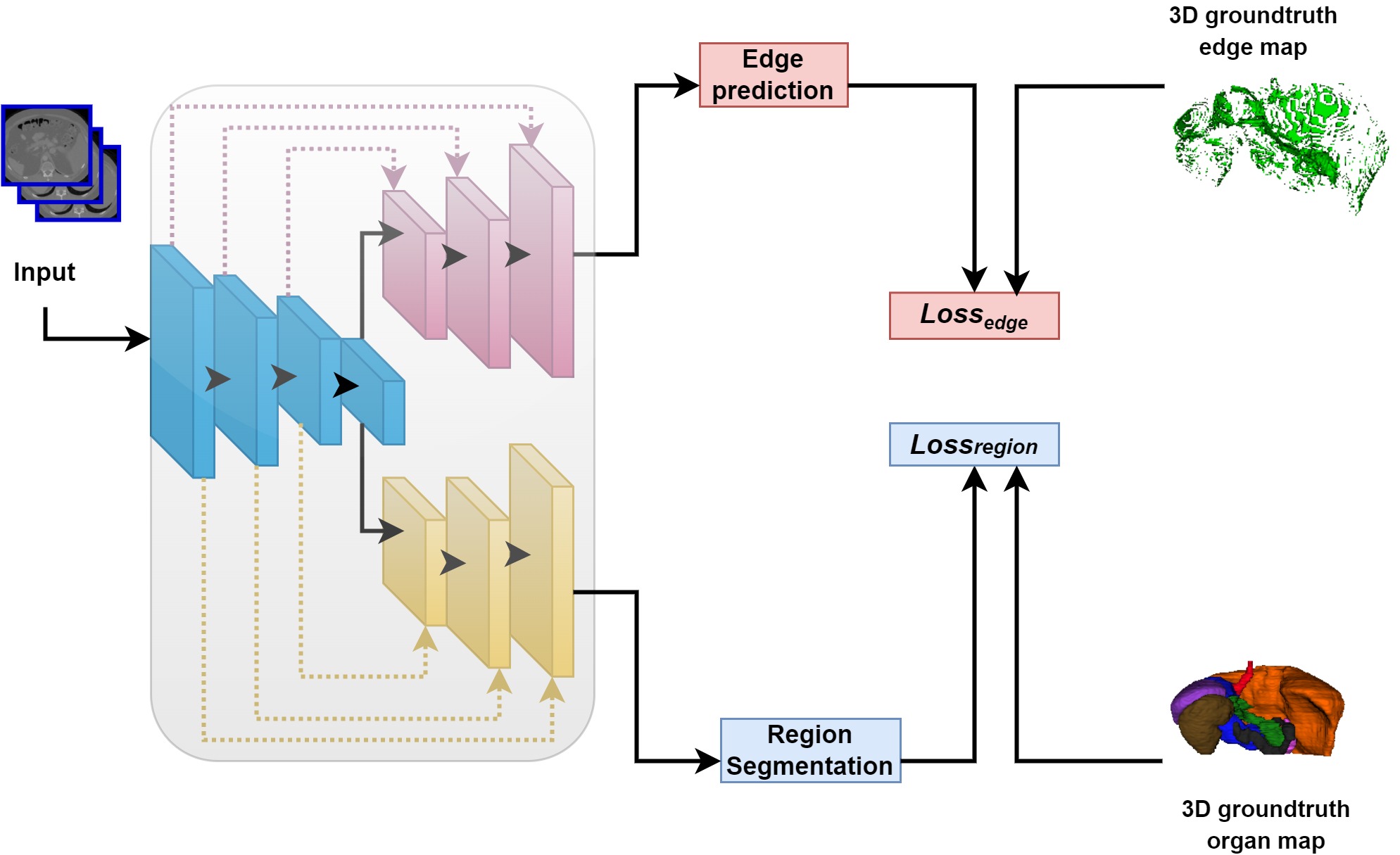}
\caption{}
\label{fig:unet_mtl_gen1k}
\end{subfigure}
\caption{Multi-task topologies of 3D boundary-constrained network. (a) Multi-task topology with shared encoder-decoder network and task-specific prediction layers, and (b) Multi-task topology with shared encoder and task-specific decoders.}
\label{fig:unet_mtl_gen2}
\end{figure*}
\section{Related Work}\label{sec:related}
Segmentation of anatomical structures from abdominal scans is a prerequisite for various high-level CT-based clinical applications. Existing computerized tools for abdominal image segmentation are either based on deep learning or non-deep learning methods. In this section, we first briefly discuss the non-deep learning methods (\autoref{sec:classic}) and then present a review of deep learning-based methods for abdominal multi-organ segmentation (\autoref{sec:dnns}). We conclude this section with a discussion on multi-task deep neural networks being employed for complementary boundary learning task to improve medical image segmentation (\autoref{sec:boundary}).
\subsection{Non-deep learning-based abdominal organs segmentation}\label{sec:classic}
Earlier methods proposed for abdominal multi-organ segmentation have primarily utilized registration-based approaches \cite{cerrolaza2015automatic},  \cite{okada2015abdominal}. Among the registration-based approaches, the widely used ones include statistical shape models \cite{cerrolaza2015automatic}, \cite{okada2015abdominal} and multi-atlas label fusion techniques \cite{xu2015efficient}, \cite{suzuki2012multi}. The development of statistical models requires registration of training images for estimating the shape or appearance of anatomical organs followed by fitting constructed models to test images for generating segmentations \cite{heimann2009statistical}, \cite{cootes2001active}. Multi atlas-based methods utilize an atlas created using multiple labelled images in the training set, and the test image is segmented by propagating the reference segmentations. Atlases are constructed by capturing the prior anatomical knowledge relevant to target organs. However, it is difficult to build an adequate model to capture the large variability of the deformable organs with limited data \cite{xu2016evaluation}. Furthermore, the performance of both these approaches is restricted by image registration accuracy. 

Registration-free approaches train a classifier using either handcrafted or learned features to segment abdominal images \cite{campadelli2009automatic}. Extraction of robust and deformation-invariant features relies on expert knowledge about abdominal organs \cite{lombaert2014laplacian}. Having the ability to learn the features automatically, FCN-based methods, have rapidly replaced the traditional solutions that require image registration or handcrafted features and have shown improved performance for abdominal CT segmentation \cite{gibson2018automatic}, \cite{bobo2018fully}, \cite{sinha2020multi}, \cite{peng2019method}. 
\subsection{Fully Convolutional Networks for abdominal multi-organ segmentation}\label{sec:dnns}
In recent years, Fully Convolution Network (FCN) and its variants (e.g., UNet \cite{ronneberger2015u}) have become a common choice for medical image segmentation. This dominancy can be attributed to their ability to learn effective task representations and efficient inference. UNet has an encoder-decoder style architecture and consists of skip connections, joining the encoding and decoding layers on the same level. Despite being trained from scratch, UNet demonstrated state-of-the-art performance for various medical image segmentation tasks \cite{chen2017rbnet}, \cite{Dong2017AutomaticBT}. Built on top of UNet, several other modified architectures were subsequently proposed, e.g., UNet$_{++}$ \cite{zhou2019unet++}, Attention-UNet \cite{schlemper2019attention}, etc. 

Existing deep learning-based studies for abdominal multi-organ segmentation have utilized 2D or 3D convolutional networks. 2D methods are less parameter-intensive; however, they cannot exploit the 3D contextual information and eventually provide sub-accurate organ-delineation performance. 3D convolutional networks are facilitated with 3D convolutions, 3D pooling, and 3D normalization to exploit the rich volumetric context and generate dense voxel-wise predictions \cite{cciccek20163d}. Advances in efficient 3D convolutional implementation and increased GPU memory have enabled the adoption of 3D convolutional models for abdominal multi-organ segmentation \cite{Zhang2020BlockLS}, \cite{hu2017automatic}.

Roth et al. \cite{Roth2018AnAO} proposed a cascaded architecture based on two 3D UNets where the first UNet is trained to separate the abdominal area from the background, and the latter utilized the output from the first UNet to simultaneously segment the abdominal organs. Peng et al. \cite{Peng2020AMO} delineated abdominal organs using 3D UNet with residual-learning based units (ResNets) to calculate patient-specific CT organ dose. In another study \cite{gibson2018automatic}, abdominal organs are segmented using a 3D FCN with dilated convolutions based densely connected units. Heinrich et al. \cite{heinrich2019obelisk} leveraged 3D deformable convolutions to spatially adapt the receptive field for abdominal multi-organ segmentation. In \cite{liu2020ct}, abdominal scans were segmented using a 3D deeply supervised patch-based UNet with grid-based attention gates to encourage the network to focus on useful salient features propagated through the skip connections. Some existing methods have employed post-processing steps, including level-sets \cite{hu2017automatic} and graph-cut \cite{article1kim} to refine initial segmentation maps obtained from 3D deep convolutional networks. 

Through the efforts mentioned above, the existing 3D methods have mostly emphasized developing better deep learning architectures and did not attempt to improve the segmentation of challenging parts of abdominal organs, e.g., voxels that belong to the contour of organs and regions within the vicinity of organ-contour. The fuzzy appearance of the boundary of organs and low contrast between the adjacent abdominal structures makes the voxels belonging to these regions more susceptible to wrong label prediction. 
\subsection{Boundary-constrained medical image segmentation}\label{sec:boundary}
Several existing deep learning-based medical image segmentation methods have utilized the boundary information of regions of interest to overcome the misprediction of boundary pixels \cite{chen2016dcan}, \cite{8363791}, \cite{8906008}, \cite{Lee_2020_CVPR}. In these methods, the networks are trained in a multi-task learning fashion to simultaneously predict the probability maps of entire organs and their corresponding boundaries. Most of these methods have resorted to the hard-parameter sharing technique, where a single network contains shared and task-specific parameters and is jointly trained to solve multiple tasks. 

Chen et al. \cite{chen2016dcan} segmented the glands and their corresponding boundaries via multi-task training. By training the model to learn the co-representations, the model achieved better gland segmentation performance than the single-task models. In \cite{Oda2018BESNetBS}, a dual-decoder-based network is presented that simultaneously detects the boundaries and predicts the semantic labels of cells. Features from the boundary-decoding path were concatenated with those learned in the entire cell region decoding path via additional skip connections. This led to the improved histopathological image segmentation performance. In \cite{Murugesan2019PsiNetSA}, boundary and distance maps were used for improved polyp and optic disk segmentation, respectively. Tan et al. \cite{8363791} proposed a multi-task medical image segmentation network consisting of a single encoder and separate dedicated arms for decoding regions and boundaries. The study was evaluated on numerous applications, including MR femur and CT kidney segmentation. Zhang et al., \cite{zhang2019net} presented a edge-based deeply supervised network for predicting the regions of interest and their corresponding boundaries. The method was validated for retinal, x-ray, and CT image segmentation. Wang et al. \cite{Wang2019AUT} proposed a two-parallel stream model in which each of the two streams was trained to segment region and detect boundary followed by fusion of contour and region prediction maps. Lee et al. \cite{Lee_2020_CVPR} proposed a framework that predicts boundary keypoint maps and makes use of adversarial loss for improved boundary preserving in medical image segmentation.

Given the challenge presented by voxels on the organs’ boundaries and the evidence in the literature that focusing on boundaries is beneficial for performance, we integrate the organs boundary prediction as an auxiliary task into the training of state-of-the-art 3D medical image segmentation networks. Since the design choice of network topology impacts the learning process, we explore two multi-task network designs and analyze their performance. The boundary co-training resulted in improved performance on abdominal CT segmentation tasks compared to the several state-of-the-art 3D fully convolutional baseline architectures.
\section{Proposed Method}\label{sec:method}
In this section, we first describe the boundary-constrained loss for training the 3D encoder-decoder network to simultaneously predict the boundaries and entire abdominal organ regions via multi-task learning (\Cref{sec:propose}), followed by an exhibition of our proposed multi-task network topologies (\Cref{sec:topo}). After that, we discuss the architecture of the 3D networks that we have as baselines in our work (\cref{sec:base}). Finally, we present the architectural design of the counterpart 3D boundary-constrained models (\Cref{sec:bound}).
\subsection{Boundary-Constrained Loss}\label{sec:propose}
Consider a 3D encoder-decoder network trained to predict the voxel labels of the abdominal CT scan with $W \times H \times Z$ dimensions, where $W$, $H$, and $Z$  denote the length, width, and depth of the scan, respectively. Such a network takes an abdominal multi-organ CT scan as an input and outputs a labelled voxel map of the same size as the input. To utilize the boundary information of abdominal organs for improved representation learning, we train the network to predict the 3D organ-semantic masks and 3D organ-boundaries in one forward propagation pass. We formulate this problem using a multi-task learning paradigm where multiple tasks are learned jointly using shared and task-specific representations. The loss $\mathcal{L}$ for this multi-task learning problem is a weighted combination of per-task losses, organ segmentation loss $\mathcal{L}_{RS}$ and organ boundary detection loss $\mathcal{L}_{BD}$. We use multi-class dice loss \cite{milletari2016v} for evaluating the performance of the multi-organ segmentation task, given as
\begin{align}\label{eq:eq3.3}
\mathcal{L}_{RS} &= \sum\limits_{c=0}^{C-1}\; \frac {2\;(\hat y_{i,c} \times y_{i,c})}{\hat y_{i,c}^{2} \; + \; y_{i,c}^{2}}
\end{align}
where, $\hat y_{i,c}$ and $y_{i,c}$ denote the 3D multi-organ probability map and ground-truth mask, respectively, of the $i^{th}$ abdominal CT scan. $C$ denotes the number of organ classes.
\begin{equation}\label{eq:eq3.4}
\hat y_{i} \; = \; \hat p(x_{i};\boldsymbol{\theta_{s}})
\end{equation}
\vspace{-6mm}
\begin{equation}\label{eq:eq3.5}
\hat p(x_{i};\boldsymbol{\theta_{s}}) \; = \; \sum\limits_{n=0}^{N-1}\;\hat p(x_{i,n};\boldsymbol{\theta_{s}}) =\;\sum\limits_{w=0}^{W-1} \sum\limits_{h=0}^{H-1} \sum\limits_{z=0}^{Z-1}\;\hat p(x_{i,w,h,z};\boldsymbol{\theta_{s}})
\end{equation}
where $\hat p(x_{i,n})$ represents the label probability of $n^{th}$ voxel in $i^{th}$ scan and $N$ refers to the total number of voxels in a scan. 

To evaluate the model's performance in predicting the boundaries, we use binary cross-entropy loss (shown in Eq. \autoref{eq:eq3.6}). Binary cross-entropy loss for predicting 3D boundaries is given as
\begin{equation}\label{eq:eq3.6}
\begin{aligned}
\mathcal{L}_{BD} &= -\sum\limits_{n=0}^{N-1}\;e_{i,n}\log(\hat e_{i,n}) + (1-e_{i,n})\log(1-\hat e_{i,n})\\
&= -\;\sum\limits_{w=0}^{W-1} \sum\limits_{h=0}^{H-1} \sum\limits_{z=0}^{Z-1}\;p(x_{i,w,h,z};\boldsymbol{\theta_{s}})\log(\hat p(x_{i,w,h,z};\boldsymbol{\theta_{s}}))\\
&+ (1-p(x_{i,w,h,z};\boldsymbol{\theta_{s}}))\log(1-\hat p(x_{i,w,h,z};\boldsymbol{\theta_{s}}))
\end{aligned}
\end{equation}
$\hat e_{i}$ and $e_{i}$ represent the edge probability map and the corresponding ground-truth. $\hat p(x_{i,w,h,z}$) represents the edge probability of the $n^{th}$ voxel in $i^{th}$ scan. $\boldsymbol{\theta_{s}}$ represents the weights of the entire deep multi-task encoder-decoder network.

The combined total loss $\mathcal{L}$ is minimized with respect to the parameters $\boldsymbol{\theta_{s}}$, as shown in \autoref{eq:eq3.1}. Thus our goal is to evaluate if a network can learn more robust features and subsequently produce improved organ segmentations by being trained to explicitly recognize the boundaries.
\begin{align}\label{eq:eq3.1}
\mathcal{L}(\boldsymbol{\theta_{s}}) \;=\; \sum\limits_{i=1}^M \mathcal{L}_{RS}\; +\; 	\lambda \; \sum\limits_{i=1}^M \mathcal{L}_{BD}
\end{align}
$M$ and $\lambda$ represents the total number of CT scans in the training set and the weight assigned to the edge detection loss in \autoref{eq:eq3.1}, respectively.

We hypothesize that the additional boundary loss ($\mathcal{L}_{BD}$) would impose a larger penalty on erroneous contour voxels, and it subsequently pushes the optimization of the segmentation network towards the solutions with more accurate boundaries. Thus, one would potentialize the ability of a boundary-constrained network to extract features that account for the semantic abdominal organ regions and boundaries. 
\subsection{Boundary-Constrained Network Topologies}\label{sec:topo}
Multi-task learning is generally formulated via hard-parameter sharing and soft-parameter sharing. In the hard-parameter sharing paradigm, multiple tasks share a subset of jointly optimized parameters, whereas task-specific parameters are optimized separately. In soft-parameter sharing, each task is parameterized using its own set of parameters which are jointly regularized using constraints \cite{Ruder2017AnOO}. In practice, hard-parameter sharing approaches incur much less parameter and computational cost. In our work, we formulate the multi-task learning problem via hard-parameter sharing to train the encoder-decoder network to do multiple tasks, i.e., organ segmentation and boundary detection. For deep neural networks, the hard-parameter sharing approach is realized by sharing some network layers between the tasks while keeping some layers task-specific.  

We explore two different network topologies to conduct multi-task training, as shown in \Cref{fig:unet_mtl_gen1,fig:unet_mtl_gen1k}. The motivation to explore multiple topologies is to investigate the impact of sharing the larger and smaller number of parameters in the network between the two tasks. We explain these multi-task topologies below.
\subsubsection{Task-Specific Output Layers (TSOL)}\label{sec:tsl}
The first multi-task topology that we explore is formulated by appending two separate prediction layers for predicting the boundaries and semantic organ masks. This topology employs an encoder-decoder network whose weights are shared between the tasks, except for the last output layers, as shown in \autoref{fig:unet_mtl_gen1}. Technically, it encourages the use of compact and tightly shared feature representations. As evident, this configuration has negligibly fewer more parameters than the single-task network. We denote this configuration as TSOL. 
\subsubsection{Task-Specific Decoders (TSD)}\label{sec:tsd}
In second mutli-task topology, we modify the 3D encoder-decoder model to have a single shared encoder but two separate decoding arms for predicting the semantic regions and boundaries. The sibling-decoding arms upsample the region and boundary maps separately. This type of formulation ensures sparse representation sharing amongst the two tasks since decoders have been parameterized separately, as shown in \autoref{fig:unet_mtl_gen1k}. The presence of two synthesis paths results in having significantly more parameters than its counterpart single-task network. We refer to this configuration as TSD.  
\begin{figure}[!hbt]
\centering
\includegraphics[scale=0.14]{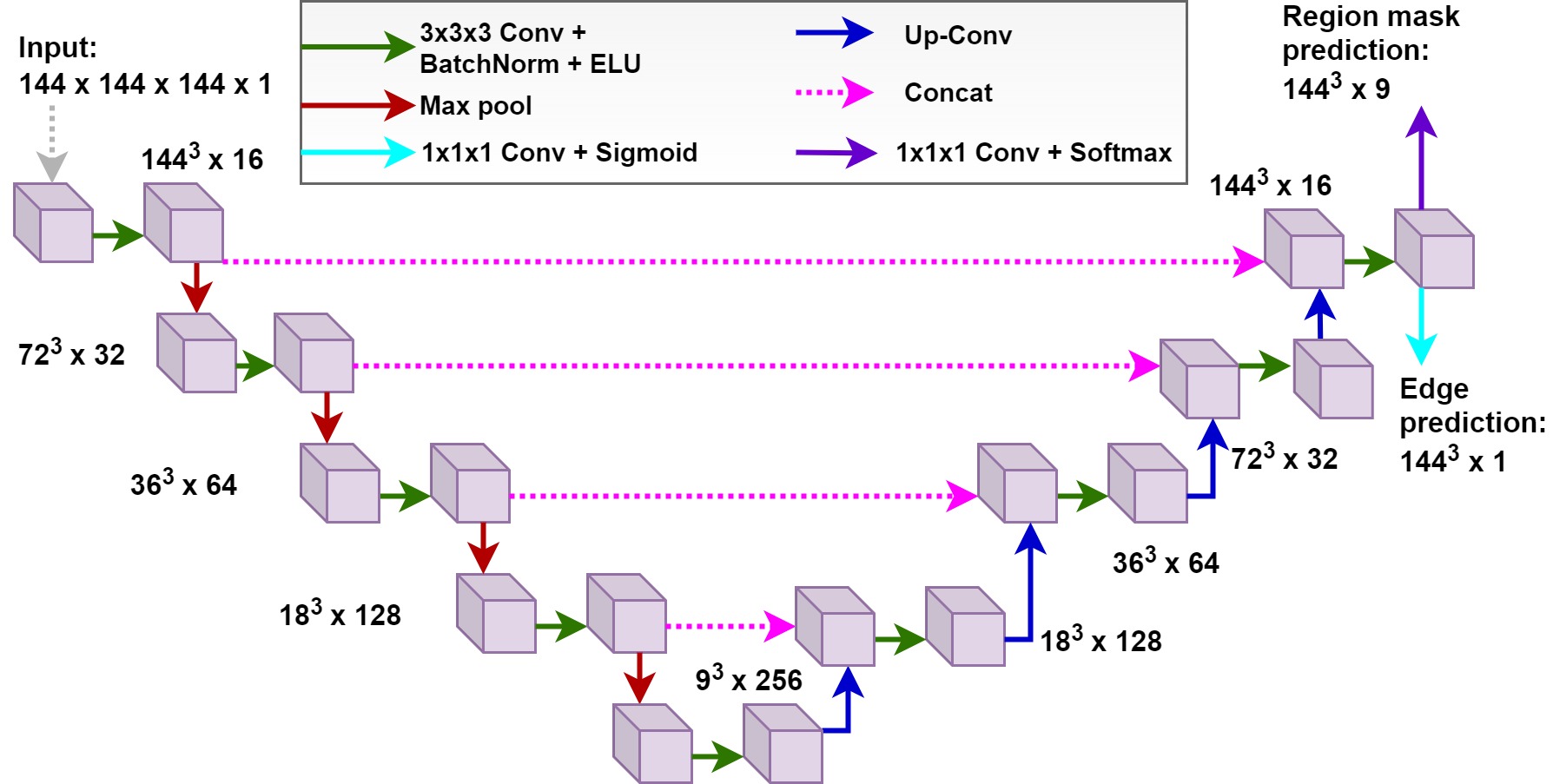} 
\caption{3D UNet-MTL-TSOL: Multi-task learning based 3D UNet with task-specific output layers for simultaneously predicting the organs and their boundaries.}
\label{fig:unetmtl1}
\end{figure}
\begin{figure}[!hbt]
\centering
\includegraphics[scale=0.13]{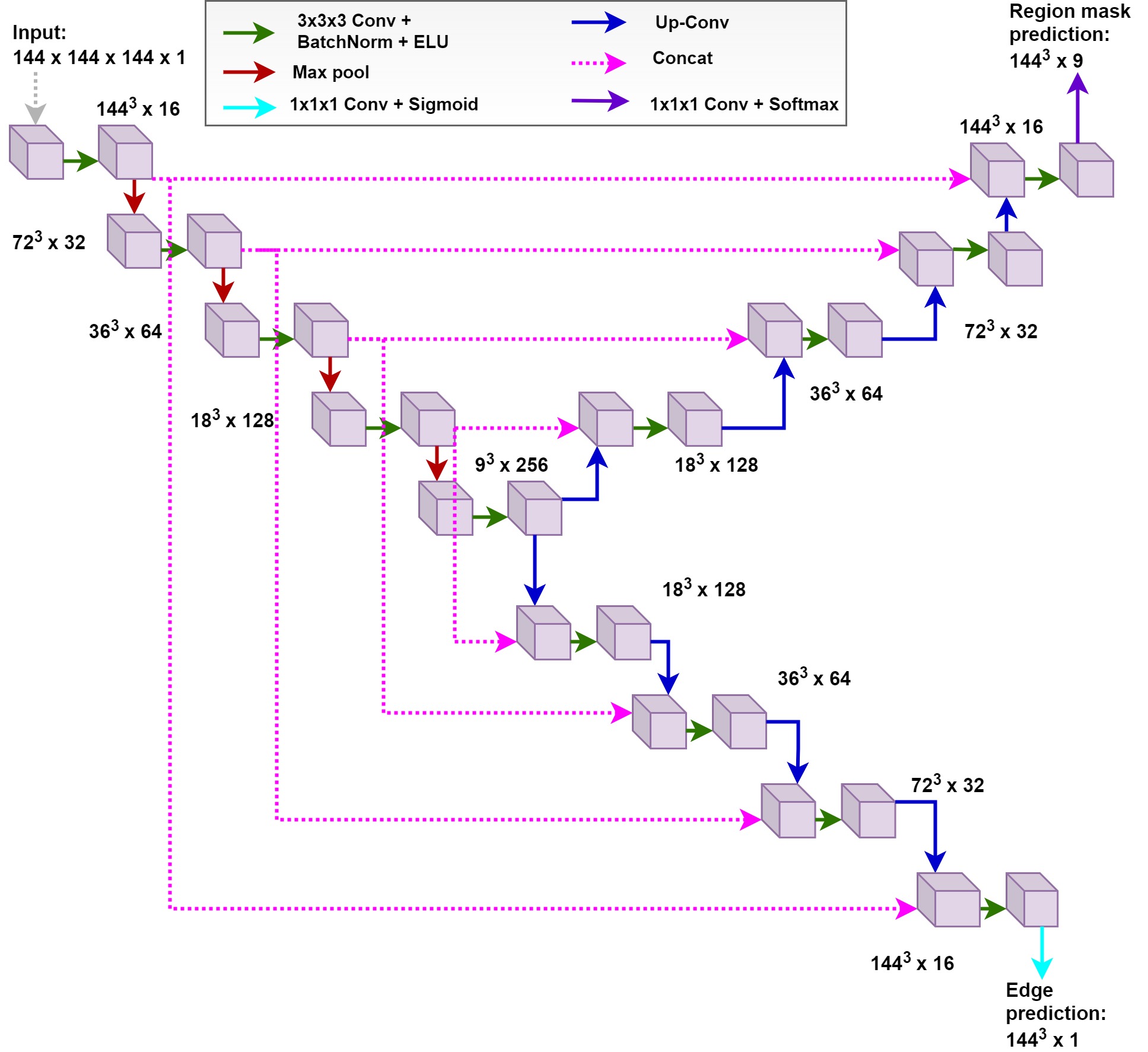} 
\caption{3D UNet-MTL-TSD: Multi-task learning based 3D UNet with task-specific decoders for simultaneously predicting the organs and their boundaries.}
\label{fig:unetmtl2}
\end{figure}
\begin{figure}[!hbt]
\centering
\includegraphics[scale=0.13]{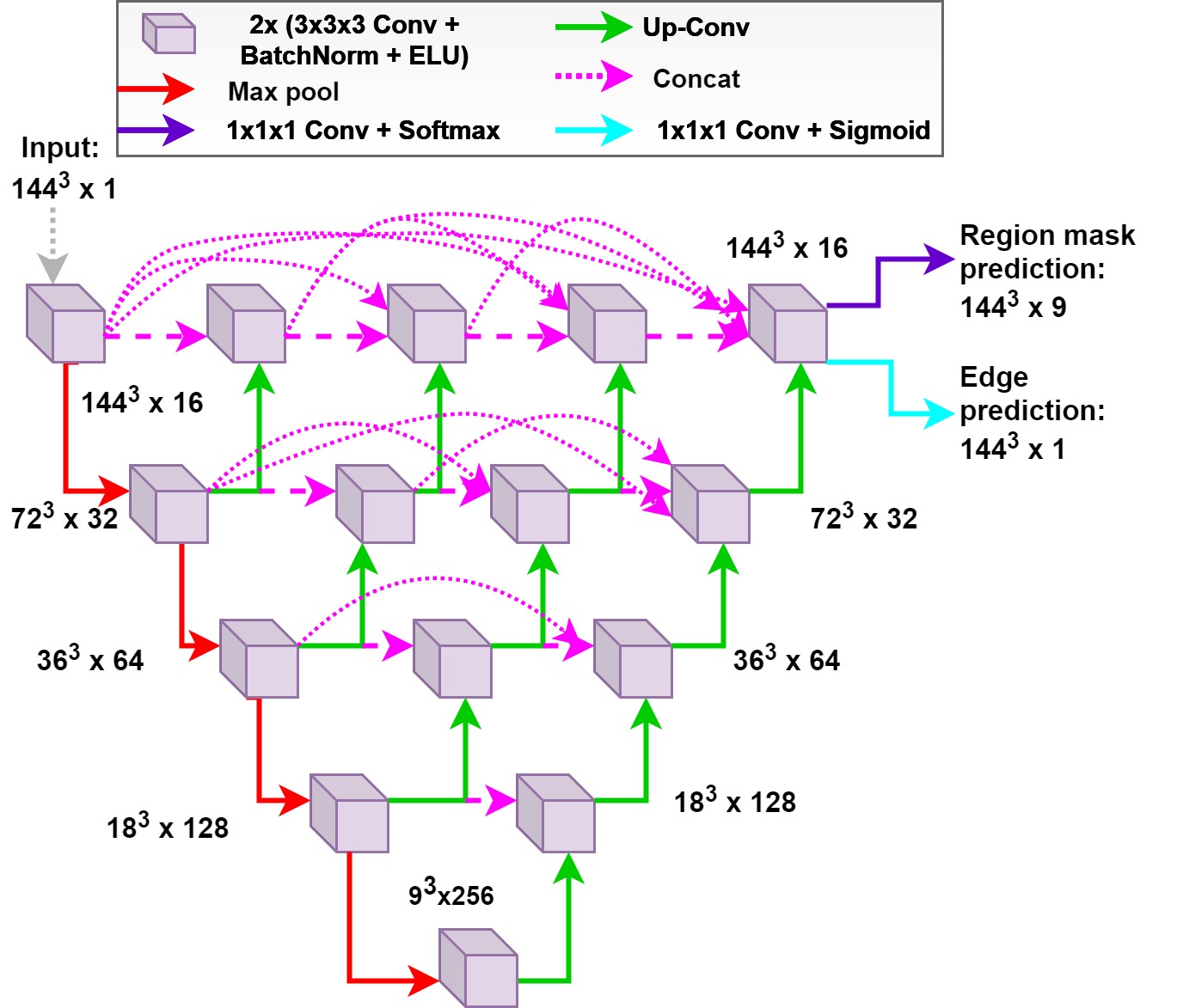} 
\caption{3D UNet$_{++}$-MTL-TSOL: Multi-task learning based 3D UNet$_{++}$ with task-specific layers for simultaneously predicting the organs and their boundaries.}
\label{fig:unetplusmtl1}
\end{figure}
\begin{figure}[!hbt]
\centering
\includegraphics[scale=0.13]{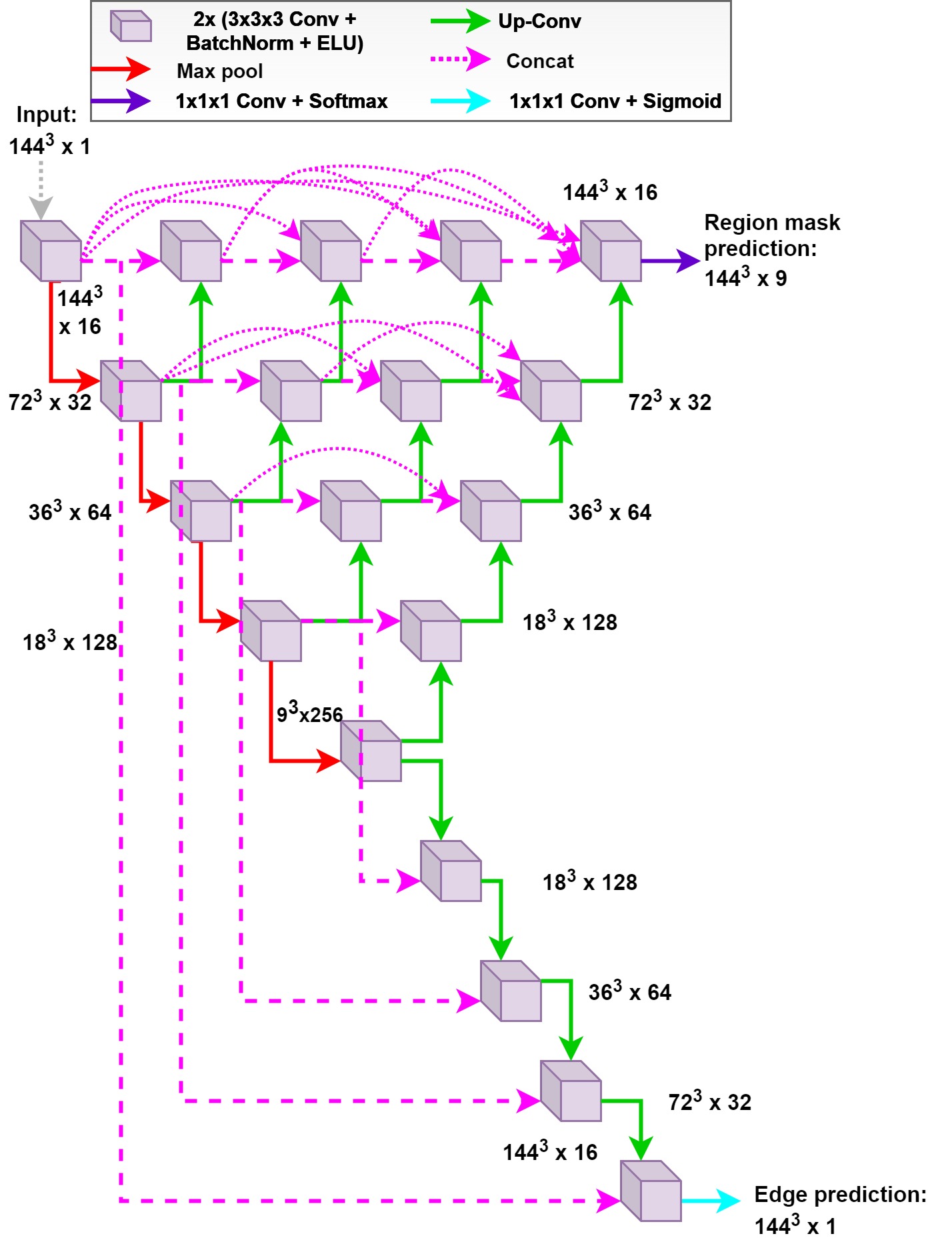} 
\caption{3D UNet$_{++}$-MTL-TSD: Multi-task learning based 3D UNet$_{++}$ with task-specific decoders for simultaneously predicting the organs and their boundaries.}
\label{fig:unetplusmtl2}
\end{figure}
\subsection{3D Baseline models}\label{sec:base}
We use UNet \cite{cciccek20163d}, UNet$_{++}$ \cite{zhou2019unet++} and Attention-UNet (Att-UNet) \cite{schlemper2019attention} as our baseline models. We illustrate these models in Supplementary material (see Figures S1 - S3). These architectures are based on encoder-decoder design and extended to segment 3D volumes by replacing the 2D convolutions, pooling, and upsampling with 3D counterparts. Each baseline model processes a 3D input scan with dimensions $W \times H \times Z$ and outputs a predicted organ-label map of the same size as input. The encoder of the model contains five convolutional blocks with pooling layers, and the decoder comprises four upsampling layers. Each convolutional block in the encoder consists of two convolutional layers with $3 \times 3 \times 3$ filters, followed by batch normalization and Exponential Linear Unit (ELU) activation \cite{DBLP:journals/corr/ClevertUH15}. We use padded convolutions to keep the output dimensions of convolutional layers the same as the input dimensions. A $2 \times 2 \times 2$ max pooling layer with a stride of two in each dimension is sandwiched between every two convolutional blocks for feature maps’ downsampling. The bilinear interpolation layers are used in the decoder to upsample the extracted feature maps in each dimension. The feature maps in the decoder are concatenated with the equal-sized representations learned in the encoder via skip connections. The concatenated feature maps are then transformed using convolutional blocks, similar to those used in the encoder. The last $1 \times 1 \times 1$ convolutional layer maps the feature channels to the class labels, followed by a \textit{softmax} activation.

The resolution of the smallest feature map is $9 \times 9 \times 9$ and the minimum and maximum feature count at the first and last encoding stage is 16 and 256, respectively. Note that the original UNet$_{++}$ model is trained with deep supervision driven by output layers of UNet with varying depths; however, we train UNet$_{++}$ without deep supervision to constrain the computational expense.
\begin{figure}[!hbt]
\centering
\includegraphics[scale=0.13]{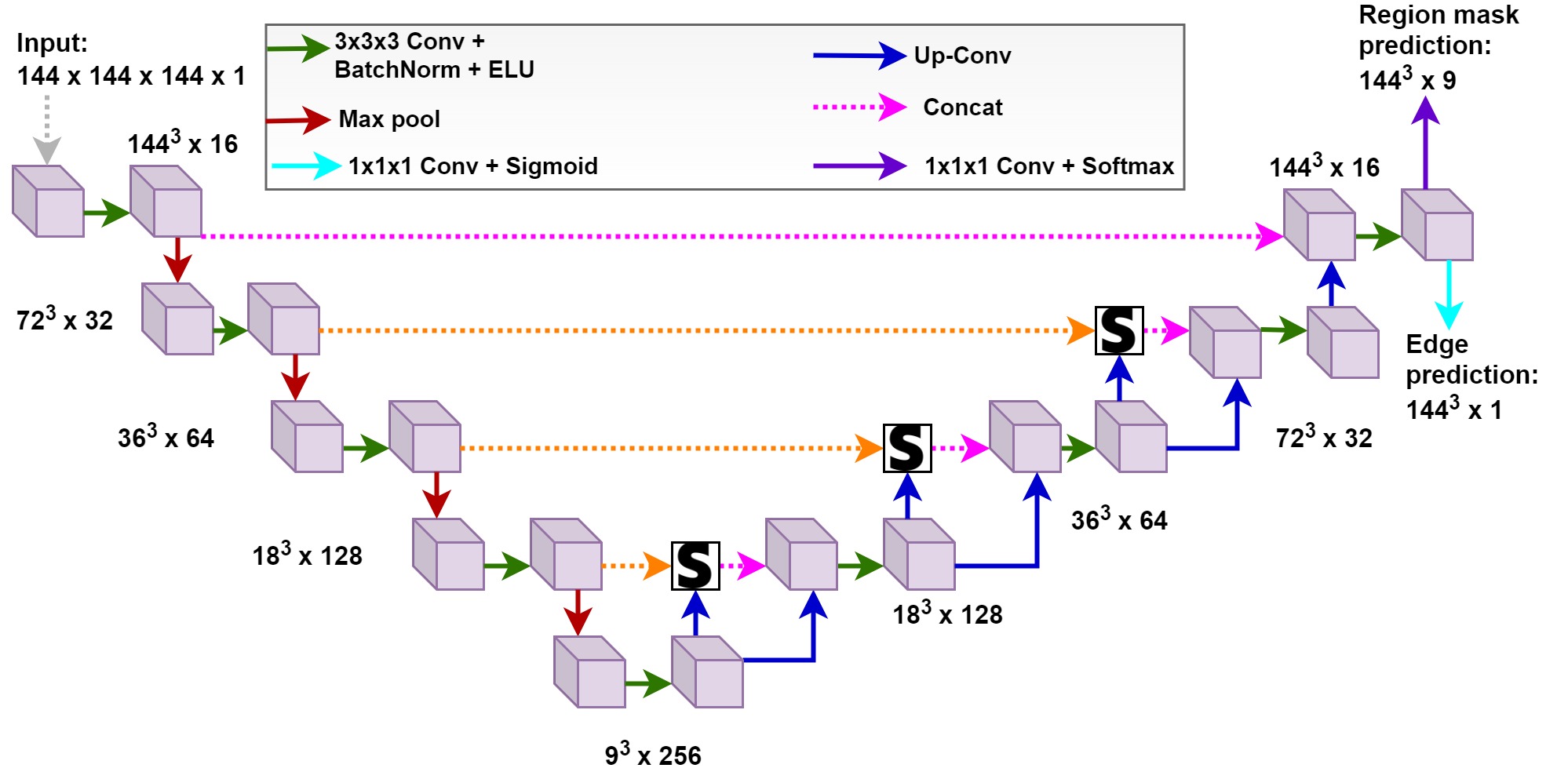} 
\caption{3D Att-UNet-MTL-TSOL: Multi-task learning based 3D Att-UNet with task-specific prediction layers for simultaneously predicting the organs and their boundaries.}
\label{fig:attenunetmtl1}
\end{figure}
\begin{figure}[!hbt]
\centering
\includegraphics[scale=0.13]{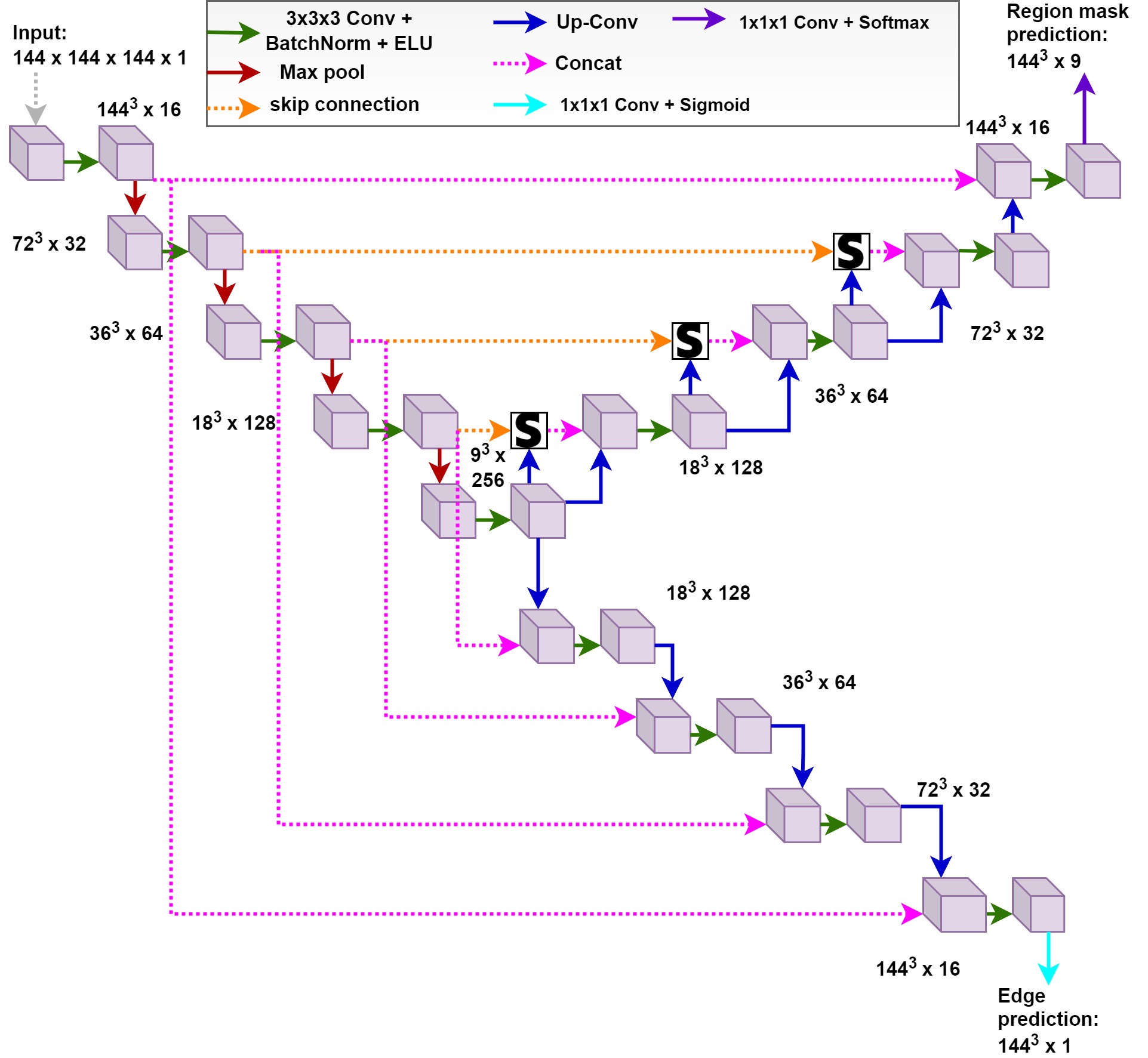} 
\caption{3D Att-UNet-MTL-TSD: Multi-task learning based 3D Att-UNet with task-specific decoders for simultaneously predicting the organs and their boundaries.}
\label{fig:attenunetmtl2}
\end{figure}
\subsection{3D Boundary-contrained models}\label{sec:bound}
To utilize the boundary information of the organs, we train the baseline models (given in \Cref{sec:base}) to predict the organ boundaries along with organs. We propose two multi-task network topologies (shown in \autoref{fig:unet_mtl_gen2}) for integrating the boundary information and for that, we modify each baseline model, i.e., 3D UNet, 3D UNet$_{++}$, and 3D Att-UNet, according to two multi-task learning-based topologies. 

To modify the baseline models according to the first multi-task topology (TSOL) (shown in \autoref{fig:unet_mtl_gen1}), we append a separate head at the end to predict boundaries along with the organs. We refer to these models as UNet-MTL-TSOL, UNet$_{++}$-MTL-TSOL, and Att-UNet-MTL-TSOL, as shown in \Cref{fig:unetmtl1,fig:unetplusmtl1,fig:attenunetmtl1}, respectively. To modify the baseline models according to the second boundary-constrained multi-task topology (TSD) (shown in \autoref{fig:unet_mtl_gen1k}), we modify the baseline models (3D UNet, 3D UNet$_{++}$, and 3D Att-UNet) to have separate decoding paths followed by prediction layers for predicting boundaries and organs. We refer to the models modified according to this topology as UNet-MTL-TSD, UNet$_{++}$-MTL-TSD, and Att-UNet-MTL-TSD and show them in \Cref{fig:unetmtl2,fig:unetplusmtl2,fig:attenunetmtl2}, respectively. Note from \autoref{fig:unetplusmtl2}, we only use the skip connections between the encoder with the greatest depth and boundary-decoder instead of utilizing feature maps extracted by nested-UNets of all depths. This design choice is made to constrain the number of parameters in UNet$_{++}$-MTL-TSD. Furthermore, observe from \autoref{fig:attenunetmtl2}, we do not employ the attention mechanism while decoding the boundary-features in Att-UNet-MTL-TSD.
\section{Experimental details}\label{sec:experiment}
This section first describes the datasets used to validate our study and the pre-processing we perform on the datasets (\Cref{sec:data}), followed by implementation details (\Cref{sec:impl}). Finally, we conclude this section by discussing the metrics used to evaluate baseline and boundary-constrained models (\Cref{sec:metrics}).
\subsection{Description of datasets and data preprocessing}\label{sec:data}
We utilize two publically available abdominal CT datasets (Pancreas-CT and BTCV) to evaluate baseline and boundary-constrained models. Abdominal scans in Pancreas-CT were acquired at the National Institutes of Health Clinical Center from pre-nephrectomy healthy kidney donors and subjects with neither major abdominal pathologies nor pancreatic cancer lesions \cite{Roth2016}. The BTCV dataset consists of abdominal scans acquired at the Vanderbilt University Medical Center from metastatic liver cancer patients or post-operative ventral hernia patients \cite{xu2016evaluation}.
\subsubsection{Pancreas-CT Dataset (TCIA-43)}
The pancreas-CT dataset \cite{Roth2016}, \cite{RothLFSLTS15} comprised 82 abdominal contrast-enhanced 3D CT scans and was initially provided with manually drawn contours of the pancreas \cite{Clark2013}, \cite{RothLFSLTS15}. Recently, 43 scans from this dataset have been re-annotated to include the segmentation of the liver, duodenum, stomach, esophagus, spleen, gallbladder, and left kidney \cite{Gibson2018}. Therefore, we use only 43 scans that have been re-annotated to incorporate labels for multiple organs. 

We first crop a region-of-interest from the CT scans and the corresponding ground-truth labels using the bounding box coordinates provided with the dataset \cite{Gibson2018}. The cropping step ensures the models are only fed with the foreground inputs without the redundant background region. The cropped region-of-interest from the CT scans and ground-truth labels are then resampled to a common dimension of $144 \times 144 \times 144$ voxels. We randomly divide the available 43 studies into 28, 5, and 10 for training, validation, and test, respectively.
\subsubsection{BTCV Dataset}
BTCV was released \cite{landmanbvc}, \cite{xu2016evaluation} as a part of a challenge held in conjunction with MICCAI 2015. The challenge compared the abdominal organs' segmentation algorithms on 3D CT scans. Our work focuses on the segmentation of eight organs from the BTCV dataset, i.e., liver, duodenum, stomach, esophagus, spleen, gallbladder, left kidney, and pancreas. 

For the BTCV dataset, we utilize the bounding box coordinates given with the dataset for cropping the region-of-interest for both the CT scans and ground-truth labels \cite{Gibson2018}. Like the Pancreas-CT dataset, the cropped region-of-interest is then resampled to a common dimension of $144 \times 144 \times 144$ voxels. Finally, we randomly divide the available 47 studies into 32, 5, and 10 for training, validation, and test, respectively. 

\begin{figure}[!hbt]
\centering
\includegraphics[scale=0.5]{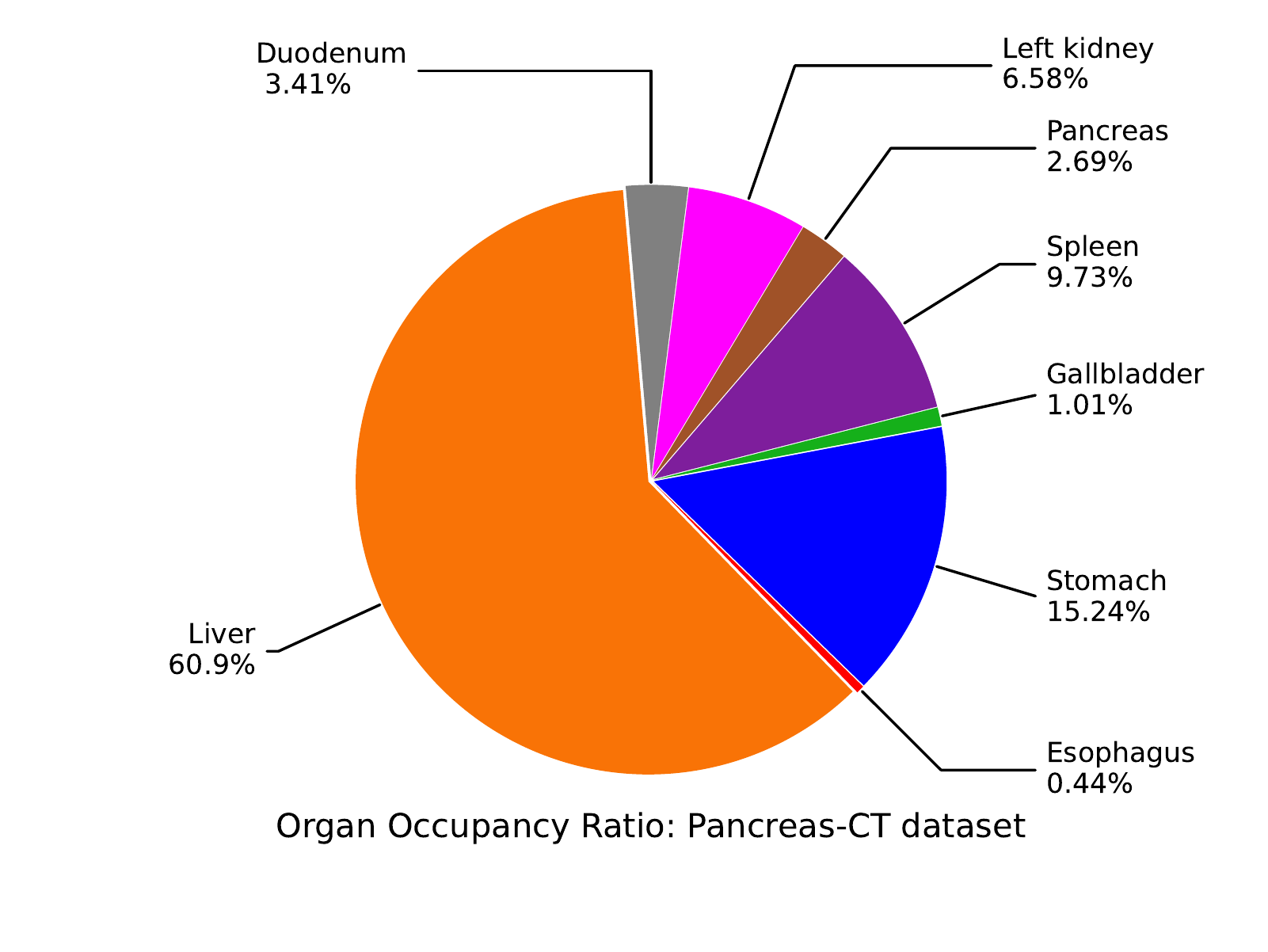} 
\caption{Organ occupancy ratio for Pancreas-ct dataset.}
\label{fig:organ_btvc}
\end{figure}

\begin{figure}[!hbt]
\centering
\includegraphics[scale=0.5]{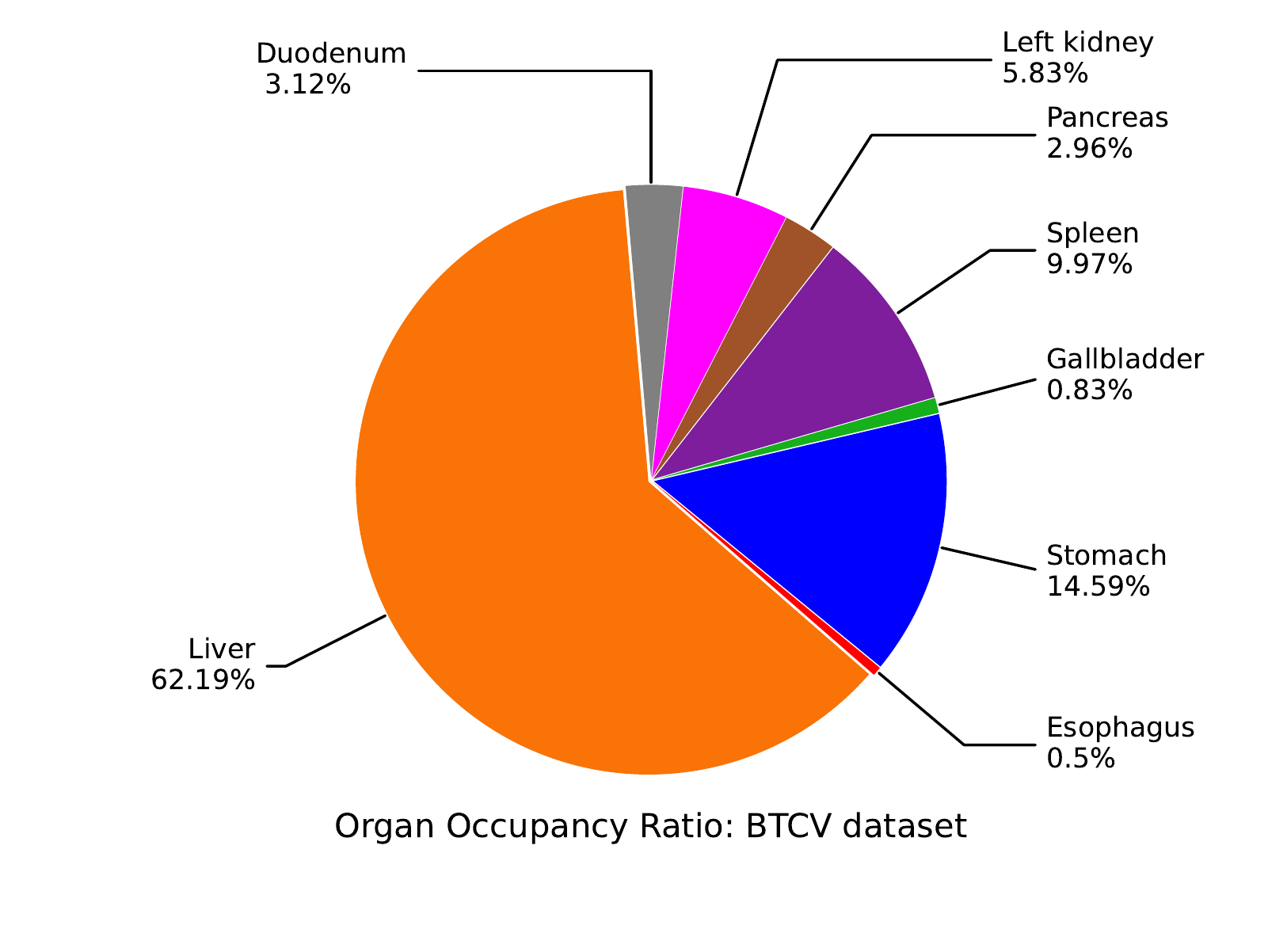} 
\caption{Organ occupancy ratio for BTCV dataset.}
\label{fig:organ_panc}
\end{figure}

For both Pancreas-CT and BTCV dataset, we applied affine random transformations to augment the data but did not observe a significant difference in segmentation performance on the validation set. Therefore, we did not use any data augmentation. We used the same dataset splits for all the experiments. To analyze the occurrence of each organ in the dataset, we present the organs' occupancy ratio in \Cref{fig:organ_btvc,fig:organ_panc}.
\subsection{Implementation details}\label{sec:impl}
All experiments are conducted using Pytorch \cite{paszke2017automatic} on two Nvidia Tesla P100, accessed through the HPC platform\footnote{\url{https://supercomputing.swin.edu.au/}}. We train all the baseline networks with a mini-batch of size 4, except for 3D UNet$_{++}$, which is trained with one batch size. These choices have been made according to available GPU memory. All baseline (single-task) networks are trained using multi-class dice loss as expressed in \Cref{eq:eq3.3,eq:eq3.4,eq:eq3.5}. For acquiring multi-class dice loss, the 3D predicted organ-segmentation maps are compared with 3D organ ground-truth maps. All the networks (baseline and boundary-constrained) are optimized via Adam optimizer \cite{kingma2014adam}. The learning rate is initially set to 0.001, decaying by a factor of 0.9 after every epoch. To assess the effect of changing values of training hyperparameters on validation segmentation performance, we conduct experiments to guide us in selecting the optimal settings for baseline models. The results for these experiments are given in Tables S1 and S2 in Supplementary section. We monitor the mean dice score on the validation set during training and utilize the model for testing that results in the highest dice coefficient on the validation set. 

We use a combination of multi-class dice loss and binary cross-entropy loss, as illustrated in \autoref{eq:eq3.1} for training boundary-constrained models. 3D organ-boundary predictions are compared with the 3D boundary ground-truths to obtain the binary cross-entropy loss. Since the datasets do not contain the boundary annotations of organs, we acquire the ground-truth boundaries by first eroding the multi-organ ground-truth labels and then taking the difference from the original ground-truth map. This process gives us boundary annotations of organs. For UNet-MTL-TSOL, UNet-MTL-TSD, Att-UNet-MTL-TSOL and Att-UNet-MTL-TSD, we use a batch of size two, whereas, for UNet$_{++}$-MTL-TSOL and UNet$_{++}$-MTL-TSD, we use a single batch size. These choices are made according to the available GPU memory. Note that the boundary predictions are only used in the training stage. During the validation stage, we only consider the organ predictions. 

We conduct a grid search (on the range from 0 to 2 with a step of 0.5) to find the optimal value of $\lambda$ (responsible for balancing the boundary detection loss). The exact value of $\lambda$ selected to balance the boundary loss for each boundary-constrained model are given in Tables S3 and S4 in Supplementary material. These tables also present the standard deviation between validation dice scores when different values of $\lambda$ are used. 
\subsection{Evaluation metrics}\label{sec:metrics}
We compare the predicted segmentation masks with the ground-truths to evaluate the segmentation performance of baseline and boundary-constrained models on test set. To do that, we utilize Dice Score, Recall, Precision, and Average Hausdorff Distance as metrics for assessing the quality of predicted segmentation masks. These metrics are calculated for each organ individually and then an average is taken across all subjects. All metrics are calculated by taking a mean of the 5 runs.
\begin{table*}[ht] 
\centering
\resizebox{\columnwidth}{!}{%
\caption{Quantitative comparison between baseline and boundary-contrained models for abdominal multi-organ segmentation: Mean values \textpm\ Std. Dev. of Dice Score, Avg. HD, Recall, and Precision are shown. Results in \textbf{BOLD} indicate the best results corresponding to each baseline.}\label{table:Table 3.1}
\begin{tabular}{lllll@{\hspace{10mm}}llll}
\hline
\mystrut &\multicolumn{4}{c}{\textbf{Pancreas-CT dataset}}&\multicolumn{4}{c}{\textbf{BTCV dataset}}\\
\cline{2-9}
\mystrut \textbf{Method} & \makecell[c]{\textbf{Dice score}}& \makecell[c]{\textbf{Avg. Hd}}& \makecell[c]{\textbf{Recall}}& \makecell[c]{\textbf{Precision}}&  \makecell[c]{\textbf{Dice score}}& \makecell[c]{\textbf{Avg. Hd}}& \makecell[c]{\textbf{Recall}}& \makecell[c]{\textbf{Precision}}\\
\cline{2-9}
\mystrut 3D UNet  &    \makecell[c]{0.799\textsubscript{\textpm\ 0.013}} & \makecell[c]{0.795\textsubscript{\textpm\ 0.088}}  & 
\makecell[c]{0.786\textsubscript{\textpm\ 0.049}} & \makecell[c]{0.850\textsubscript{\textpm\ 0.082}}  &
\makecell[c]{0.753\textsubscript{\textpm\ 0.008}} & \makecell[c]{1.085\textsubscript{\textpm\ 0.098}}&
\makecell[c]{0.760\textsubscript{\textpm\ 0.085}} & \makecell[c]{0.769\textsubscript{\textpm\ 0.116}}\\
\mystrut 3D UNet-MTL-TSOL &   \makecell[c]{0.813\textsubscript{\textpm\ 0.009}}& \makecell[c]{0.748\textsubscript{\textpm\ 0.130}}   &  
\makecell[c]{\textbf{0.807\textsubscript{\textpm\ 0.055}}}& \makecell[c]{0.849\textsubscript{\textpm\ 0.079}}         &                                        
\makecell[c]{0.775\textsubscript{\textpm\ 0.004}} & \makecell[c]{0.989\textsubscript{\textpm\ 0.095}}&
\makecell[c]{\textbf{0.770\textsubscript{\textpm\ 0.086}}} & \makecell[c]{0.799\textsubscript{\textpm\ 0.106}}\\
 \mystrut 3D UNet-MTL-TSD&   \makecell[c]{\textbf{0.814\textsubscript{\textpm\ 0.005}}} & \makecell[c]{\textbf{0.703\textsubscript{\textpm\ 0.059}}} &      
\makecell[c]{0.798\textsubscript{\textpm\ 0.053}}& \makecell[c]{\textbf{0.859\textsubscript{\textpm\ 0.079}}}&
\makecell[c]{\textbf{0.776\textsubscript{\textpm\ 0.002}}} & \makecell[c]{\textbf{0.918\textsubscript{\textpm\ 0.047}}}&
\makecell[c]{0.764\textsubscript{\textpm\ 0.081}} & \makecell[c]{\textbf{0.802\textsubscript{\textpm\ 0.107}}}\\\hline
\mystrut 3D UNet$_{++}$ &    \makecell[c]{\textbf{0.772\textsubscript{\textpm\ 0.009}}}&\makecell[c]{1.369\textsubscript{\textpm\ 0.114}}&                                    
\makecell[c]{\textbf{0.760\textsubscript{\textpm\ 0.089}}}&\makecell[c]{0.825\textsubscript{\textpm\ 0.106}}&
\makecell[c]{0.715\textsubscript{\textpm\ 0.011}} & \makecell[c]{1.419\textsubscript{\textpm\ 0.088}}&
\makecell[c]{0.707\textsubscript{\textpm\ 0.078}} & \makecell[c]{0.781\textsubscript{\textpm\ 0.136}}\\
\mystrut 3D UNet$_{++}$-MTL-TSOL &   \makecell[c]{\textbf{0.772\textsubscript{\textpm\ 0.014}}}	&\makecell[c]{\textbf{1.170\textsubscript{\textpm\ 0.161}}}&                                                       \makecell[c]{0.754\textsubscript{\textpm\ 0.073}}&\makecell[c]{0.829\textsubscript{\textpm\ 0.110}}&
\makecell[c]{\textbf{0.741\textsubscript{\textpm\ 0.011}}} & \makecell[c]{\textbf{1.189\textsubscript{\textpm\ 0.099}}}&
\makecell[c]{\textbf{0.733\textsubscript{\textpm\ 0.080}}} & \makecell[c]{\textbf{0.789\textsubscript{\textpm\ 0.121}}}\\
\mystrut 3D UNet$_{++}$-MTL-TSD &  \makecell[c]{0.771\textsubscript{\textpm\ 0.008}}   &  \makecell[c]{1.308\textsubscript{\textpm\ 0.269}}&                                                  
\makecell[c]{0.751\textsubscript{\textpm\ 0.077}}   &  \makecell[c]{\textbf{0.832\textsubscript{\textpm\ 0.111}}}&
\makecell[c]{0.723\textsubscript{\textpm\ 0.008}} & \makecell[c]{1.349\textsubscript{\textpm\ 0.027}}&
\makecell[c]{0.708\textsubscript{\textpm\ 0.084}} & \makecell[c]{0.783\textsubscript{\textpm\ 0.127}}\\\hline
\mystrut 3D Att-UNet  &  \makecell[c]{0.792\textsubscript{\textpm\ 0.015}} & \makecell[c]{0.825\textsubscript{\textpm\ 0.082}} &                 
\makecell[c]{0.788\textsubscript{\textpm\ 0.054}} & \makecell[c]{0.831\textsubscript{\textpm\ 0.093}}&
\makecell[c]{0.752\textsubscript{\textpm\ 0.008}} & \makecell[c]{1.313\textsubscript{\textpm\ 0.238}}&
\makecell[c]{0.754\textsubscript{\textpm\ 0.091}} & \makecell[c]{0.778\textsubscript{\textpm\ 0.112}}\\
\mystrut 3D Att-UNet-MTL-TSOL&  \makecell[c]{\textbf{0.820\textsubscript{\textpm\ 0.004}}}    & \textbf{\makecell[c]{0.673\textsubscript{\textpm\ 0.035}}}&                                                          \makecell[c]{\textbf{0.822\textsubscript{\textpm\ 0.050}}}    & \makecell[c]{0.839\textsubscript{\textpm\ 0.076}}&
\makecell[c]{0.769\textsubscript{\textpm\ 0.008}} & \makecell[c]{1.065\textsubscript{\textpm\ 0.175}}&
\makecell[c]{\textbf{0.769\textsubscript{\textpm\ 0.086}}} & \makecell[c]{0.790\textsubscript{\textpm\ 0.095}}\\
\mystrut 3D Att-UNet-MTL-TSD&  \makecell[c]{0.807\textsubscript{\textpm\ 0.005}}  &\makecell[c]{0.732\textsubscript{\textpm\ 0.049}}&                                                
 \makecell[c]{0.797\textsubscript{\textpm\ 0.050}} &\textbf{\makecell[c]{0.847\textsubscript{\textpm\ 0.080}}}&
\makecell[c]{\textbf{0.778\textsubscript{\textpm\ 0.004}}} & \makecell[c]{\textbf{0.919\textsubscript{\textpm\ 0.052}}}&
\makecell[c]{0.759\textsubscript{\textpm\ 0.077}} & \makecell[c]{\textbf{0.810\textsubscript{\textpm\ 0.104}}}\\\hline
\end{tabular}
}
\end{table*}
\begin{table}[ht] 
\centering
\resizebox{\columnwidth}{!}{%
\caption{Comparison of parameter-cost and computational time.}\label{table:Table 3.2}
\begin{tabular}{lll}
\hline
Method & \makecell[c]{No. of parameters}& \makecell[c]{Inference time (ms)}\\\hline
3D UNet  &\makecell[c]{5.89M (-)}&\makecell[c]{44}\\
3D UNet-MTL-TSOL &   \makecell[c]{5.89M (17)}	&\makecell[c]{53}\\
3D UNet-MTL-TSD& \makecell[c]{8.24M (2.35M)}	&\makecell[c]{68}\\\hline
3D UNet$_{++}$ &    \makecell[c]{6.87M}	&\makecell[c]{63} \\
3D UNet$_{++}$ - MTL-TSOL & \makecell[c]{6.87M (17)}	&\makecell[c]{79}\\
3D UNet$_{++}$ - MTL-TSD& \makecell[c]{9.22M (2.35M)}	&\makecell[c]{81}\\\hline
3D Att-UNet  &  \makecell[c]{6.47M}	&\makecell[c]{77}\\
3D Att-UNet-MTL-TSOL&  \makecell[c]{6.47M (17)}	&\makecell[c]{78}\\
3D Att-MTL-TSD& \makecell[c]{8.82M (2.35M)}	&\makecell[c]{89}\\\hline
\end{tabular}
}
\end{table}
\begin{figure*}[!hbt]
\captionsetup[subfigure]{justification=centering}
\centering
\rotatebox[origin=c]{90}{\makebox[2in]{\textbf{Pancreas-CT dataset}}}%
\begin{subfigure}{.3\textwidth}
\centering
\includegraphics[scale=0.3]{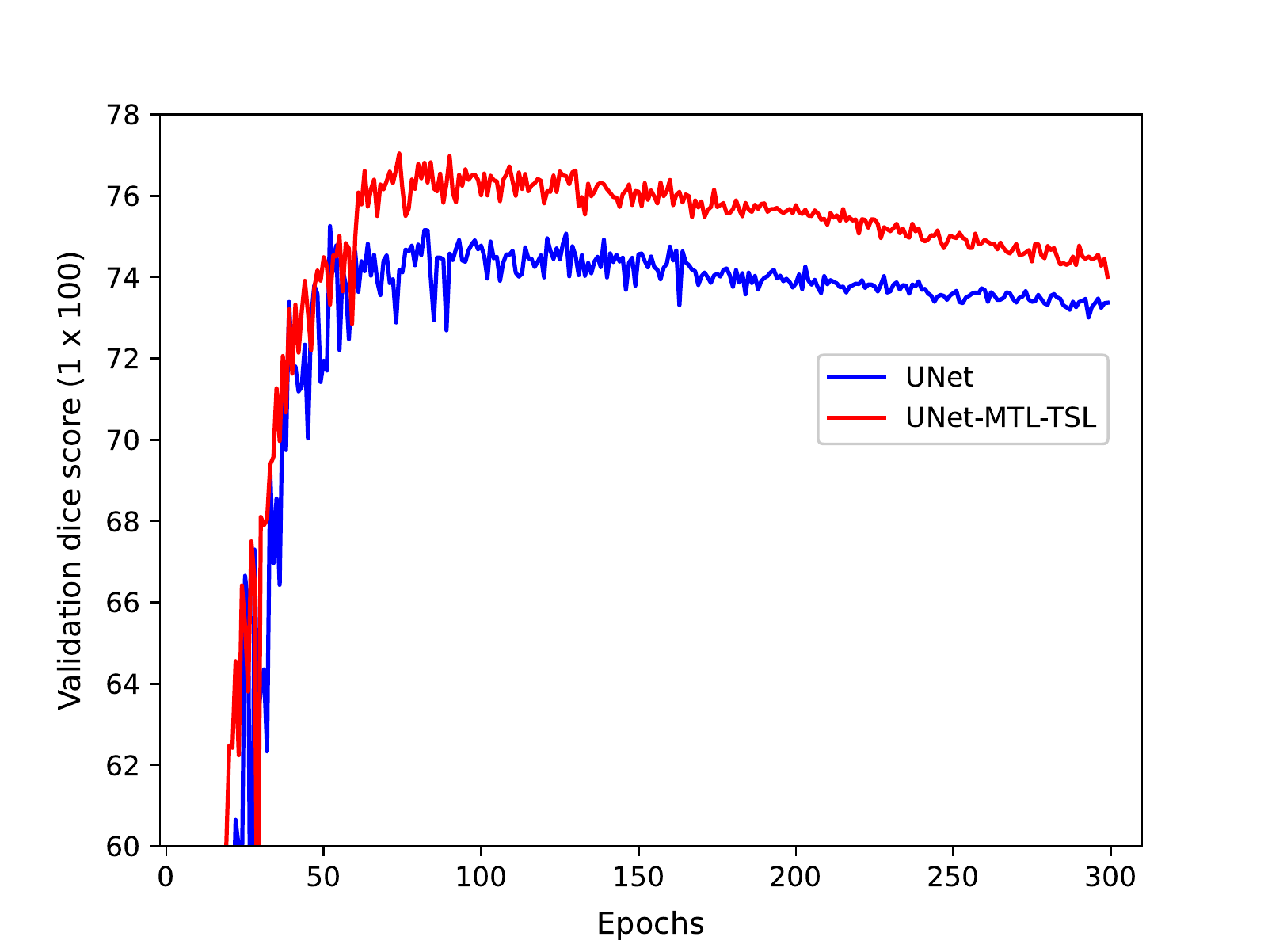} 
\hfill
\includegraphics[scale=0.3]{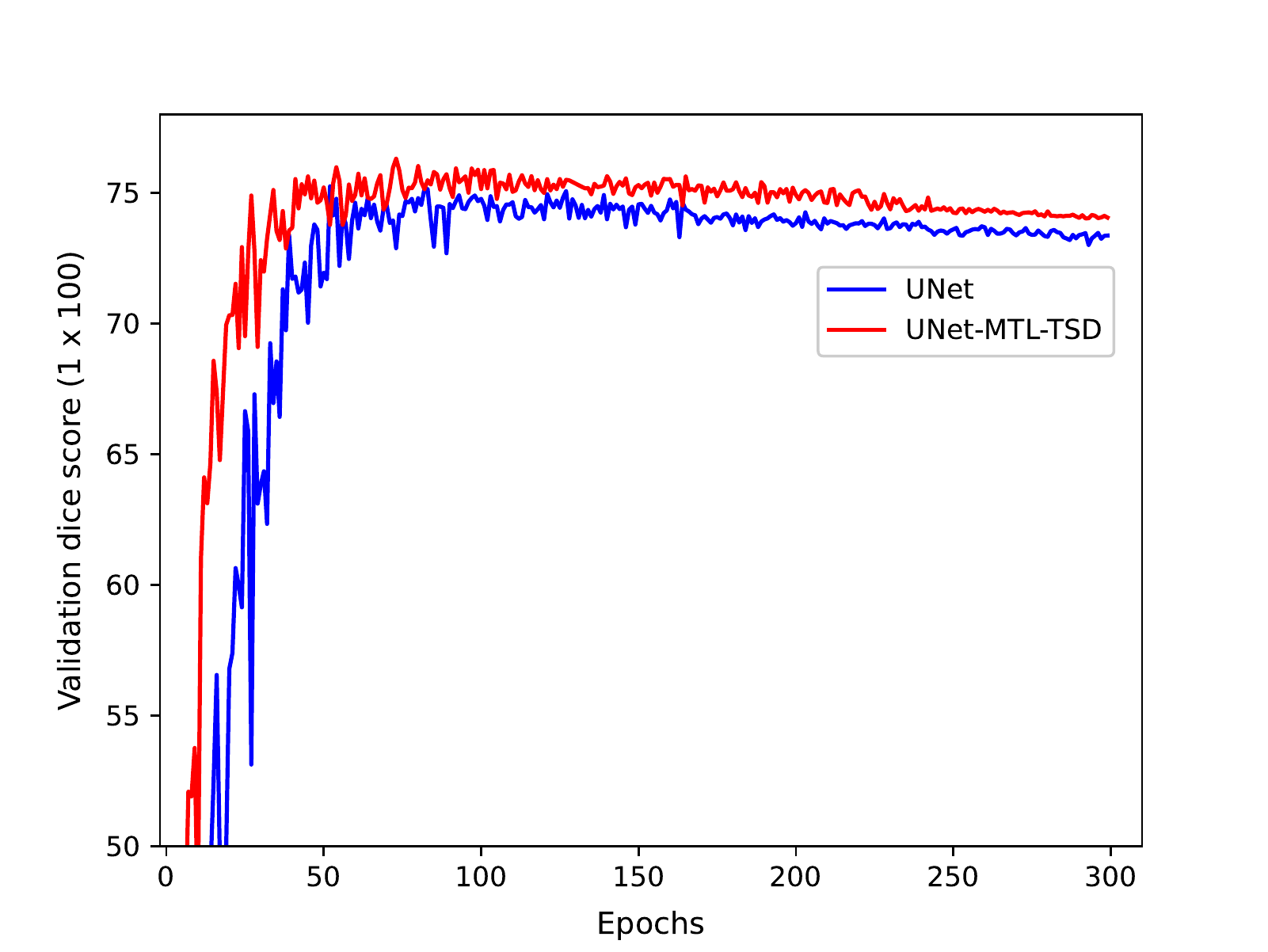}
\caption{}
\label{fig:sub20}
\end{subfigure}%
\begin{subfigure}{.3\textwidth}
\centering
\includegraphics[scale=0.3]{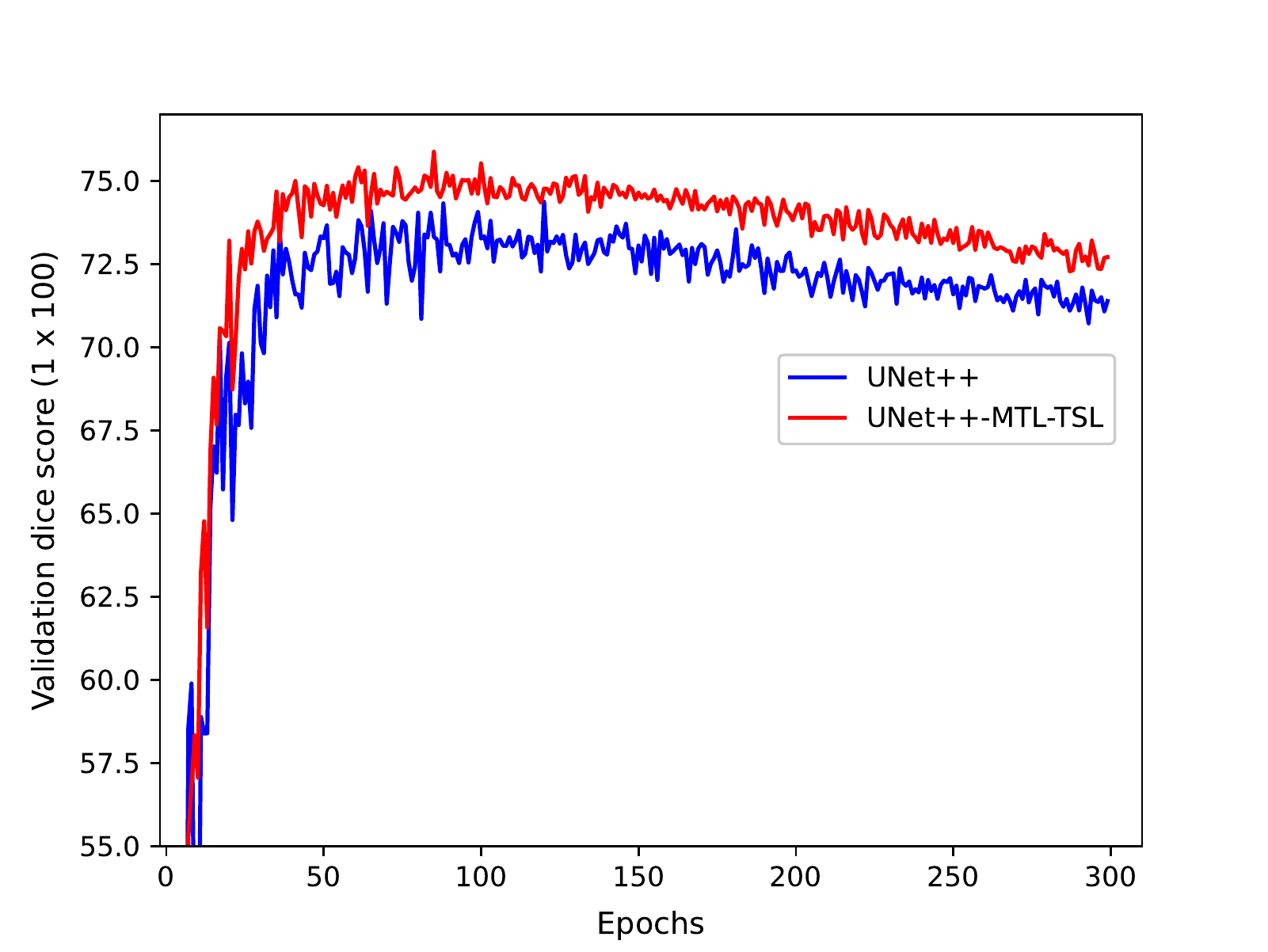}
\hfill
\includegraphics[scale=0.3]{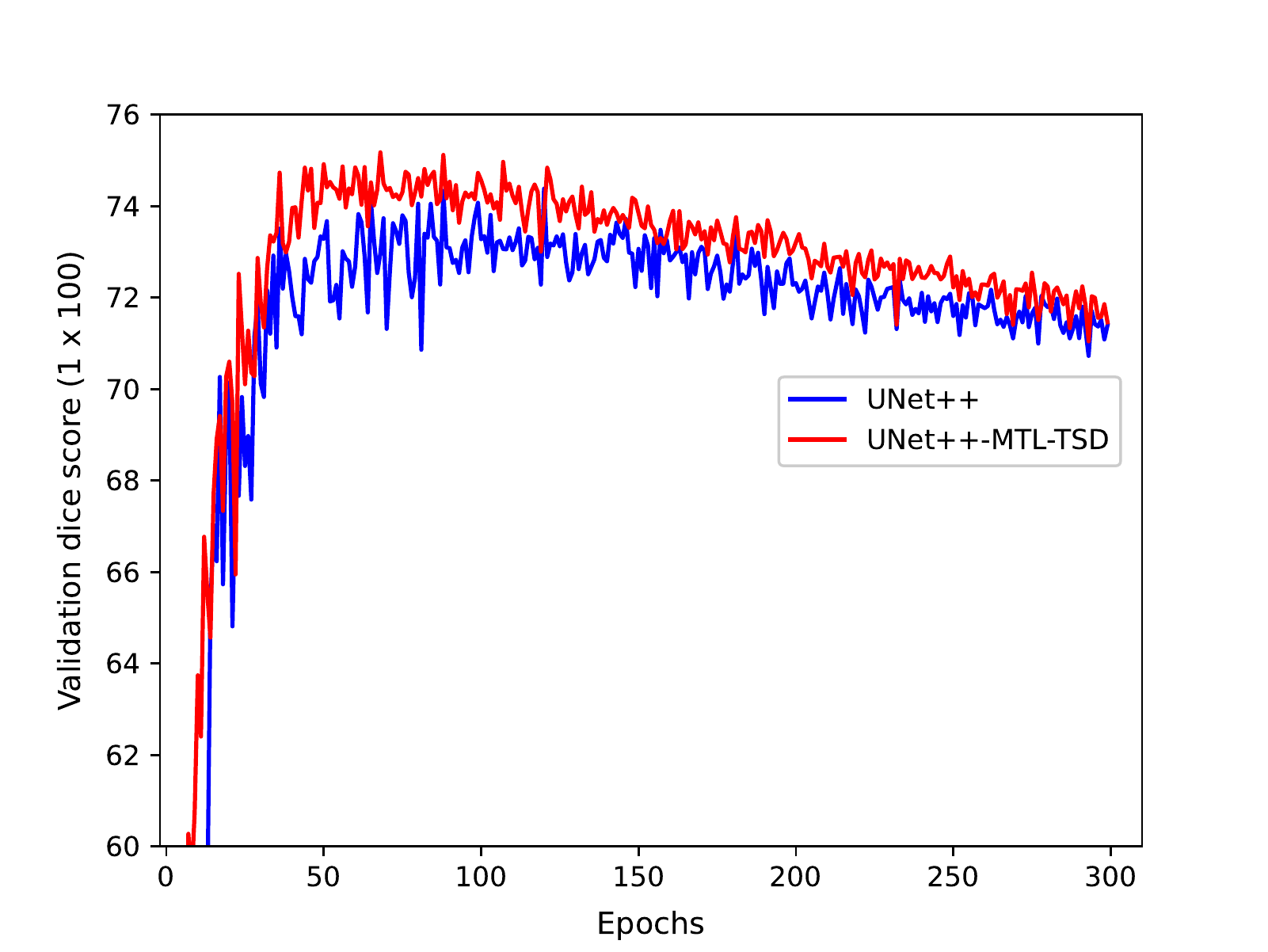}
\caption{}
\label{fig:sub21}
\end{subfigure}%
\begin{subfigure}{.3\textwidth}
\centering
\includegraphics[scale=0.3]{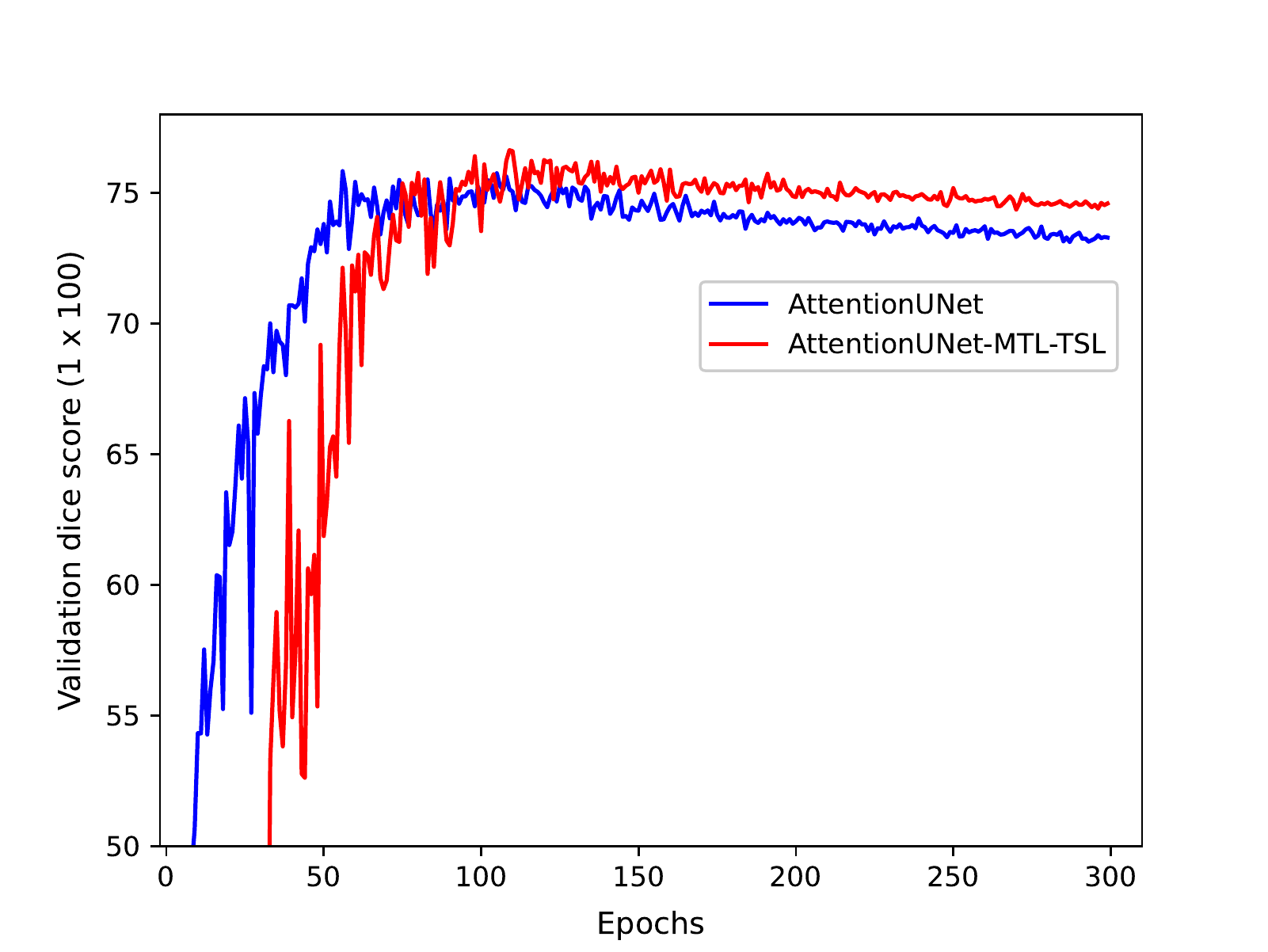}
\hfill
\includegraphics[scale=0.3]{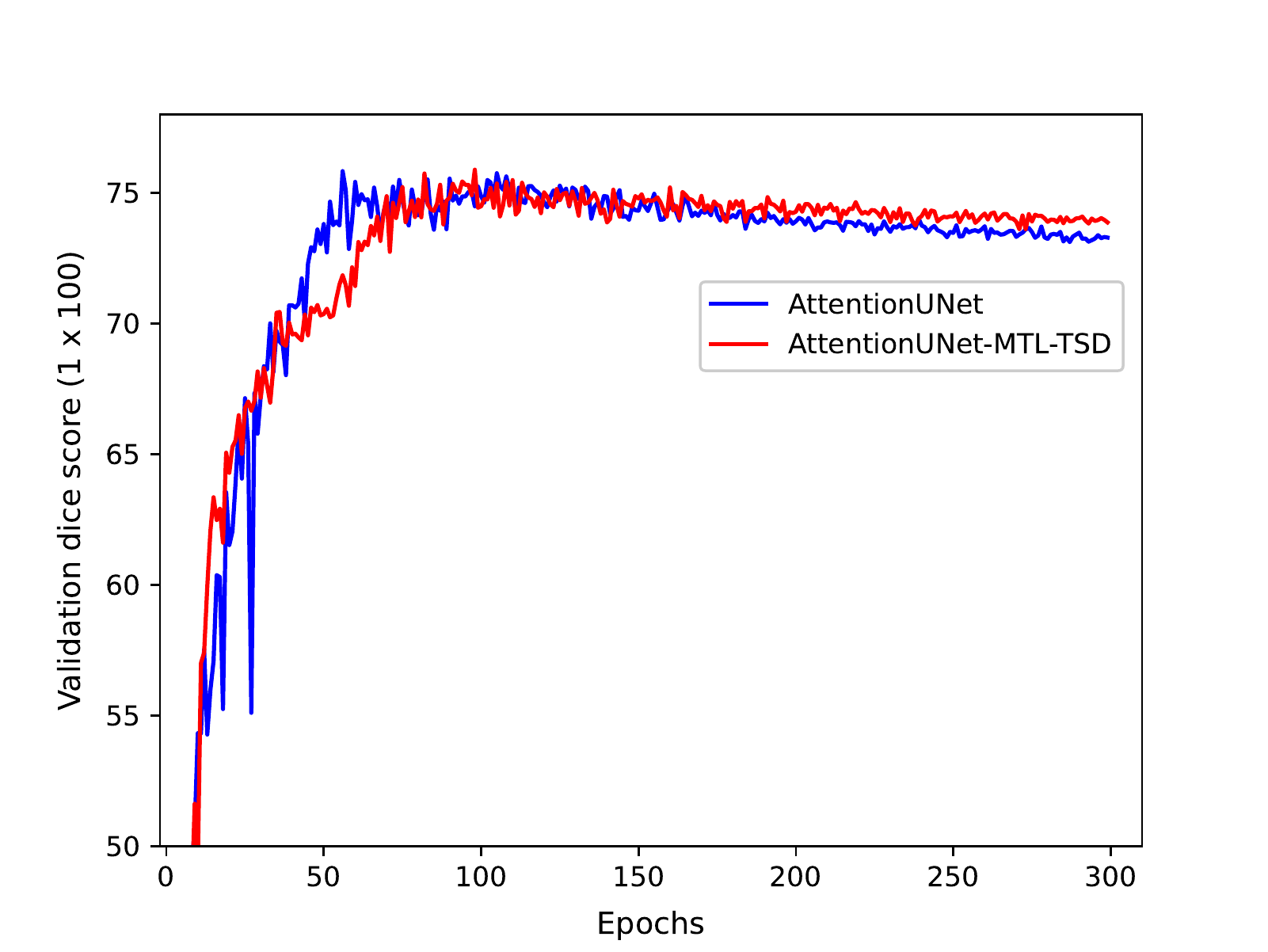} 
\caption{}
\label{fig:sub22}
\end{subfigure}
\rotatebox[origin=c]{90}{\makebox[2in]{\textbf{BTCV dataset}}}%
\begin{subfigure}{.3\textwidth}
\centering
\includegraphics[scale=0.3]{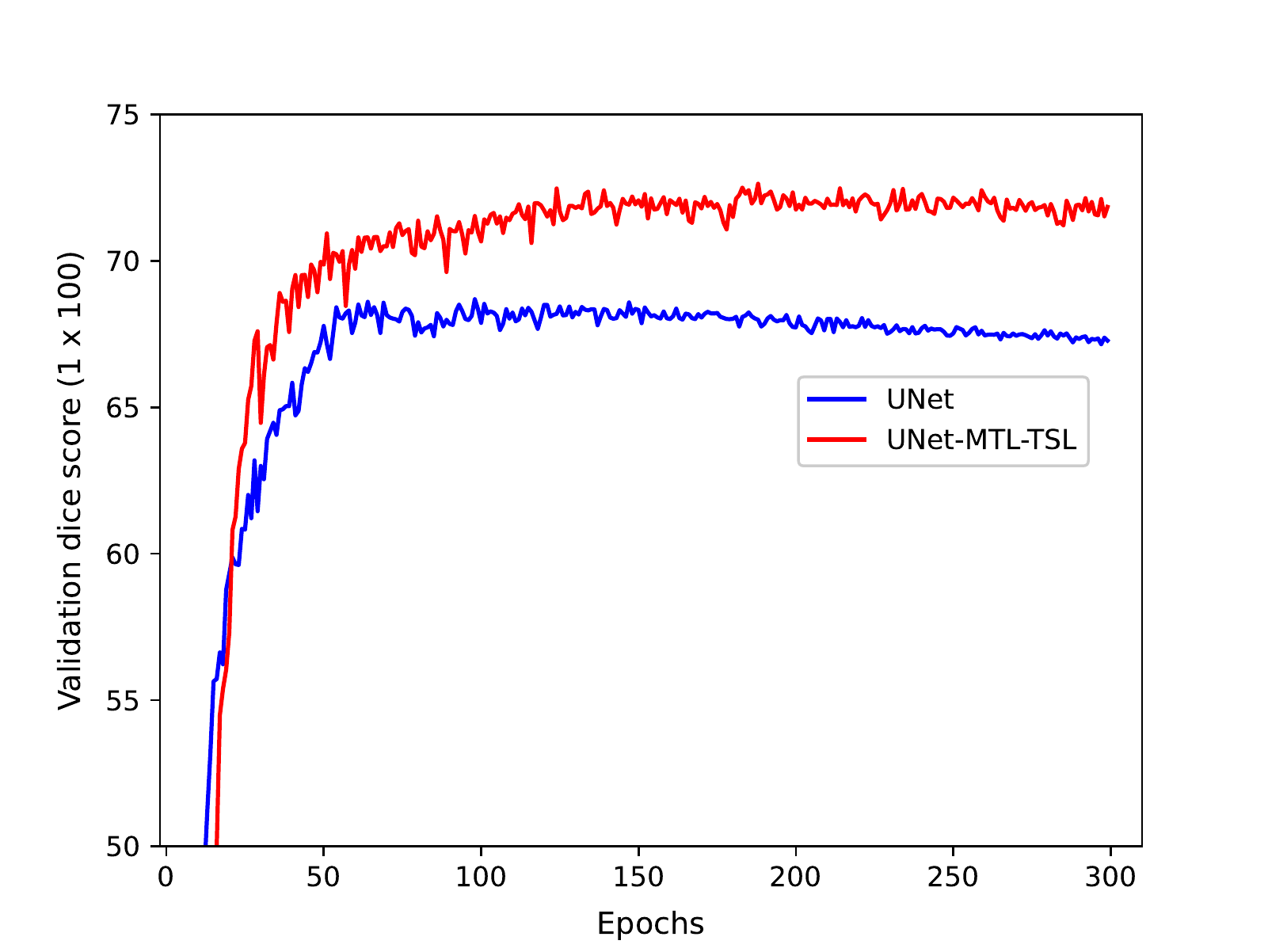}
\hfill
\includegraphics[scale=0.3]{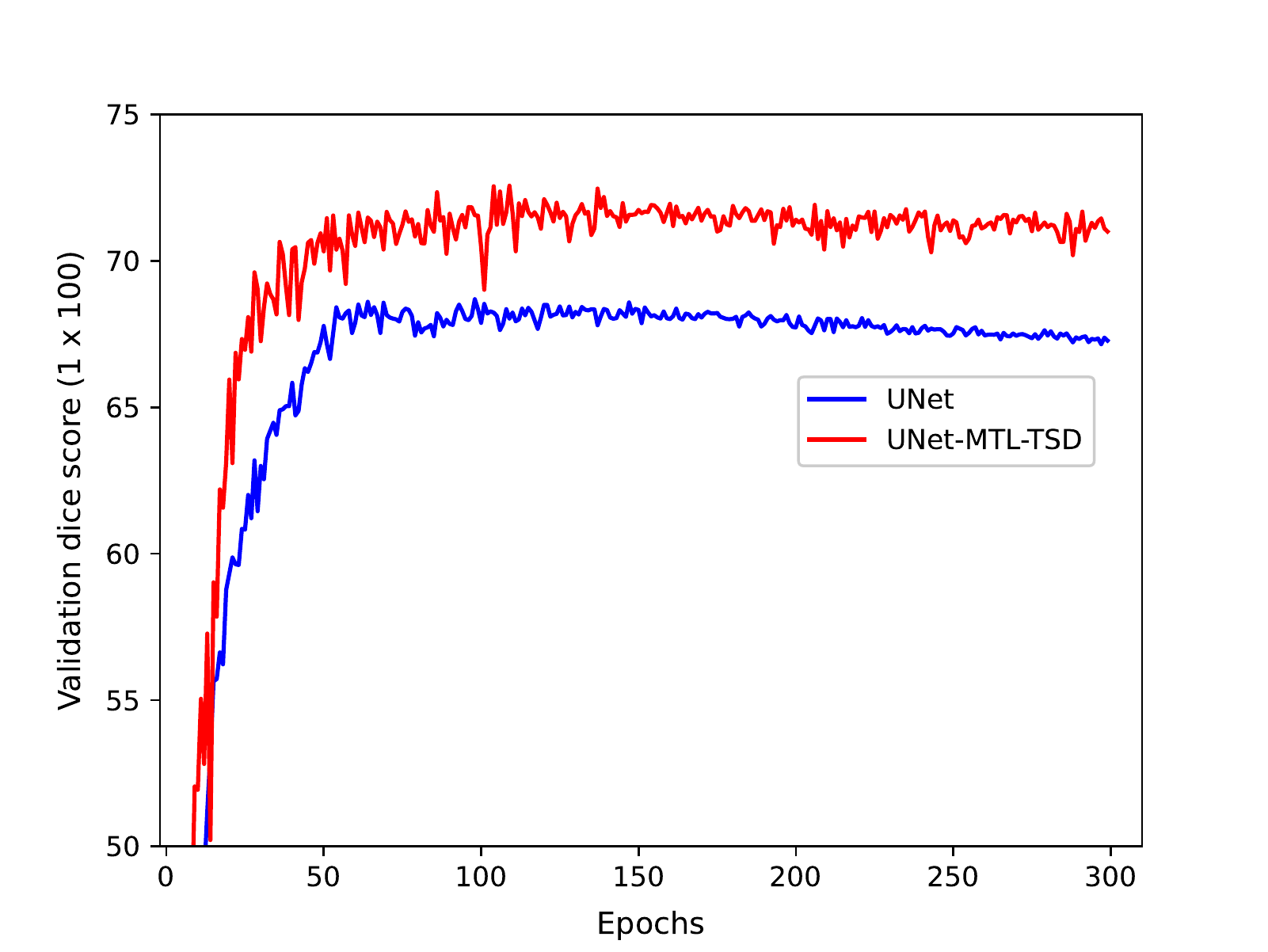}
\caption{}
\label{fig:sub23}
\end{subfigure}%
\begin{subfigure}{.3\textwidth}
\centering
\includegraphics[scale=0.3]{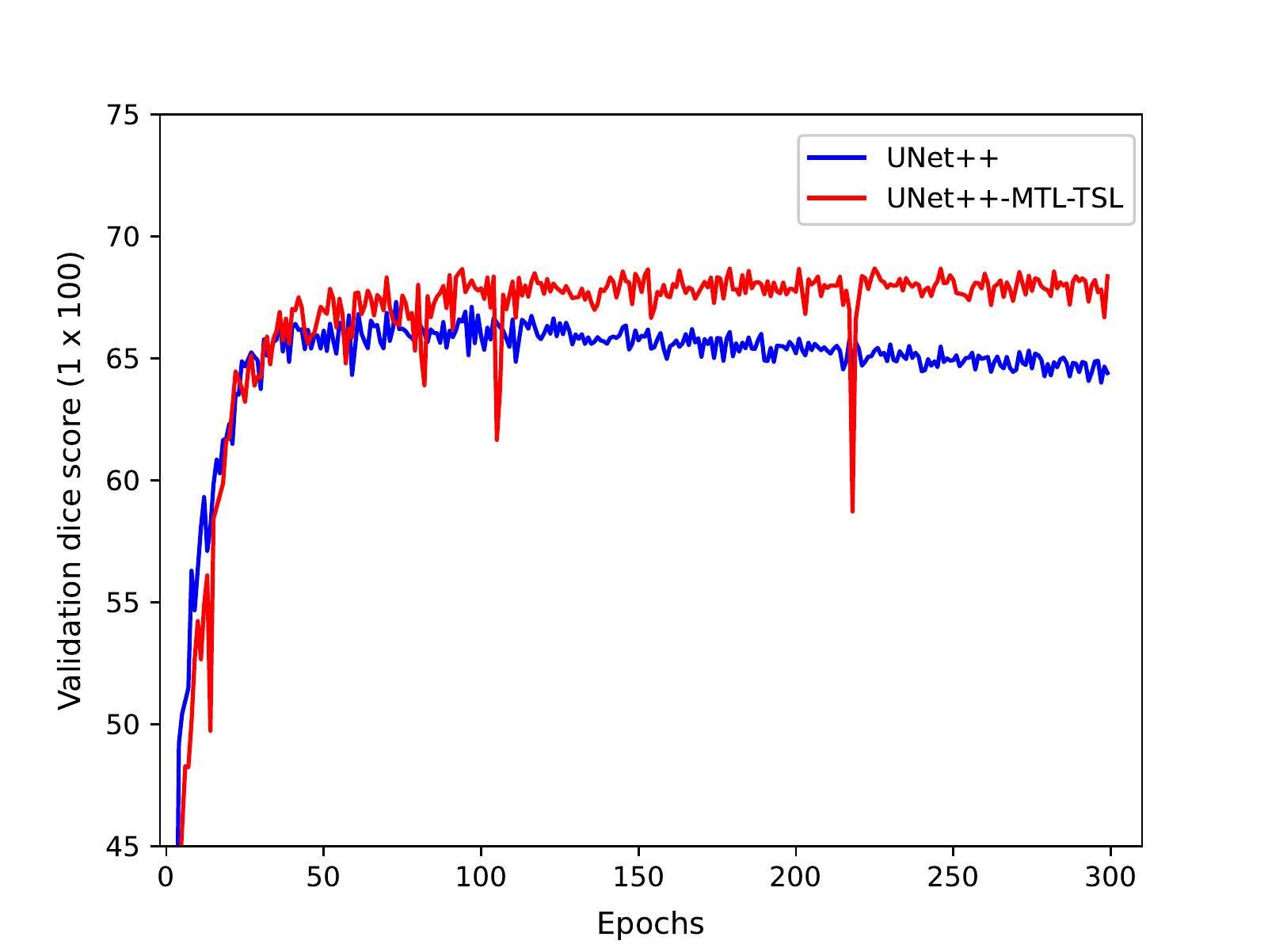}
\hfill
\includegraphics[scale=0.3]{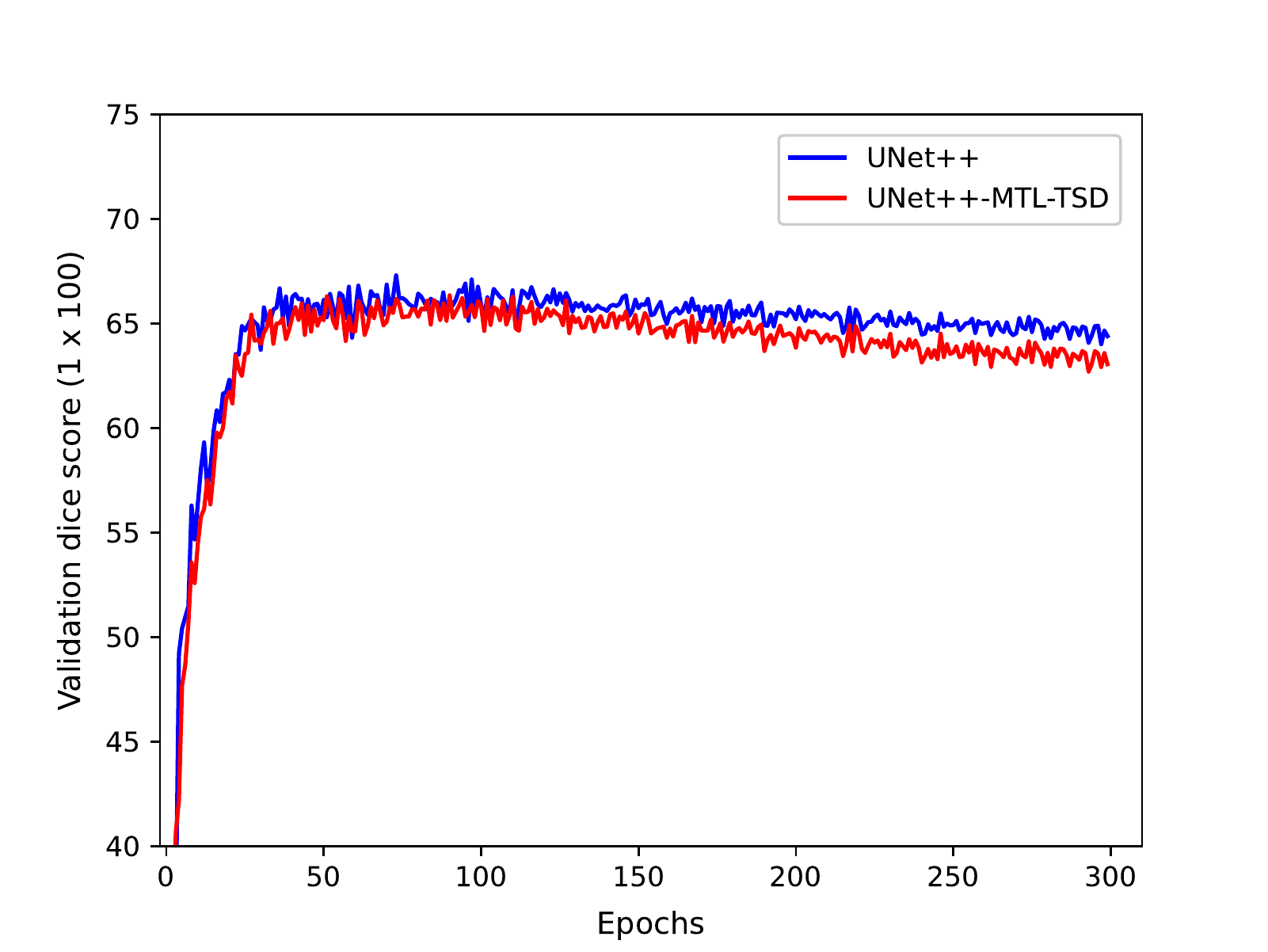}
\caption{}
\label{fig:sub24}
\end{subfigure}
\begin{subfigure}{.3\textwidth}
\centering
\includegraphics[scale=0.3]{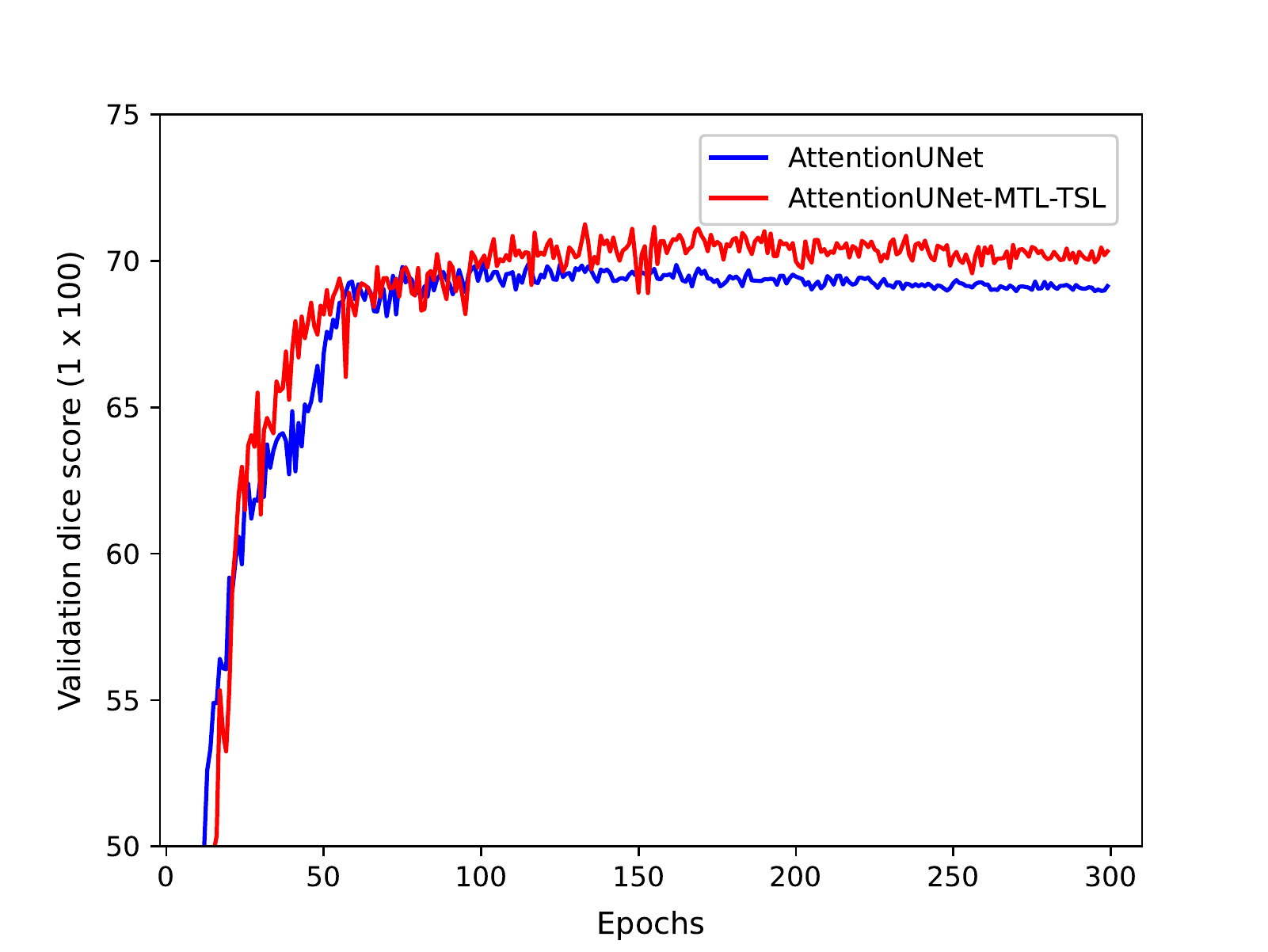}
\hfill
\includegraphics[scale=0.3]{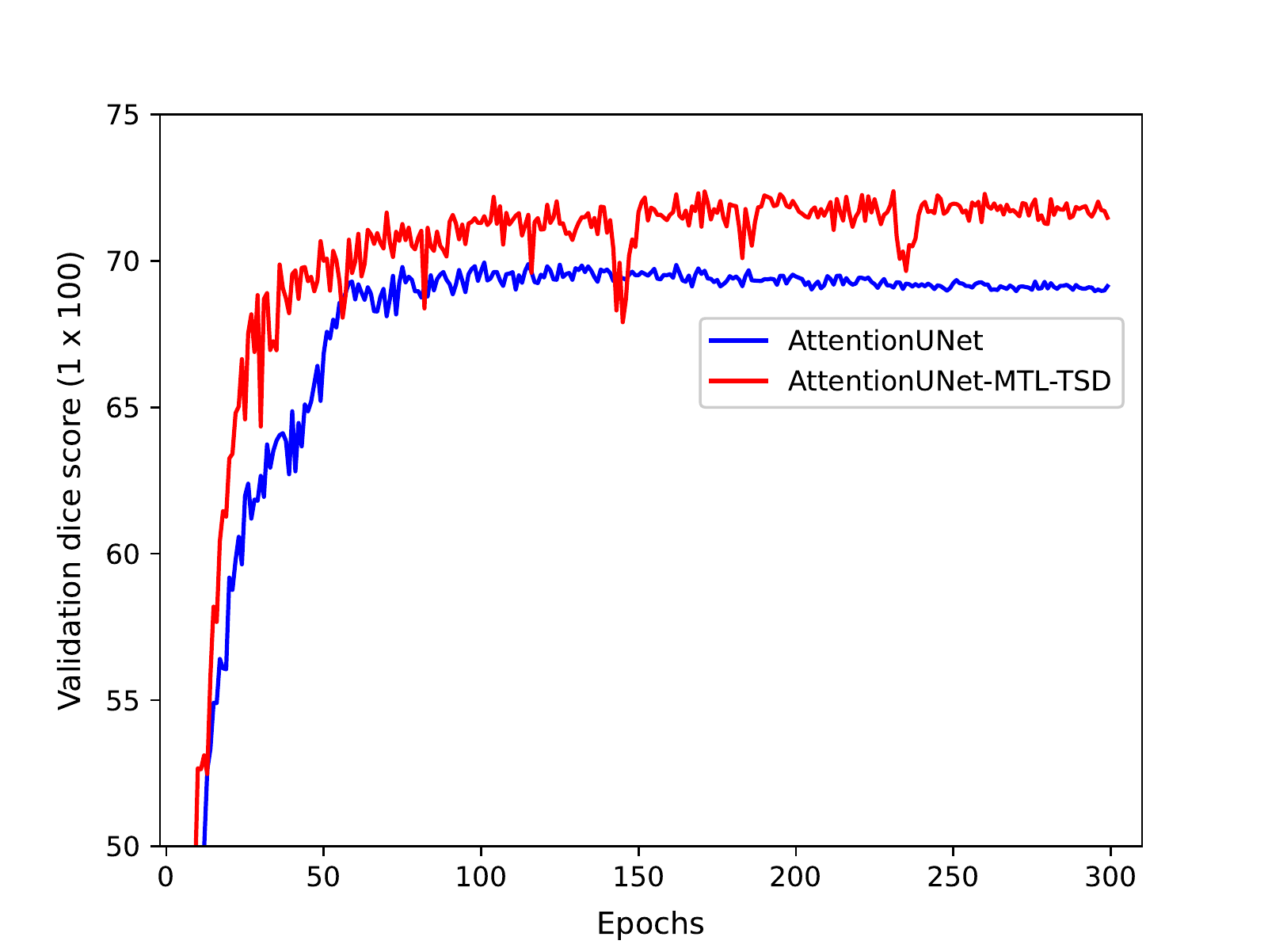}
\caption{}
\label{fig:sub25}
\end{subfigure}
\caption{Examining the mean dice score computed by comparing predicted multi-organ segmentation masks with the ground-truth on validation set: First two rows (a-c) and last two rows (d-f) correspond to the dice score curves for Pancreas-ct dataset and BTVC dataset, respectively. Each figure compares the mean DC of the baseline (\textcolor{blue}{Blue}) and counterpart boundary-constrained model (\textcolor{red}{Red}) on the validation set, computed after each epoch. The baseline models are trained to predict the masks of the abdominal regions, whereas the boundary-constrained models are trained to predict the labels of organs and boundaries.}
\label{fig:valcurve}
\end{figure*}
\section{Results and Analysis}\label{sec:result}
This section demonstrates the experimental results obtained from boundary-constrained abdominal segmentation models and compares them against the performance given by baseline models. For the sake of brevity, we denote the baseline and boundary-constrained models with the abbreviations below (shown in \textbf{bold}).
\begin{enumerate}[label=(\alph*)]
\item \textbf{3D UNet}: 3D UNet shown in Fig. S1 in Supplementary material.
\item \textbf{3D UNet-MTL-TSOL:} Boundary constrained 3D UNet with task specific output layers shown in \autoref{fig:unetmtl1}.
\item \textbf{3D UNet-MTL-TSD:} Boundary constrained 3D UNet with task specific decoders shown in \autoref{fig:unetmtl2}.
\item \textbf{3D UNet$_{++}$}: 3D UNet$_{++}$ shown in Fig. S2 in Supplementary material.
\item \textbf{3D UNet$_{++}$-MTL-TSOL:} Boundary constrained 3D UNet$_{++}$ with task specific output layers shown in \autoref{fig:unetplusmtl1}.
\item \textbf{3D UNet$_{++}$-MTL-TSD:} Boundary constrained 3D UNet$_{++}$ with task specific decoders shown in \autoref{fig:unetplusmtl2}.
\item \textbf{3D Att-UNet}: 3D Att-UNet shown in Fig. S3 in Supplementary material.
\item \textbf{3D Att-UNet-MTL-TSOL:} Boundary constrained 3D Att-UNet with task specific output layers shown in \autoref{fig:attenunetmtl1}.
\item \textbf{3D Att-UNet-MTL-TSD:} Boundary constrained 3D Att-UNet with task specific decoders shown in \autoref{fig:attenunetmtl2}.
\end{enumerate}
\subsection{Quantitative Results}
\autoref{table:Table 3.1} summarizes the segmentation results for the Pancreas-CT and BTCV datasets. We report the mean Dice Score, mean Average Hausdorff Distance (Avg. HD), mean Recall, and mean Precision computed by comparing the predicted segmentation against the ground-truth. These measures are calculated on test sets of each dataset. \autoref{table:Table 3.1} shows that the boundary-constrained models achieve improved multi-organ segmentation on the abdominal CT scans. Firstly, the mean Dice scores are improved by 1.8\% (UNet vs. UNet-MTL-TSOL) and 3.5\% (Att-UNet vs. Att-UNet-MTL-TSOL), for Pancreas-CT dataset. The corresponding values of mean dice scores for BTCV dataset are improved by 3.1\% (UNet vs. UNet-MTL-TSD), 3.6\% (UNet$_{++}$ vs. UNet$_{++}$-MTL-TSOL), and 3.5\% (Att-UNet vs. Att-UNet-MTL-TSD). The improved overlap between predicted segmentations and manually annotated masks can be attributed to the enhanced semantic representations learned by boundary-constrained models. 

Secondly, we observe that boundary-constrained models achieve a lower Avg. HDs for all the datasets than those obtained from baseline models as shown in \autoref{table:Table 3.1}. For example, after adding boundary information, the Avg. HD values of UNet, UNet$_{++}$, and Att-UNet are decreased by 11.5\%, 14.5\%, and 18.4\%, respectively, for the Pancreas-CT dataset. Likewise, a decrease of 15.4\%, 16.2\%, and 30\%, respectively, is seen for the BTCV dataset. Furthermore, we notice that the Avg. HD is still lower for the cases where boundary-constrained models obtained lower or equivalent mean Dice score, e.g., UNet$_{++}$-MTL-TSOL vs. UNet$_{++}$ and UNet$_{++}$-MTL-TSD vs. UNet$_{++}$, for Pancreas-CT results. This indicates that even for the equivalent dice overlap, the performance of the boundary-constrained models in accurately predicting the boundaries is improved.

Thirdly, our boundary-constrained models achieve higher values of mean Recall and mean Precision for all the models and datasets except for UNet$_{++}$ corresponding to the Pancreas-CT dataset, as shown in  \autoref{table:Table 3.1}. The utilization of boundary information has caused a decrease in false-negative rates and false-positive rates. Specifically, the mean recall is increased by 1.3\% (UNet vs. UNet-MTL-TSOL) and 4.3\% (Att-UNet vs. Att-UNet-MTL-TSOL) for Pancreas-CT dataset. For BTCV dataset, an increase of 1.3\% (UNet vs. UNet-MTL-TSOL), 3.7\% (UNet$_{++}$-MTL-TSOL vs. UNet$_{++}$), and 1.9\% (Att-UNet vs. Att-UNet-MTL-TSOL) is observed in Mean Recalls. Finally, we see a maximum improvement of 1.9\% and 4.3\% in values of mean Precision for Pancreas-CT and BTCV datasets, respectively. The improvment in mean Recalls and mean Precisions show the capability of boundary-constrained models in addressing the issues of under-segmentation and over-segmentation. 
\begin{table*}[ht] 
\centering
\resizebox{\columnwidth}{!}{%
\caption{Organ-wise Mean Dice scores \textpm\ Std. Dev. obtained from baseline and best performing boundary-constrained models: \textbf{Bold} shows the highest value corresponding to each baseline.}\label{table:Table 3.3}
\vspace{-1mm}
\begin{tabular}{lllllllll}
\hline
Method&\makecell[c]{\raisebox{0.2em}{\fcolorbox{Purple}{Purple}{\rule{0pt}{0pt}\rule{0pt}{0pt}}}}& \makecell[c]{\raisebox{0.2em}{\fcolorbox{RawSienna}{RawSienna}{\rule{0pt}{0pt}\rule{0pt}{0pt}}}}&\makecell[c]{\raisebox{0.2em}{\fcolorbox{Rhodamine}{Rhodamine}{\rule{0pt}{0pt}\rule{0pt}{0pt}}}} &\makecell[c]{\raisebox{0.2em}{\fcolorbox{st}{st}{\rule{0pt}{0pt}\rule{0pt}{0pt}}}}&\makecell[c]{\raisebox{0.2em}{\fcolorbox{red}{red}{\rule{0pt}{0pt}\rule{0pt}{0pt}}}}&\makecell[c]{\raisebox{0.2em}{\fcolorbox{BurntOrange}{BurntOrange}{\rule{0pt}{0pt}\rule{0pt}{0pt}}}}&\makecell[c]{\raisebox{0.2em}{\fcolorbox{lv}{lv}{\rule{0pt}{0pt}\rule{0pt}{0pt}}}}&\makecell[c]{\raisebox{0.2em}{\fcolorbox{du}{du}{\rule{0pt}{0pt}\rule{0pt}{0pt}}}}\\
\hline
&\multicolumn{8}{c}{\textbf{Pancreas-CT dataset}}\\
\cline{2-9}
\mystrut 3D UNet &     \makecell[c]{\textbf{0.935\textsubscript{\textpm\ 0.008}}} 	&\makecell[c]{0.700\textsubscript{\textpm\ 0.035}} &	\makecell[c]{0.927\textsubscript{\textpm\ 0.014}	}&\makecell[c]{0.793\textsubscript{\textpm\ 0.030}}	&\makecell[c]{0.714\textsubscript{\textpm\ 0.024}}&	\makecell[c]{0.934\textsubscript{\textpm\ 0.009}}&	\makecell[c]{0.829\textsubscript{\textpm\ 0.024}}&	\makecell[c]{0.566\textsubscript{\textpm\ 0.031}}\\
3D UNet-MTL-TSD&      \makecell[c]{0.931\textsubscript{\textpm\ 0.008}} 
&\makecell[c]{\textbf{0.723\textsubscript{\textpm\ 0.028}}}&	
\makecell[c]{\textbf{0.937\textsubscript{\textpm\ 0.004}}}
&\makecell[c]{\textbf{0.820\textsubscript{\textpm\ 0.028}}}
&	\makecell[c]{\textbf{0.718\textsubscript{\textpm\ 0.014}}}
&	\makecell[c]{\textbf{0.947\textsubscript{\textpm\ 0.003}}}&	
\makecell[c]{\textbf{0.854\textsubscript{\textpm\ 0.015}}}&	
\makecell[c]{\textbf{0.583\textsubscript{\textpm\ 0.021}}}\\\hline
\mystrut 3D UNet$_{++}$ & 
\makecell[c]{\textbf{0.917\textsubscript{\textpm\ 0.006}}}
&	\makecell[c]{0.613\textsubscript{\textpm\ 0.053}} 
&	\makecell[c]{0.897\textsubscript{\textpm\ 0.019}}
&	\makecell[c]{\textbf{0.802\textsubscript{\textpm\ 0.026}}}
&	\makecell[c]{\textbf{0.712\textsubscript{\textpm\ 0.013}}} 
&	\makecell[c]{0.899\textsubscript{\textpm\ 0.012}}	
&\makecell[c]{\textbf{0.818\textsubscript{\textpm\ 0.014}}}
&	\makecell[c]{0.516\textsubscript{\textpm\ 0.042}} \\
3D UNet$_{++}$ - MTL-TSOL  & \makecell[c]{0.915\textsubscript{\textpm\ 0.005}}
&\makecell[c]{\textbf{0.621\textsubscript{\textpm\ 0.042}}}&
\makecell[c]{\textbf{0.915\textsubscript{\textpm\ 0.013}}}
&	\makecell[c]{0.774\textsubscript{\textpm\ 0.051}}&	
\makecell[c]{0.697\textsubscript{\textpm\ 0.013}}&
\makecell[c]{\textbf{0.913\textsubscript{\textpm\ 0.005}}}
&	\makecell[c]{\textbf{0.818\textsubscript{\textpm\ 0.032}}}
&	\makecell[c]{\textbf{0.519\textsubscript{\textpm\ 0.035}}}\\\hline
3D Att-UNet&\makecell[c]{0.927\textsubscript{\textpm\ 0.007}}
&	\makecell[c]{0.688\textsubscript{\textpm\ 0.021}}
&\makecell[c]{\textbf{0.927\textsubscript{\textpm\ 0.013}}}
&	\makecell[c]{0.772\textsubscript{\textpm\ 0.037}}
& \makecell[c]{0.700\textsubscript{\textpm\ 0.021}}
&	 \makecell[c]{0.929\textsubscript{\textpm\ 0.016}}
&\makecell[c]{0.827\textsubscript{\textpm\ 0.029}}
&\makecell[c]{0.565\textsubscript{\textpm\ 0.036}}\\
3D Att-UNet-MTL-TSOL&
\makecell[c]{\textbf{0.937\textsubscript{\textpm\ 0.008}}}
&	\makecell[c]{\textbf{0.724\textsubscript{\textpm\ 0.010}}}
&\makecell[c]{\textbf{0.927\textsubscript{\textpm\ 0.009}}}
&	\makecell[c]{\textbf{0.858\textsubscript{\textpm\ 0.017}}}
& \makecell[c]{\textbf{0.734\textsubscript{\textpm\ 0.005}}}
&	 \makecell[c]{\textbf{0.949\textsubscript{\textpm\ 0.004}}}
&\makecell[c]{\textbf{0.847\textsubscript{\textpm\ 0.018}}}
&\makecell[c]{\textbf{0.589\textsubscript{\textpm\ 0.023}}}\\\hline
&\multicolumn{8}{c}{\textbf{BTCV dataset}}\\
\cline{2-9}
3D UNet &     
\makecell[c]{0.878\textsubscript{\textpm\ 0.016}} 	
&\makecell[c]{0.659\textsubscript{\textpm\ 0.018}} 
&	\makecell[c]{0.897\textsubscript{\textpm\ 0.014}}	&
\makecell[c]{0.567\textsubscript{\textpm\ 0.020}}	&
\makecell[c]{0.698\textsubscript{\textpm\ 0.009}}&	
\makecell[c]{0.932\textsubscript{\textpm\ 0.008}}&
	\makecell[c]{0.813\textsubscript{\textpm\ 0.012}}&
		\makecell[c]{0.576\textsubscript{\textpm\ 0.023}}\\
3D UNet-MTL-TSD&      \makecell[c]{\textbf{0.897\textsubscript{\textpm\ 0.011}}}
&	\makecell[c]{\textbf{0.682\textsubscript{\textpm\ 0.009}}}
&	\makecell[c]{\textbf{0.912\textsubscript{\textpm\ 0.004}}}
&\makecell[c]{\textbf{0.608\textsubscript{\textpm\ 0.019}}}
&	\makecell[c]{\textbf{0.732\textsubscript{\textpm\ 0.016}}}
&	\makecell[c]{\textbf{0.937\textsubscript{\textpm\ 0.007}}}
&	\makecell[c]{\textbf{0.837\textsubscript{\textpm\ 0.008}}}
&	\makecell[c]{\textbf{0.599\textsubscript{\textpm\ 0.011}}}\\\hline
3D UNet$_{++}$ & 
\makecell[c]{0.846\textsubscript{\textpm\ 0.016}}&	
\makecell[c]{0.559\textsubscript{\textpm\ 0.032}} &
	\makecell[c]{0.851\textsubscript{\textpm\ 0.030}}&
		\makecell[c]{0.552\textsubscript{\textpm\ 0.023}}&	
		\makecell[c]{0.691\textsubscript{\textpm\ 0.014}} &
			\makecell[c]{0.882\textsubscript{\textpm\ 0.008}}	&
			\makecell[c]{0.799\textsubscript{\textpm\ 0.015}}&	
			\makecell[c]{0.538\textsubscript{\textpm\ 0.028}} \\
3D UNet$_{++}$ - MTL-TSOL  &         
\makecell[c]{\textbf{0.853\textsubscript{\textpm\ 0.023}}}
&\makecell[c]{\textbf{0.625\textsubscript{\textpm\ 0.018}}}
&	\makecell[c]{\textbf{0.891\textsubscript{\textpm\ 0.019}}}
&\makecell[c]{\textbf{0.575\textsubscript{\textpm\ 0.019}}}
&\makecell[c]{\textbf{0.694\textsubscript{\textpm\ 0.014}}}
&	\makecell[c]{\textbf{0.918\textsubscript{\textpm\ 0.004}}}
&	\makecell[c]{\textbf{0.829\textsubscript{\textpm\ 0.018}}}
&	\makecell[c]{\textbf{0.549\textsubscript{\textpm\ 0.032}}}\\\hline
3D Att-UNet & 
\makecell[c]{0.881\textsubscript{\textpm\ 0.020}} &
	\makecell[c]{0.668\textsubscript{\textpm\ 0.016}} &	
	\makecell[c]{0.882\textsubscript{\textpm\ 0.036}}&
	\makecell[c]{0.561\textsubscript{\textpm\ 0.009}}	&
	\makecell[c]{0.704\textsubscript{\textpm\ 0.023}}&	
	\makecell[c]{0.939\textsubscript{\textpm\ 0.006}}&	
	\makecell[c]{0.805\textsubscript{\textpm\ 0.007}}&	
	\makecell[c]{0.575\textsubscript{\textpm\ 0.007}}\\
3D Att-UNet-MTL-TSD&                             \makecell[c]{\textbf{0.913\textsubscript{\textpm\ 0.009}}}
&	\makecell[c]{\textbf{0.674\textsubscript{\textpm\ 0.006}}}
&\makecell[c]{\textbf{0.920\textsubscript{\textpm\ 0.015}}}
&	\makecell[c]{\textbf{0.603\textsubscript{\textpm\ 0.034}}}
&	\makecell[c]{\textbf{0.720\textsubscript{\textpm\ 0.009}}}
&\makecell[c]{\textbf{0.947\textsubscript{\textpm\ 0.007}}}
&\makecell[c]{\textbf{0.832\textsubscript{\textpm\ 0.022}}}
&	\makecell[c]{\textbf{0.608\textsubscript{\textpm\ 0.025}}}\\\hline
\end{tabular}
}
\end{table*}

\autoref{fig:valcurve} shows the performance curves based on the mean Dice scores, computed by comparing the predicted multi-organ segmentation with the ground-truth masks on the validation set. \Cref{fig:sub20,fig:sub21,fig:sub22} correspond to performance curves for the Pancreas-CT and \Cref{fig:sub23,fig:sub24,fig:sub25} for the BTCV dataset. It can be seen that the incorporation of boundary information has improved the mean Dice score as the training progresses.
\begin{table*}[ht] 
\centering
\resizebox{\columnwidth}{!}{%
\caption{Organ-wise Mean Avg. HDs \textpm\ Std. Dev. obtained from baseline and best performing boundary-constrained models: \textbf{Bold} shows the highest value corresponding to each baseline.}\label{table:Table 3.4}
\begin{tabular}{lllllllll}
\hline
Method&\makecell[c]{\raisebox{0.2em}{\fcolorbox{Purple}{Purple}{\rule{0pt}{0pt}\rule{0pt}{0pt}}}}& \makecell[c]{\raisebox{0.2em}{\fcolorbox{RawSienna}{RawSienna}{\rule{0pt}{0pt}\rule{0pt}{0pt}}}}&\makecell[c]{\raisebox{0.2em}{\fcolorbox{Rhodamine}{Rhodamine}{\rule{0pt}{0pt}\rule{0pt}{0pt}}}} &\makecell[c]{\raisebox{0.2em}{\fcolorbox{st}{st}{\rule{0pt}{0pt}\rule{0pt}{0pt}}}}&\makecell[c]{\raisebox{0.2em}{\fcolorbox{red}{red}{\rule{0pt}{0pt}\rule{0pt}{0pt}}}}&\makecell[c]{\raisebox{0.2em}{\fcolorbox{BurntOrange}{BurntOrange}{\rule{0pt}{0pt}\rule{0pt}{0pt}}}}&\makecell[c]{\raisebox{0.2em}{\fcolorbox{lv}{lv}{\rule{0pt}{0pt}\rule{0pt}{0pt}}}}&\makecell[c]{\raisebox{0.2em}{\fcolorbox{du}{du}{\rule{0pt}{0pt}\rule{0pt}{0pt}}}}\\\hline
&\multicolumn{8}{c}{\textbf{Pancreas-CT dataset}}\\
\cline{2-9}
\mystrut 3D UNet &     
\makecell[c]{\textbf{0.261\textsubscript{\textpm\ 0.085}}} 	&
\makecell[c]{0.936\textsubscript{\textpm\ 0.279}} &	
\makecell[c]{0.184\textsubscript{\textpm\ 0.143}}	&
\makecell[c]{0.803\textsubscript{\textpm\ 0.356}}	&
\makecell[c]{0.788\textsubscript{\textpm\ 0.275}}&	
\makecell[c]{\textbf{0.234\textsubscript{\textpm\ 0.063}}}&
	\makecell[c]{\textbf{0.834\textsubscript{\textpm\ 0.234}}}&	
	\makecell[c]{2.321\textsubscript{\textpm\ 0.316}}\\
3D UNet-MTL-TSD&      \makecell[c]{0.317\textsubscript{\textpm\ 0.083}} 
&	\makecell[c]{\textbf{0.911\textsubscript{\textpm\ 0.244}}}
&	\makecell[c]{\textbf{0.150\textsubscript{\textpm\ 0.036}}}
&	\makecell[c]{\textbf{0.389\textsubscript{\textpm\ 0.074}}}
&	\makecell[c]{\textbf{0.683\textsubscript{\textpm\ 0.072}}}
&	\makecell[c]{0.259\textsubscript{\textpm\ 0.122}}
&	\makecell[c]{0.879\textsubscript{\textpm\ 0.304}}
&\makecell[c]{\textbf{2.038\textsubscript{\textpm\ 0.204}}}\\\hline
\mystrut 3D UNet$_{++}$ & 
\makecell[c]{0.419\textsubscript{\textpm\ 0.069}}&	
\makecell[c]{1.852\textsubscript{\textpm\ 0.554}} &	
\makecell[c]{0.265\textsubscript{\textpm\ 0.153}}&	
\makecell[c]{1.569\textsubscript{\textpm\ 0.649}}&	
\makecell[c]{1.445\textsubscript{\textpm\ 0.494}} &	
\makecell[c]{0.711\textsubscript{\textpm\ 0.218}}	&
\makecell[c]{1.371\textsubscript{\textpm\ 0.457}}&	
\makecell[c]{3.315\textsubscript{\textpm\ 0.637}} \\
3D UNet$_{++}$ - MTL-TSOL  &     
 \makecell[c]{\textbf{0.389\textsubscript{\textpm\ 0.055}}	}&
\makecell[c]{\textbf{1.718\textsubscript{\textpm\ 0.638}}}&	
\makecell[c]{\textbf{0.167\textsubscript{\textpm\ 0.048}}}&	
\makecell[c]{\textbf{0.873\textsubscript{\textpm\ 0.229}}}&	
\makecell[c]{\textbf{1.187\textsubscript{\textpm\ 0.328}}}&	
\makecell[c]{\textbf{0.675\textsubscript{\textpm\ 0.104}}}&	
\makecell[c]{\textbf{1.296\textsubscript{\textpm\ 0.483}}}&	
\makecell[c]{\textbf{3.058\textsubscript{\textpm\ 0.948}}}\\\hline
3D Att-UNet & 
\makecell[c]{0.363\textsubscript{\textpm\ 0.063}} &	
\makecell[c]{\textbf{0.928\textsubscript{\textpm\ 0.085}}} &	
\makecell[c]{\textbf{0.174\textsubscript{\textpm\ 0.106}}}&
\makecell[c]{0.527\textsubscript{\textpm\ 0.099}}	&
\makecell[c]{0.809\textsubscript{\textpm\ 0.277}}&	
\makecell[c]{0.442\textsubscript{\textpm\ 0.175}}&	
\makecell[c]{0.969\textsubscript{\textpm\ 0.336}}&	
\makecell[c]{2.384\textsubscript{\textpm\ 0.466}}\\
3D Att-UNet-MTL-TSOL&                         \makecell[c]{\textbf{0.255\textsubscript{\textpm\ 0.041}}}
&	\makecell[c]{0.993\textsubscript{\textpm\ 0.315}}	
&\makecell[c]{0.192\textsubscript{\textpm\ 0.033}}
&	\makecell[c]{\textbf{0.305\textsubscript{\textpm\ 0.075}}}
&	\makecell[c]{\textbf{0.559\textsubscript{\textpm\ 0.022}}}
&	 \makecell[c]{\textbf{0.154\textsubscript{\textpm\ 0.054}}}
&\makecell[c]{\textbf{0.766\textsubscript{\textpm\ 0.274}}}
&	\makecell[c]{\textbf{2.162\textsubscript{\textpm\ 0.434}}}\\\hline
&\multicolumn{8}{c}{\textbf{BTCV dataset}}\\
\cline{2-9}
3D UNet &     
\makecell[c]{0.474\textsubscript{\textpm\ 0.161}} 	&
\makecell[c]{\textbf{1.741\textsubscript{\textpm\ 0.179}}} &	
\makecell[c]{0.337\textsubscript{\textpm\ 0.077}}	&
\makecell[c]{1.750\textsubscript{\textpm\ 0.334}}	&
\makecell[c]{0.770\textsubscript{\textpm\ 0.140}}&	
\makecell[c]{0.636\textsubscript{\textpm\ 0.206}}&	
\makecell[c]{0.941\textsubscript{\textpm\ 0.155}}&	
\makecell[c]{2.032\textsubscript{\textpm\ 0.149}}\\
3D UNet-MTL-TSD&      
\makecell[c]{\textbf{0.329\textsubscript{\textpm\ 0.105}}}
&	\makecell[c]{1.787\textsubscript{\textpm\ 0.326}} 
&\makecell[c]{\textbf{0.253\textsubscript{\textpm\ 0.059}}}
&	\makecell[c]{\textbf{1.076\textsubscript{\textpm\ 0.136}}}
& \makecell[c]{\textbf{0.648\textsubscript{\textpm\ 0.168}}}
&	 \makecell[c]{\textbf{0.619\textsubscript{\textpm\ 0.226}}}
&	\makecell[c]{\textbf{0.778\textsubscript{\textpm\ 0.123}}}
&	 \makecell[c]{\textbf{1.849\textsubscript{\textpm\ 0.199}}}\\\hline
3D UNet$_{++}$
& \makecell[c]{\textbf{0.780\textsubscript{\textpm\ 0.220}}}&	
\makecell[c]{2.423\textsubscript{\textpm\ 0.510}} &	
\makecell[c]{0.574\textsubscript{\textpm\ 0.282}}&	
\makecell[c]{2.103\textsubscript{\textpm\ 0.284}}&
	\makecell[c]{1.034\textsubscript{\textpm\ 0.187}} &	
	\makecell[c]{\textbf{0.984\textsubscript{\textpm\ 0.136}}}	&
	\makecell[c]{1.153\textsubscript{\textpm\ 0.148}}&
		\makecell[c]{2.306\textsubscript{\textpm\ 0.206}} \\
3D UNet$_{++}$ - MTL-TSOL  &         
\makecell[c]{0.794\textsubscript{\textpm\ 0.089}}	
&\makecell[c]{\textbf{1.842\textsubscript{\textpm\ 0.518}}}
&\makecell[c]{\textbf{0.328\textsubscript{\textpm\ 0.087}}}
& \makecell[c]{\textbf{1.630\textsubscript{\textpm\ 0.479}}}
&	 \makecell[c]{\textbf{0.908\textsubscript{\textpm\ 0.221}}}
&	\makecell[c]{1.080\textsubscript{\textpm\ 0.063}}
&	 \makecell[c]{\textbf{0.860\textsubscript{\textpm\ 0.248}}}
&	\makecell[c]{\textbf{2.073\textsubscript{\textpm\ 0.265}}}\\\hline
3D Att-UNet& 
\makecell[c]{0.577\textsubscript{\textpm\ 0.245}} &	
\makecell[c]{2.017\textsubscript{\textpm\ 0.418}} &	
\makecell[c]{0.404\textsubscript{\textpm\ 0.154}}&
\makecell[c]{2.889\textsubscript{\textpm\ 1.387}}	&
\makecell[c]{0.981\textsubscript{\textpm\ 0.659}}&	
\makecell[c]{0.435\textsubscript{\textpm\ 0.162}}&	
\makecell[c]{1.011\textsubscript{\textpm\ 0.133}}&
	\makecell[c]{2.193\textsubscript{\textpm\ 0.249}}\\
3D Att-UNet-MTL-TSD&                             	
\makecell[c]{\textbf{0.273\textsubscript{\textpm\ 0.055}}}&	
 \makecell[c]{\textbf{1.377\textsubscript{\textpm\ 0.123}}}
& \makecell[c]{\textbf{0.191\textsubscript{\textpm\ 0.062}}}
& \makecell[c]{\textbf{1.631\textsubscript{\textpm\ 0.456}}}
&	 \makecell[c]{\textbf{0.887\textsubscript{\textpm\ 0.248}}	}
&  \makecell[c]{\textbf{0.318\textsubscript{\textpm\ 0.202}}}
	&  \makecell[c]{\textbf{0.789\textsubscript{\textpm\ 0.149}}}
	&	 \makecell[c]{\textbf{1.888\textsubscript{\textpm\ 0.318}}}\\\hline
\end{tabular}
}
\end{table*}
\subsection{Computational Complexity and Architectural Analysis} 
\Cref{table:Table 3.2} reports parameter count (in million M) and each method's time to segment a single CT abdominal volume in the test phase. We also highlight the increase in parameter count (given in brackets) compared with the baseline model. Among the single-task baseline models, UNet$_{++}$ is most parameter-extensive with 6.87 $\times 10^6$ parameters. The parameter count of boundary-constrained models with TSOL topology have only 17 parameters more than the baseline models; however, these models showed significantly considerable improvements in the segmentation of abdominal organs. The shared encoder and decoder in the TSOL design enable the small parameter count while capacitating the segmentation algorithm to learn the masks and boundaries using separate task-specific output layers. Observing the second multi-task topology TSD, our model requires approximately 2.4 M more parameters than the baseline. 

Furthermore, we can see (from \Cref{table:Table 3.1}) that the boundary-constrained models take more time to segment a single volume at the inference stage. This behavior can be associated with the extended parameter size of boundary-constrained models since they are trained to predict a 3D boundary of organs in addition to region masks.

Analyzing the relationship between the multi-task network design and the segmentation performance from \Cref{table:Table 3.1}, we note there is not a single/fixed topology that led to the maximum improvement. For example, the TSD showed maximum improvement in mean DC and Avg HD corresponding to the Pancreas-CT dataset. Hd over the baseline UNet. However, when the performance of Att-UNet is compared with boundary-constrained models, we notice that the TSOL demonstrated the best results. This reveals that the multi-task network configuration that offers the best results varies depending on each baseline architecture. All in all, we found that integration of boundary information improved the multi-organ segmentation, independent of the network topology.
\subsection{Assesment of organ-wise segmentation performance}
To assess which specific organs benefitted greatly from incorporating boundary information for the segmentation task, we examine the mean Dice scores and mean Avg. HDs achieved by baseline, and best performing boundary constrained models (from \autoref{table:Table 3.1}) for each abdominal organ. We compute the Dice scores and Average Hausdorff distances for each organ individually and then average across all subjects. From \autoref{table:Table 3.3}, we can see both baseline and boundary-constrained models have yielded the highest mean Dice scores for liver ( \raisebox{0.3em}{\fcolorbox{BurntOrange}{BurntOrange}{\rule{0pt}{0pt}\rule{0pt}{0pt}}} ), spleen ( \raisebox{0.2em}{\fcolorbox{Purple}{Purple}{\rule{0pt}{0pt}\rule{0pt}{0pt}}} ), and kidney ( \raisebox{0.2em}{\fcolorbox{Rhodamine}{Rhodamine}{\rule{0pt}{0pt}\rule{0pt}{0pt}}} ) and lowest for duodenum ( \raisebox{0.3em}{\fcolorbox{du}{du}{\rule{0pt}{0pt}\rule{0pt}{0pt}}} ). However, boundary-based models have led to the maximum relative improvement for the gallbladder ( \raisebox{0.3em}{\fcolorbox{st}{st}{\rule{0pt}{0pt}\rule{0pt}{0pt}}} ), pancreas ( \raisebox{0.3em}{\fcolorbox{RawSienna}{RawSienna}{\rule{0pt}{0pt}\rule{0pt}{0pt}}} ), and duodenum ( \raisebox{0.3em}{\fcolorbox{du}{du}{\rule{0pt}{0pt}\rule{0pt}{0pt}}} ). From \Cref{table:Table 3.4}, we observe that the boundary-constrained models have significantly improved the mean Avg. HD distances for the spleen ( \raisebox{0.2em}{\fcolorbox{Purple}{Purple}{\rule{0pt}{0pt}\rule{0pt}{0pt}}} ), kidney ( \raisebox{0.2em}{\fcolorbox{Rhodamine}{Rhodamine}{\rule{0pt}{0pt}\rule{0pt}{0pt}}} ), and gallbladder ( \raisebox{0.3em}{\fcolorbox{st}{st}{\rule{0pt}{0pt}\rule{0pt}{0pt}}} ). Finally, relating the increase in dice overlap to the organ occurrence (shown in Figures \ref{fig:organ_btvc} and \ref{fig:organ_panc}), we observe that the most significant improvement has occurred for the underweighted classes. In contrast, the boundary distances are maximally decreased for both small (e.g., gallbladder ( \raisebox{0.3em}{\fcolorbox{st}{st}{\rule{0pt}{0pt}\rule{0pt}{0pt}}} ) and large structures like spleen ( \raisebox{0.2em}{\fcolorbox{Purple}{Purple}{\rule{0pt}{0pt}\rule{0pt}{0pt}}} ). Furthermore, the boundary-constrained models led to lower standard deviations of the mean Dice scores and Avg. HDs across different subjects which show the robustness of proposed models.
\begin{figure}[!hbt]
\begin{subfigure}[b]{.23\textwidth}
\includegraphics[width=\textwidth]{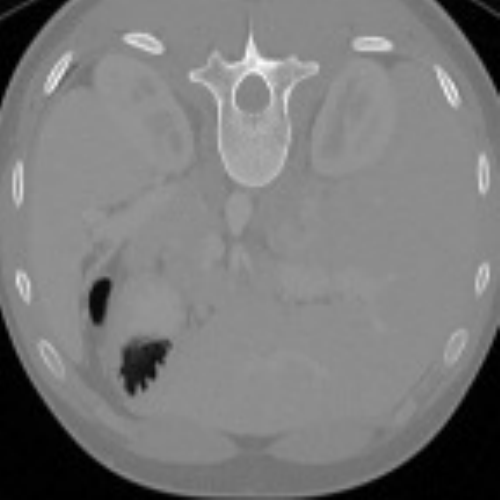} 
\caption{}
\label{fig:origtrimap}
\end{subfigure}
\begin{subfigure}[b]{.23\textwidth}
\includegraphics[width=\textwidth]{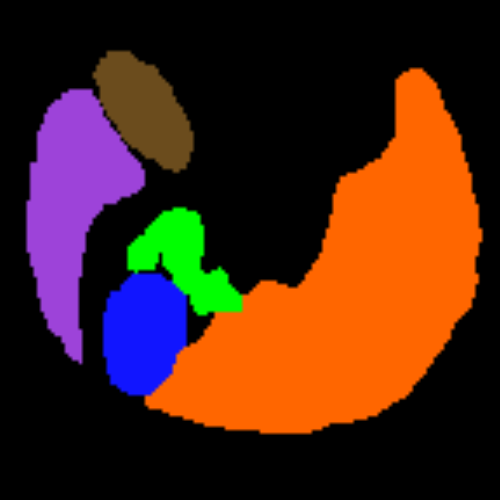}
\caption{}
\label{fig:labelmap}
\end{subfigure}
\begin{subfigure}[b]{.23\textwidth}
\includegraphics[width=\textwidth]{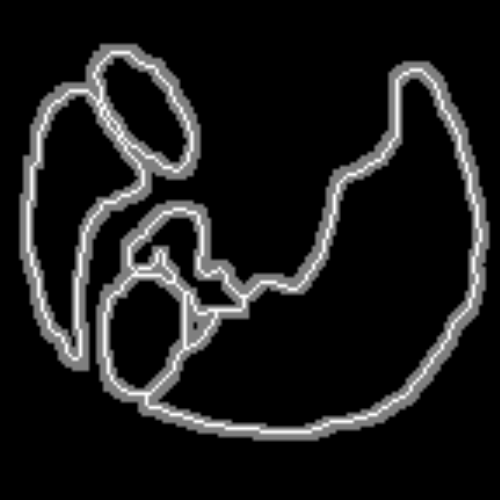}
\caption{}
\label{fig:trmap4}
\end{subfigure}
\begin{subfigure}[b]{.23\textwidth}
\includegraphics[width=\textwidth]{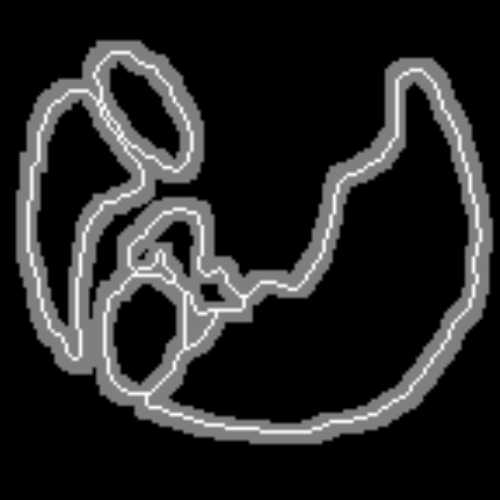}
\caption{}
\label{fig:trmap6}
\end{subfigure}
\caption{Trimap regions along boundary of organs shown on 2D abdominal images. (a) Original abdominal image, (b) Organ label-maps corresponding to 2D image, (c) Trimap region of 5 voxel width around boundaries of organs shown in gray color, and (d) Trimap region of 7 voxel width shown in gray color.
}\label{fig:trimap}
\end{figure}
\begin{figure}[ht]
\begin{subfigure}{.28\textwidth}
\includegraphics[scale=0.20]{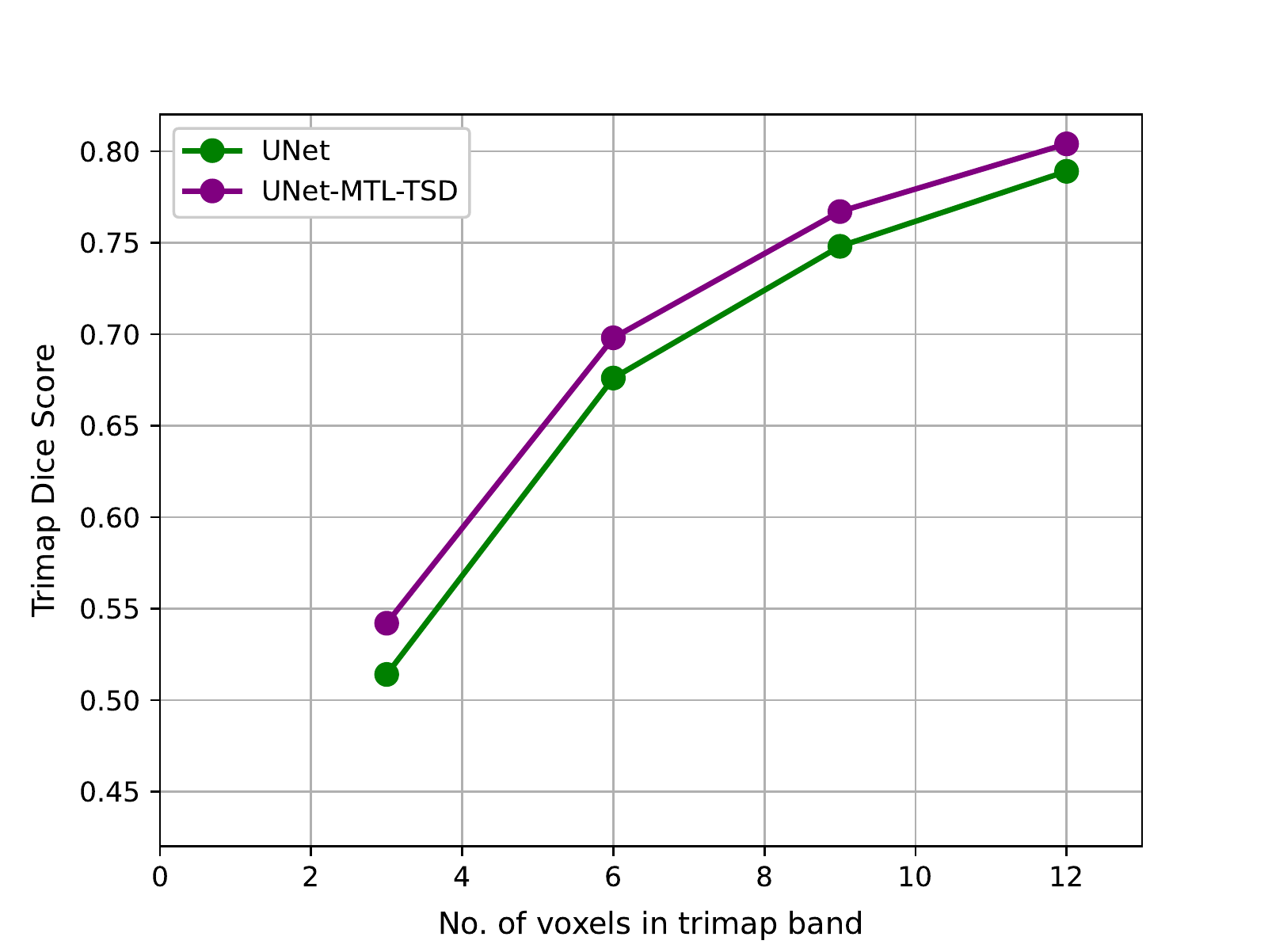}
\hfill
\includegraphics[scale=0.20]{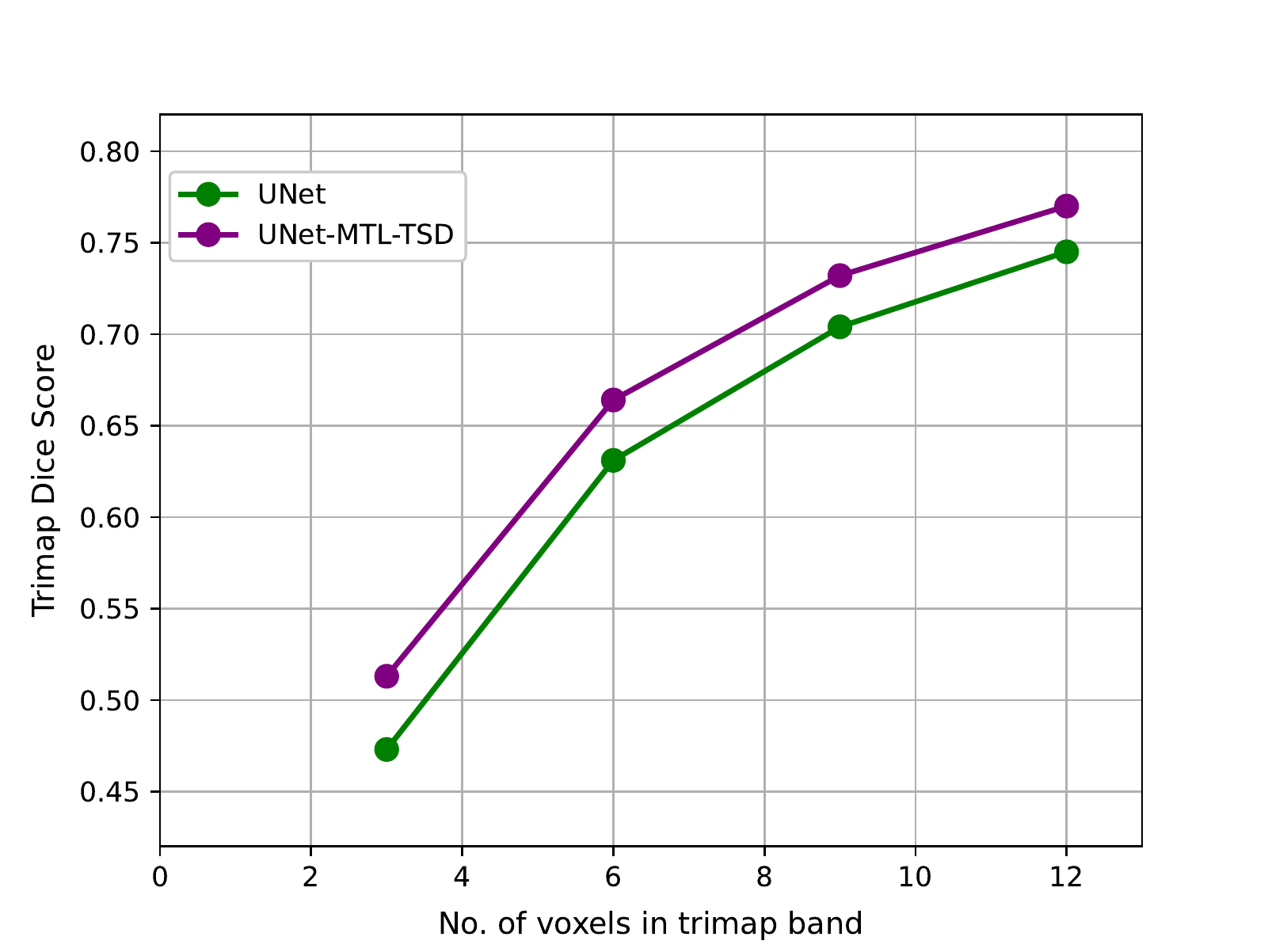}
\caption{}
\end{subfigure}
\hspace*{\fill}
\begin{subfigure}{.28\textwidth}
\includegraphics[scale=0.20]{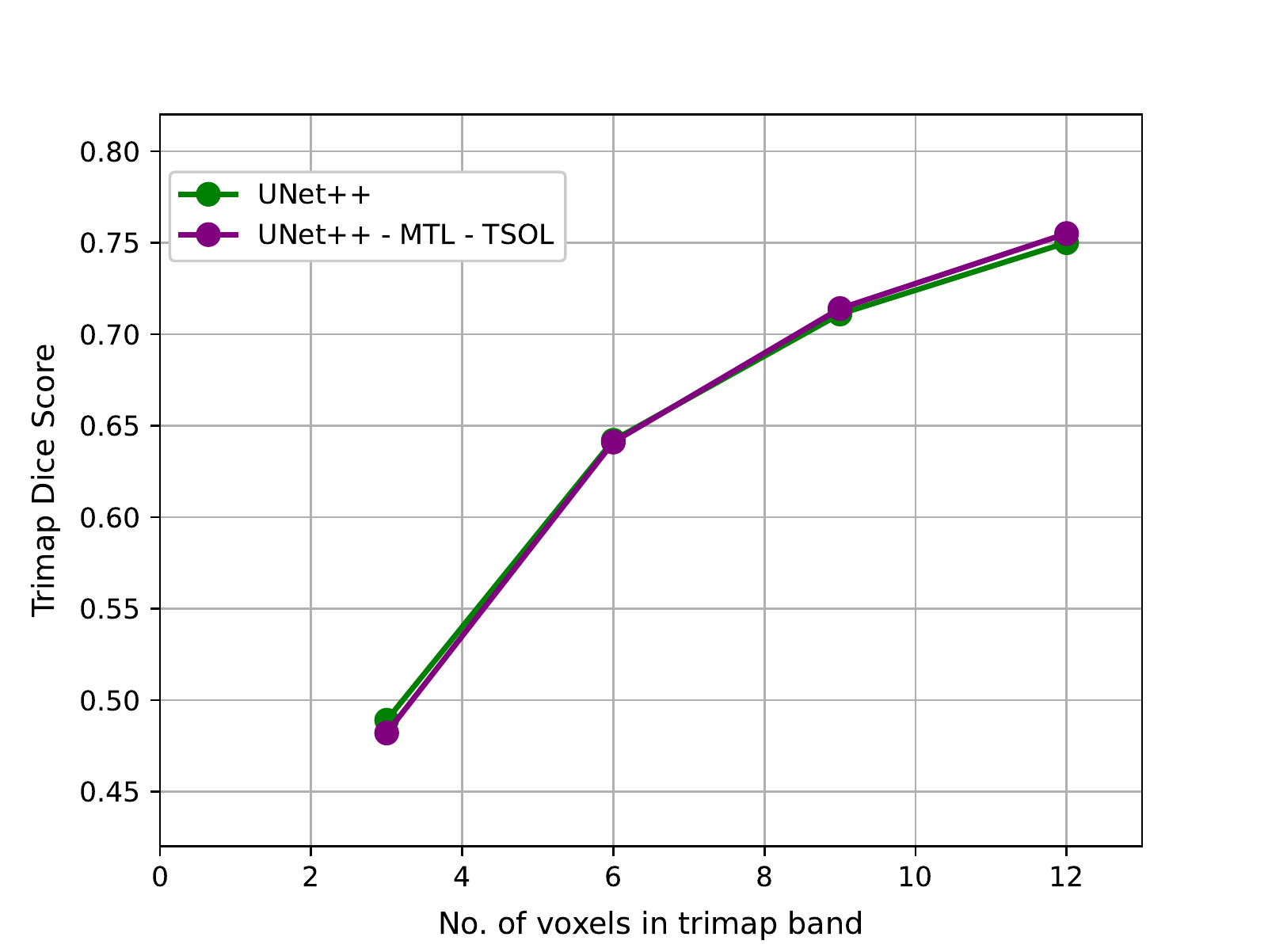}
\hfill
\includegraphics[scale=0.20]{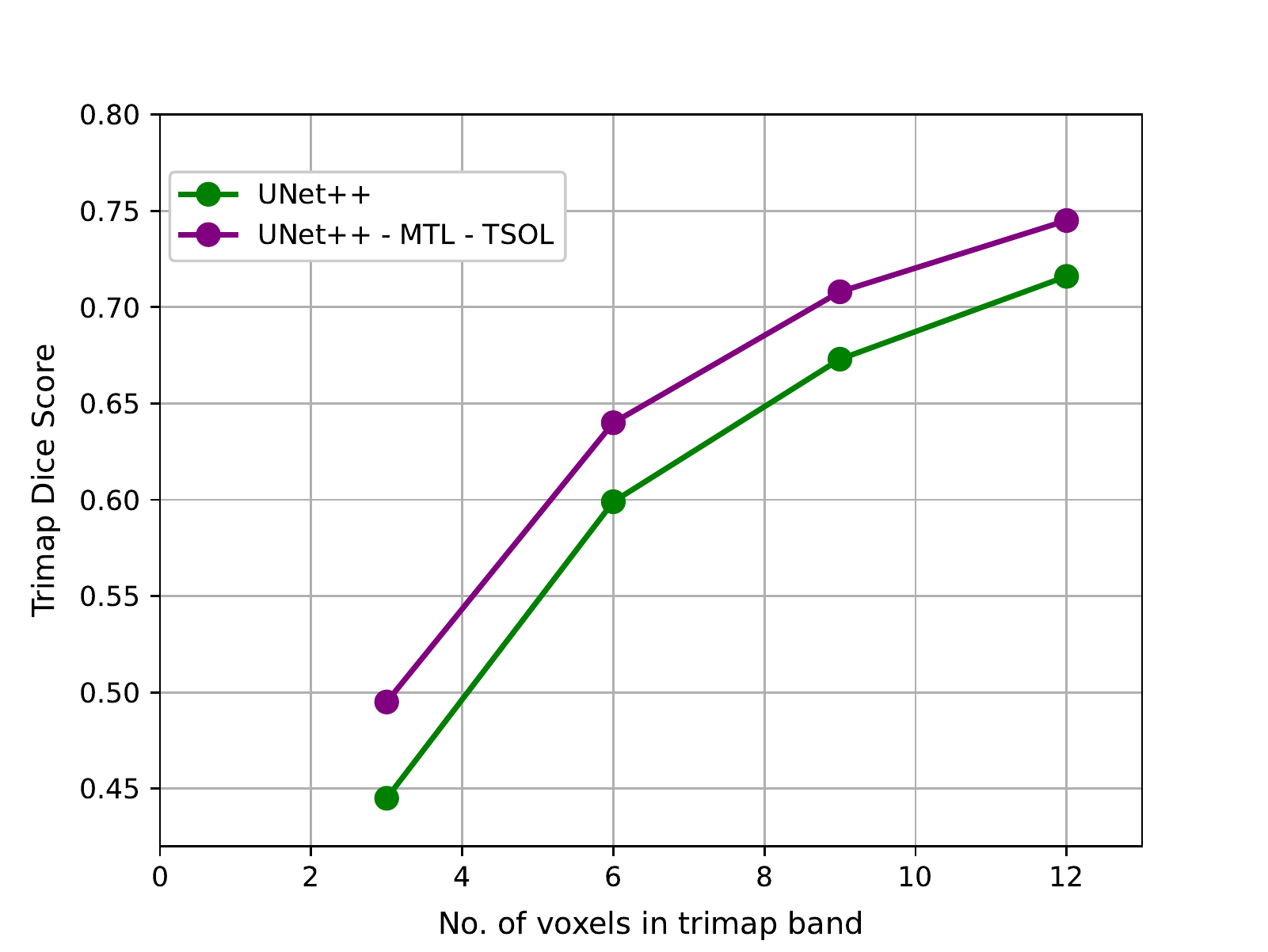}
\caption{}
\end{subfigure}
\hspace*{\fill}
\begin{subfigure}{.28\textwidth}
\includegraphics[scale=0.20]{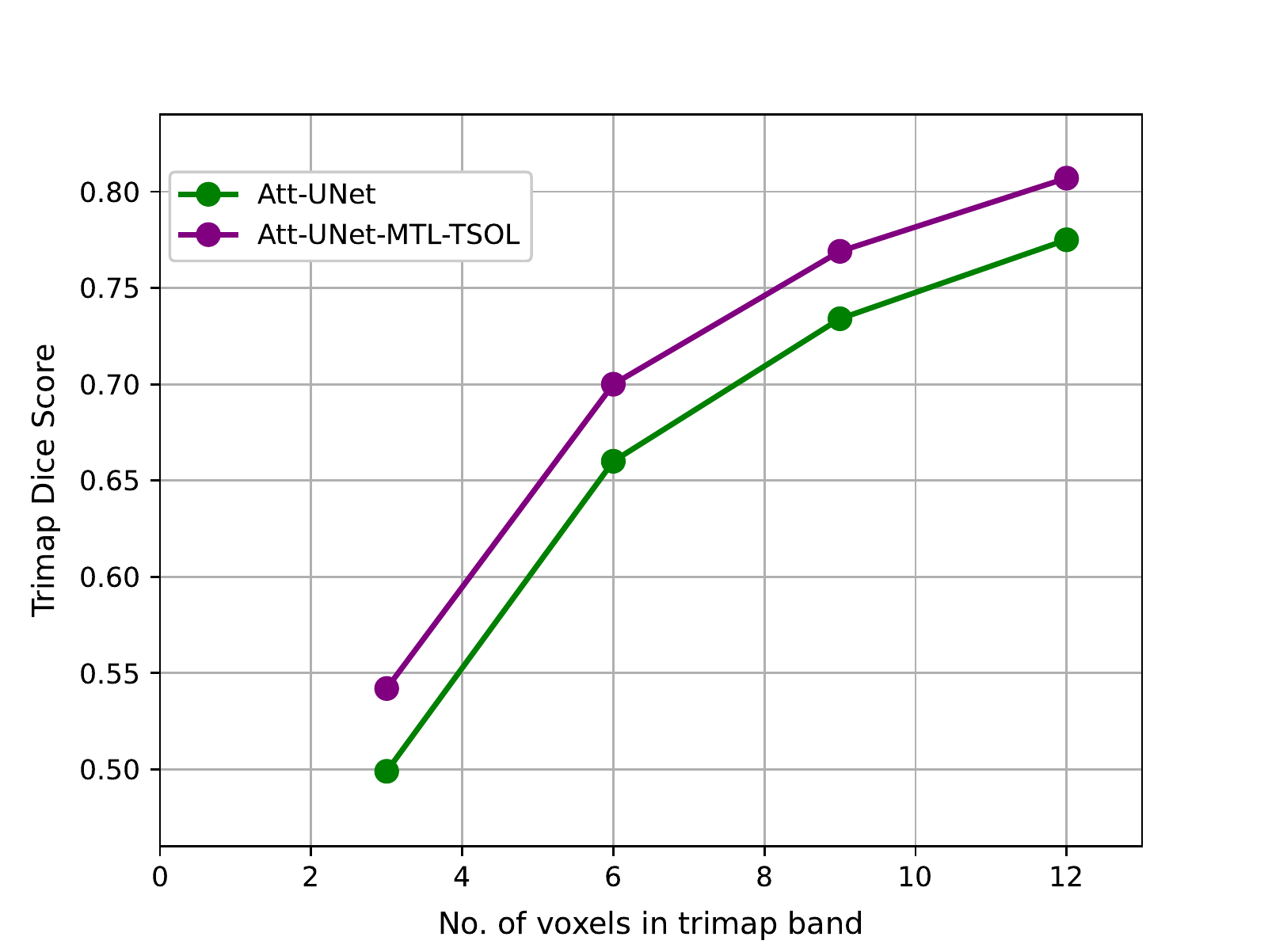}
\hfill
\includegraphics[scale=0.20]{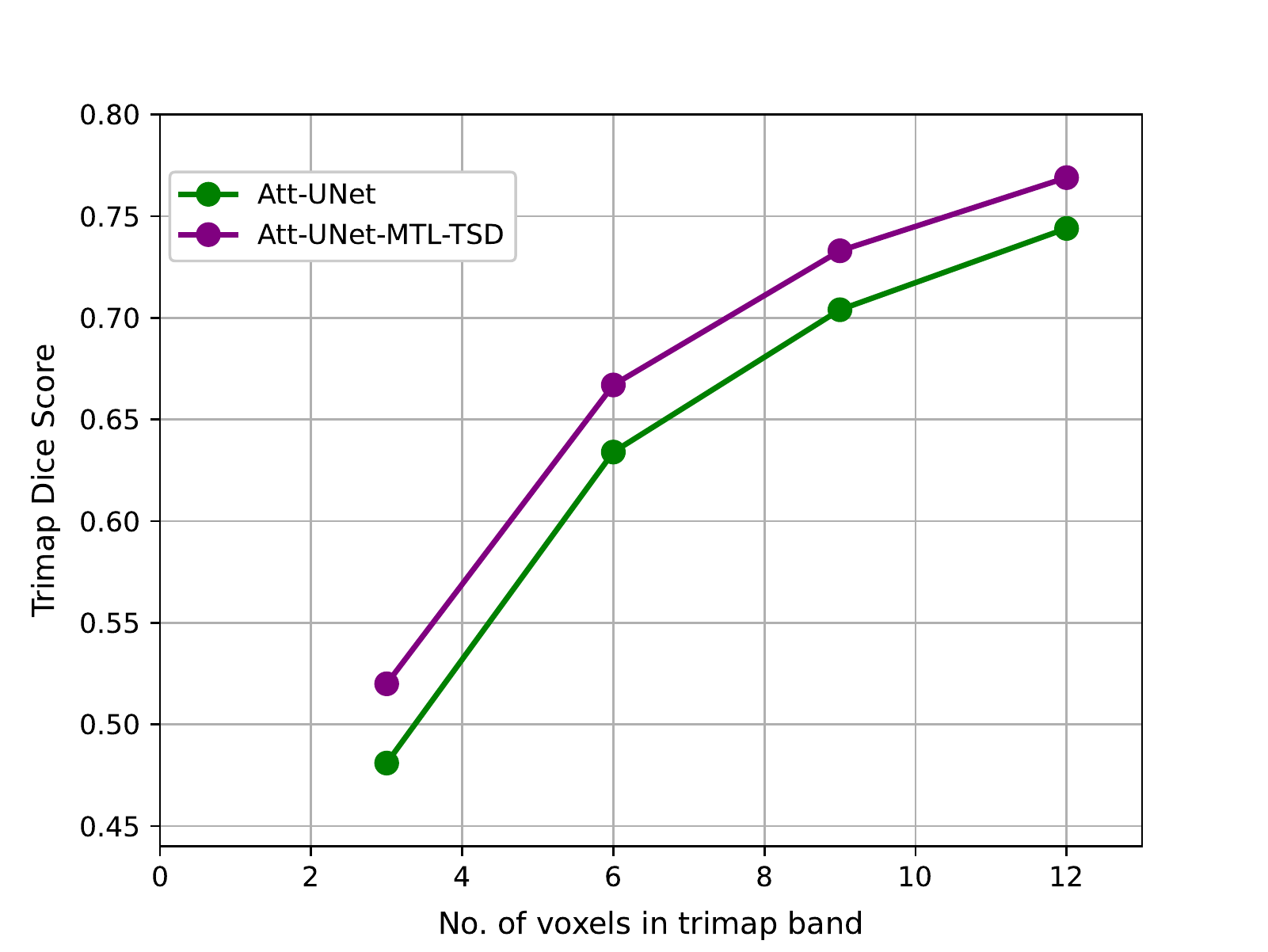}
\caption{}
\end{subfigure}
\hspace*{\fill}
\caption{Evaluation of boundary segmentation via trimaps. In each subfigure, we have plotted the dice scores (y-axis) obtained by comparing the predicted and groundtruth segmentations in trimap regions with varying number of voxels (x-axis). The dice scores for both the baseline (\textcolor{st}{green}) and boundary-constrained counterparts (\textcolor{Purple}{purple}) are plotted. Top row of (a-c) corresponds to trimap segmentation comparison for Pancreas-CT dataset and bottom row corresponds to BTCV dataset.}
\label{fig:tri}
\end{figure}
\subsection{Segmentation performance along boundary of organs}
Unclear boundaries of organs and low inter-organ contrast prevent accurate segmentation of challenging regions around the organ boundaries on abdominal CT scans. To assess if the incorporation of boundary information has improved the segmentation of voxels within the close vicinity of organ boundaries, we evaluate the quality of predicted voxel-labels within these regions and compare them against the ones acquired via baseline methods. To do this, we generate trimaps \cite{4587417} \cite{Cheng2021BoundaryII} with different voxel-bands surrounding the boundaries of organs. Trimap is a narrow region along the boundary of an object which is utilized to evaluate the quality of segmentation within a specific distance from the object's contour. First, we generate trimap regions around the boundary of organs for predicted and ground-truth segmentations, and then, we compare the 3D trimaps by computing mean Dice scores between them. We show the exemplary trimap regions on 2D abdominal axial slices in \autoref{fig:trimap}. For the sake of compendious presentation, we have computed the trimap Dice scores only for TSOL network topology. \autoref{fig:tri} shows the mean Dice score plotted against the number of voxels the trimap contains. The top row shows trimap plots for the Pancreas-CT dataset, whereas the bottom row shows the trimap plots for the BTCV dataset. Our proposed boundary-constrained models consistently perform better than the baseline models in predicting the semantic labels within the vicinity of organ-boundaries, except for one case, i.e., (3D UNet$_{++}$ vs 3D UNet$_{++}$-MTL-TSOL). 
\subsection{Qualitative Results}
\begin{figure*}[!hbt]
\centering
\begin{subfigure}{.23\textwidth}
\centering
\includegraphics[scale=0.25]{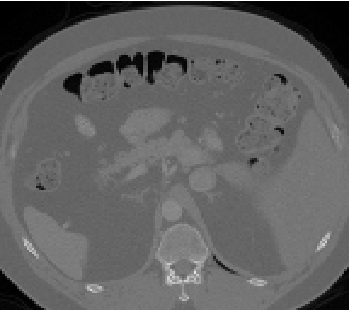}
\hfill
\includegraphics[scale=0.25]{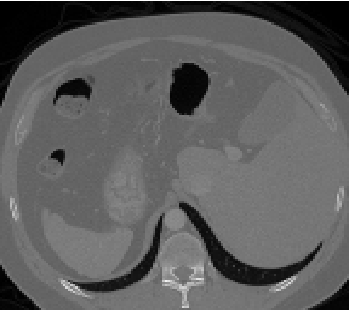}
\hfill
\includegraphics[scale=0.25]{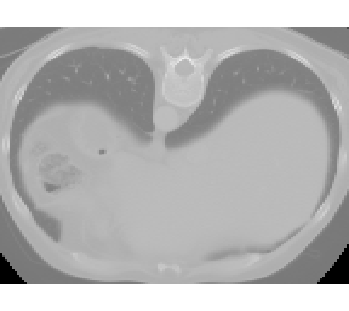}
\hfill
\includegraphics[scale=0.25]{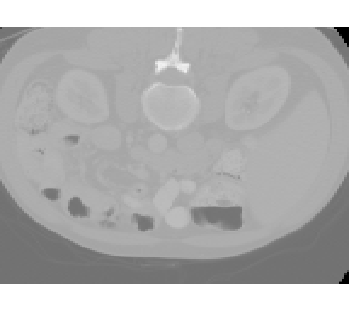}
\phantomcaption
\label{fig:sub30}
\end{subfigure}%
\begin{subfigure}{.23\textwidth}
\centering
\includegraphics[scale=0.25]{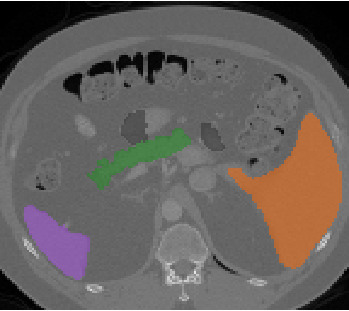}
\hfill
\includegraphics[scale=0.25]{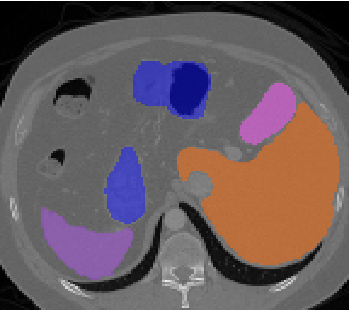}
\hfill
\includegraphics[scale=0.25]{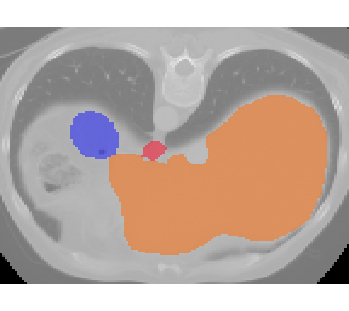}
\hfill
\includegraphics[scale=0.25]{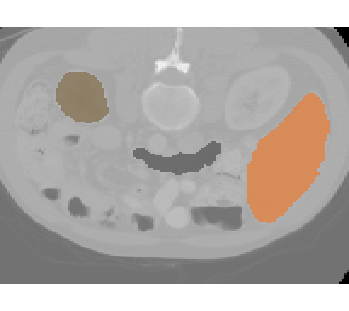}
\phantomcaption
\label{fig:sub31}
\end{subfigure}
\begin{subfigure}{.23\textwidth}
\centering
\includegraphics[scale=0.25]{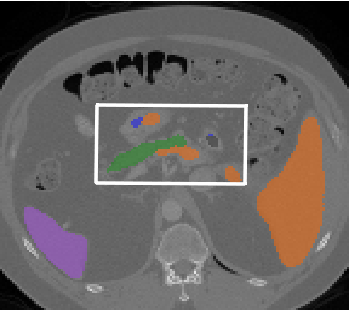}
\hfill
\includegraphics[scale=0.25]{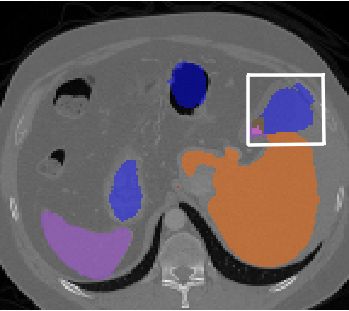}
\hfill
\includegraphics[scale=0.25]{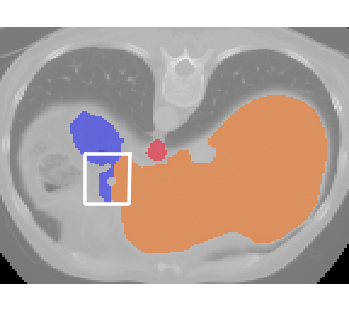}
\hfill
\includegraphics[scale=0.25]{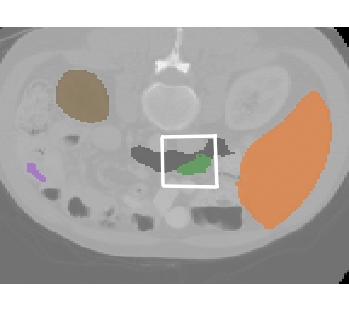}
\phantomcaption
\label{fig:sub32}
\end{subfigure}
\begin{subfigure}{.23\textwidth}
\centering
\includegraphics[scale=0.25]{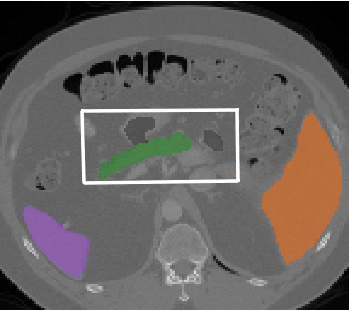}
\hfill
\includegraphics[scale=0.25]{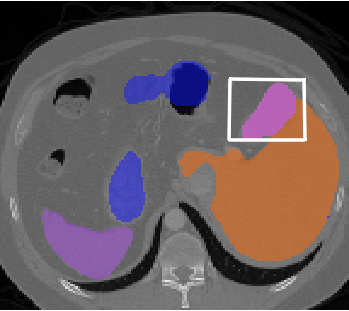}
\hfill 
\includegraphics[scale=0.25]{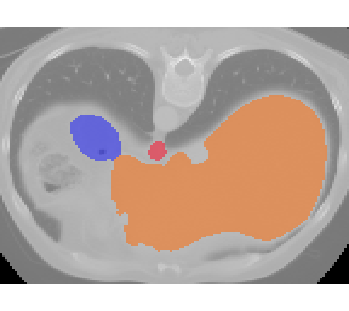}
\hfill
\includegraphics[scale=0.25]{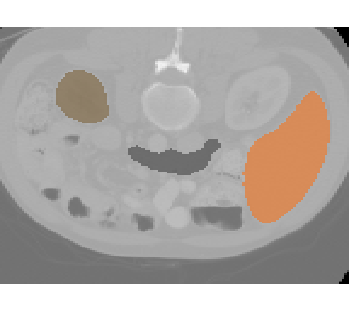}
\phantomcaption
\label{fig:sub33}
\end{subfigure}
\caption{Qualitative results for multi-organ segmentation between baseline and boundary-constrained models are shown. Rows 1-2 correspond to results for the Pancreas-CT dataset, whereas Rows 3-4 show results for the BTCV dataset. Columns 1-2 show the original image and the corresponding ground-truth mask overlayed on the 2D image, i.e., spleen ( \raisebox{0.3em}{\fcolorbox{Purple}{Purple}{\rule{0pt}{0pt}\rule{0pt}{0pt}}} ), left kidney ( \raisebox{0.3em}{\fcolorbox{Rhodamine}{Rhodamine}{\rule{0pt}{0pt}\rule{0pt}{0pt}}} ), gallbladder ( \raisebox{0.3em}{\fcolorbox{st}{st}{\rule{0pt}{0pt}\rule{0pt}{0pt}}} ), liver ( \raisebox{0.3em}{\fcolorbox{BurntOrange}{BurntOrange}{\rule{0pt}{0pt}\rule{0pt}{0pt}}} ), stomach ( \raisebox{0.3em}{\fcolorbox{lv}{lv}{\rule{0pt}{0pt}\rule{0pt}{0pt}}} ), and duodenum ( \raisebox{0.3em}{\fcolorbox{du}{du}{\rule{0pt}{0pt}\rule{0pt}{0pt}}} ). Columns 3-4 illustrate the segmentation results related to baseline UNet and UNet-MTL-TSOL counterparts. White boxes indicate the segmented regions improved by incorporation of boundary information.}
\label{fig:visual}
\end{figure*}

\begin{figure}[!hbt]
\begin{subfigure}[b]{.27\textwidth}
\hspace{-2cm}
\includegraphics[trim={2cm 2cm 2cm 0.5cm},clip,width=\textwidth]{Figures/panc_ct6_3d.jpg}
\hfill
\includegraphics[trim={2cm 2cm 2cm 2cm},clip,width=\textwidth]{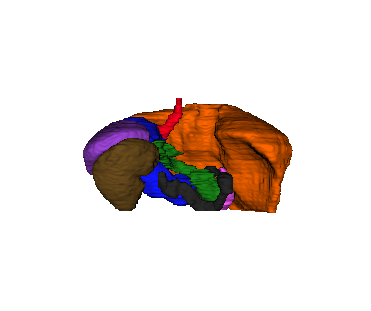}
\phantomcaption
\label{fig:sub34}
\end{subfigure}%
\begin{subfigure}[b]{.27\textwidth}
\centering
\includegraphics[trim={2cm 2cm 2cm 0.5cm},clip,scale=0.23]{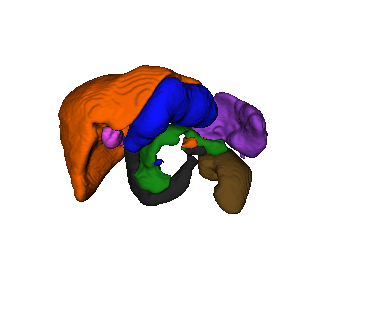}
\hfill
\includegraphics[trim={2cm 2cm 2cm 1.6cm},clip,width=\textwidth]{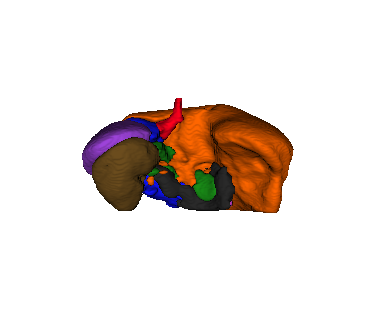}
\phantomcaption
\label{fig:sub35}
\end{subfigure}
\begin{subfigure}[b]{.27\textwidth}
\centering
\includegraphics[trim={2cm 2cm 2cm 0.5cm},clip,scale=0.23]{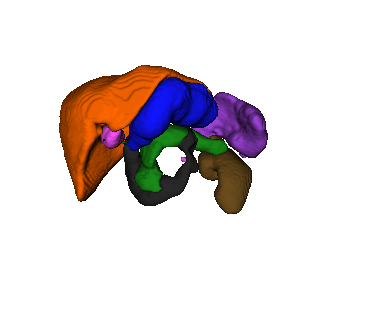}
\hfill
\includegraphics[trim={2cm 2cm 2cm 1.6cm},clip,width=\textwidth]{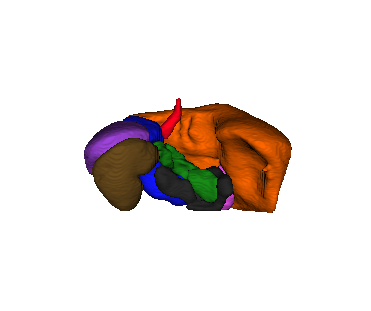}
\phantomcaption
\label{fig:sub36}
\end{subfigure}
\caption{Visualization of 3D segmentation predictions using Pancreas-CT and BTCV dataset. First column corresponds to abdominal groundtruth annotations, i.e., spleen ( \raisebox{0.3em}{\fcolorbox{Purple}{Purple}{\rule{0pt}{0pt}\rule{0pt}{0pt}}} ), pancreas ( \raisebox{0.3em}{\fcolorbox{RawSienna}{RawSienna}{\rule{0pt}{0pt}\rule{0pt}{0pt}}} ), left kidney ( \raisebox{0.3em}{\fcolorbox{Rhodamine}{Rhodamine}{\rule{0pt}{0pt}\rule{0pt}{0pt}}} ), gallbladder ( \raisebox{0.3em}{\fcolorbox{st}{st}{\rule{0pt}{0pt}\rule{0pt}{0pt}}} ), esophagus ( \raisebox{0.3em}{\fcolorbox{red}{red}{\rule{0pt}{0pt}\rule{0pt}{0pt}}} ), liver ( \raisebox{0.3em}{\fcolorbox{BurntOrange}{BurntOrange}{\rule{0pt}{0pt}\rule{0pt}{0pt}}} ), stomach ( \raisebox{0.3em}{\fcolorbox{lv}{lv}{\rule{0pt}{0pt}\rule{0pt}{0pt}}} ), and duodenum ( \raisebox{0.3em}{\fcolorbox{du}{du}{\rule{0pt}{0pt}\rule{0pt}{0pt}}} ), whereas second and third column corresponds to volume labels predicted by UNet and UNet-MTL-TSD.}
\label{fig:sub37}
\end{figure}

\autoref{fig:visual} shows semantic segmentation predictions on a single 2D axial slice of the 3D scans. The first and second row correspond to segmentation results from the Pancreas-CT dataset whereas, the third and fourth row corresponds to results from the BTCV dataset. Each column (from left to right) illustrates the original abdominal 2D images (\autoref{fig:sub30}), ground-truth masks (\autoref{fig:sub31}), baseline model (UNet and Att-UNet) segmentations (\autoref{fig:sub32}), and segmentations acquired from the boundary-constrained counterparts (UNet-MTL-TSD and Att-UNet-MTL-TSOL) in (\autoref{fig:sub33}). We observe that the baseline models led to under-segmentations and over-segmentations, indicated in white boxes in \autoref{fig:sub32}. Furthermore, the segmentations generated by single-task baseline models show isolated and biologically implausible organs' parts. Moreover, comparing with the corresponding boundary-constrained segmentations (\autoref{fig:sub33}), the incorporation of boundary information has prevented the issue of mispredictions near boundary of organs and led to generation of biologically plausible segmentations. \autoref{fig:sub37} presents the 3D segmentations generated by baseline (UNet) and boundary-constrained model (UNet-MTL-TSD) along with the ground-truths from the Pancreas-CT (first row) and BTCV dataset (second row). Notice that the boundary-constrained segmentations (third column) are more similar to the ground-truth masks (first column) as compared to the baseline segmentations (second column). These qualitative results show the improvements induced by the use of organs borders in training the 3D fully convolutional models for abdominal organs segmentation.
\section{Discussion}\label{sec:discuss}
Accurate segmentation of abdominal organs from CT scans is required for numerous advanced clinical procedures such as computer-assisted surgery and organ transplantation. The low-contrasted appearance and weak edges of abdominal organs in CT scans adversely affect the accurate segmentation. In this paper, we propose to leverage boundary information of organs as an additional cue for improved 3D abdominal multi-organ segmentation. The boundary-constrained encoder-decoder network simultaneously learns to delineate the semantic abdominal regions and detect the boundaries of organs. This multi-task learning model exploits the statistics from more than one ground-truth source and subsequently retains features shared between the tasks. The boundary annotations of abdominal organs can be easily generated from the ground-truth masks and provide cost-free additional knowledge about the organs.

As reported by quantitative results in \autoref{table:Table 3.1}, the proposed boundary-constrained 3D encoder-decoder models achieve improved multi-organ segmentation across the majority of the baselines (3D UNet, 3D UNet$_{++}$, and 3D Att-UNet) and datasets (Pancreas-CT and BTCV). We have shown that the significant improvement in segmentation performance evaluated via Dice Score, Avg. HD, Recall, and Precision is caused by the improved segmentation of organs' boundaries and regions surrounding boundaries. Furthermore, significant improvements with a negligible increase in parameter count (0.0002\% -TSOL topology) reveal the benefit of regularizing the existing encoder-decoder segmentation models using boundary information.  

The reduction in Avg. HD for both datasets across all baselines depicts the advantage of informing the model about the vulnerable regions of the organ. The dramatic decrease in Avg. HD, ranging from 18\% to 30\%, shows that the model learned feature combinations that were expressive about the entire appearance of organs. We believe that training the models with auxiliary knowledge encourages learning more generalizable and discriminative features. Notably, the additional experiments that we have conducted to precisely assess the improvements in segmentation of regions in the vicinity of organ-boundaries further verify the superiority of training the segmentation model with complementary boundary information (shown in \autoref{fig:sub36}).

Since there may exist several different configurations through which a fully convolutional architecture can be designed under a multi-task learning paradigm, we explore two network topologies based on the extent of parameter-sharing between the tasks. Through our extensive comparison, we notice that an overly-shared multi-task network (TSOL) performs on par with the network designed to have an increased number of task-specific layers (TSD) (\autoref{table:Table 3.1}). This indicates that models with many parameters do not necessarily correspond to higher performance. Most importantly, we also found that incorporation of boundary information improved the multi-organ segmentation performance, regardless of the network topology. One of the critical challenges in designing 3D multi-task deep learning models is determining which layers should be shared while keeping the computational expense reasonable. In the future, we aim to investigate other network topologies for training the encoder-decoder model in a multi-task learning fashion. Another valuable extension of our work is to develop a mechanism/policy that can automatically dictate the sharing pattern of network layers between the two tasks.

As reported in \Cref{table:Table 3.3,table:Table 3.4}, the improvement in the segmentation performance of organs rarely occurring in the dataset reveals the efficacy of boundary-regularized models in compensating for their rare presence. The qualitative results shown in \Cref{fig:visual,fig:sub37} verify the positive impact of making the model aware of organ-boundaries during training. The ability of our model to simultaneously learn the improved representations of multiple organs is indicated by the qualitative examples in \Cref{fig:visual,fig:sub37}, where the boundary-constrain has reduced the occurrence of over- and under-segmented organs.
\section{Conclusion}\label{sec:conc}
In this paper, we leverage organ boundary information for an improved 3D abdominal multi-organ segmentation by addressing the challenge of unclear boundaries in low-contrasted CT scans. We demonstrate that boundary information can be seamlessly introduced in the training of 3D encoder-decoder models through different multi-task configurations. The experimental results show the boundary-constrained multi-organ segmentation outperforms the ones obtained from several FCN-based baseline models, including 3D UNet, 3D UNet$_{++}$, 3D Attention-UNet. Furthermore, we found that the multi-task topology that shows maximum improvement is not fixed and varies depending on the baseline architecture. This insight shows that the optimal utilization of auxiliary information cannot always be harvested through a single deep multi-task design but instead requires the exploration of different multi-task topologies. Our findings also reveal that leveraging organs boundary features improves the segmentation of underweighted organs like the gallbladder, pancreas, and duodenum with a negligible parameter-overhead. Additionally, the experimental results disclose that the boundary-constrained models improve the labelling of weak sub-parts of organs in the vicinity of boundaries. We believe the proposed 3D boundary-constrained models would be valuable for enhancing abdominal organ segmentation and utilizing those segmentations in relevant clinical applications.\\\\
\textbf{Conflict of interest statement:} On behalf of all authors, the corresponding author states that there is no conflict of interest.
\bibliographystyle{elsarticle-num} 
\bibliography{elsarticle_arxiv}

\begin{thebibliography}{10}
\expandafter\ifx\csname url\endcsname\relax
  \def\url#1{\texttt{#1}}\fi
\expandafter\ifx\csname urlprefix\endcsname\relax\def\urlprefix{URL }\fi
\expandafter\ifx\csname href\endcsname\relax
  \def\href#1#2{#2} \def\path#1{#1}\fi

\bibitem{okada2012multi}
T.~Okada, M.~G. Linguraru, M.~Hori, Y.~Suzuki, R.~M. Summers, N.~Tomiyama,
  Y.~Sato, Multi-organ segmentation in abdominal ct images, in: 2012 Annual
  International Conference of the IEEE Engineering in Medicine and Biology
  Society, IEEE, 2012, pp. 3986--3989.

\bibitem{gibson2018automatic}
E.~Gibson, F.~Giganti, Y.~Hu, E.~Bonmati, S.~Bandula, K.~Gurusamy, B.~Davidson,
  S.~P. Pereira, M.~J. Clarkson, D.~C. Barratt, Automatic multi-organ
  segmentation on abdominal ct with dense v-networks, IEEE transactions on
  medical imaging 37~(8) (2018) 1822--1834.

\bibitem{hu2017automatic}
P.~Hu, F.~Wu, J.~Peng, Y.~Bao, F.~Chen, D.~Kong, Automatic abdominal
  multi-organ segmentation using deep convolutional neural network and
  time-implicit level sets, International journal of computer assisted
  radiology and surgery 12~(3) (2017) 399--411.

\bibitem{article1kim}
H.~Kim, J.~Jung, J.~Kim, B.~Cho, J.~Jang, S.-w. Lee, J.-G. Lee, S.~Yoon,
  Abdominal multi-organ auto-segmentation using 3d-patch-based deep
  convolutional neural network, Scientific Reports 10 (2020) 6204.
\newblock \href {https://doi.org/10.1038/s41598-020-63285-0}
  {\path{doi:10.1038/s41598-020-63285-0}}.

\bibitem{xu2015efficient}
Z.~Xu, R.~P. Burke, C.~P. Lee, R.~B. Baucom, B.~K. Poulose, R.~G. Abramson,
  B.~A. Landman, Efficient multi-atlas abdominal segmentation on clinically
  acquired ct with simple context learning, Medical image analysis 24~(1)
  (2015) 18--27.

\bibitem{suzuki2012multi}
M.~Suzuki, M.~G. Linguraru, K.~Okada, Multi-organ segmentation with missing
  organs in abdominal ct images, in: International Conference on Medical Image
  Computing and Computer-Assisted Intervention, Springer, 2012, pp. 418--425.

\bibitem{cerrolaza2015automatic}
J.~J. Cerrolaza, M.~Reyes, R.~M. Summers, M.~{\'A}. Gonz{\'a}lez-Ballester,
  M.~G. Linguraru, Automatic multi-resolution shape modeling of multi-organ
  structures, Medical image analysis 25~(1) (2015) 11--21.

\bibitem{okada2015abdominal}
T.~Okada, M.~G. Linguraru, M.~Hori, R.~M. Summers, N.~Tomiyama, Y.~Sato,
  Abdominal multi-organ segmentation from ct images using conditional
  shape--location and unsupervised intensity priors, Medical image analysis
  26~(1) (2015) 1--18.

\bibitem{campadelli2009automatic}
P.~Campadelli, E.~Casiraghi, S.~Pratissoli, G.~Lombardi, Automatic abdominal
  organ segmentation from ct images, ELCVIA: electronic letters on computer
  vision and image analysis 8~(1) (2009) 1--14.

\bibitem{Selver2014SegmentationOA}
M.~A. Selver, Segmentation of abdominal organs from ct using a multi-level,
  hierarchical neural network strategy, Computer methods and programs in
  biomedicine 113 3 (2014) 830--52.

\bibitem{heinrich2019obelisk}
M.~P. Heinrich, O.~Oktay, N.~Bouteldja, Obelisk-net: Fewer layers to solve 3d
  multi-organ segmentation with sparse deformable convolutions, Medical image
  analysis 54 (2019) 1--9.

\bibitem{bobo2018fully}
M.~F. Bobo, S.~Bao, Y.~Huo, Y.~Yao, J.~Virostko, A.~J. Plassard, I.~Lyu,
  A.~Assad, R.~G. Abramson, M.~A. Hilmes, et~al., Fully convolutional neural
  networks improve abdominal organ segmentation, in: Medical Imaging 2018:
  Image Processing, Vol. 10574, International Society for Optics and Photonics,
  2018, p. 105742V.

\bibitem{sinha2020multi}
A.~Sinha, J.~Dolz, Multi-scale self-guided attention for medical image
  segmentation, IEEE Journal of Biomedical and Health Informatics (2020).

\bibitem{Lu2022}
H.~Lu, S.~Tian, L.~Yu, L.~Liu, J.~Cheng, W.~Wu, X.~Kang, D.~Zhang, {DCACNet}:
  Dual context aggregation and attention-guided cross deconvolution network for
  medical image segmentation, Computer Methods and Programs in Biomedicine 214
  (2022) 106566.
\newblock \href {https://doi.org/10.1016/j.cmpb.2021.106566}
  {\path{doi:10.1016/j.cmpb.2021.106566}}.

\bibitem{ronneberger2015u}
O.~Ronneberger, P.~Fischer, T.~Brox, U-net: Convolutional networks for
  biomedical image segmentation, in: International Conference on Medical image
  computing and computer-assisted intervention, Springer, 2015, pp. 234--241.

\bibitem{Roth2018AnAO}
H.~R. Roth, H.~Oda, X.~Zhou, N.~Shimizu, Y.~Yang, Y.~Hayashi, M.~Oda,
  M.~Fujiwara, K.~Misawa, K.~Mori, An application of cascaded 3d fully
  convolutional networks for medical image segmentation, Computerized medical
  imaging and graphics : the official journal of the Computerized Medical
  Imaging Society 66 (2018) 90--99.

\bibitem{Zhou2018PerformanceEO}
X.~Zhou, K.~Yamada, T.~Kojima, R.~Takayama, S.~Wang, X.~Zhou, T.~Hara,
  H.~Fujita, Performance evaluation of 2d and 3d deep learning approaches for
  automatic segmentation of multiple organs on ct images, in: Medical Imaging,
  2018.

\bibitem{Guo2021}
W.~Guo, G.~Zhang, Z.~Gong, Q.~Li, Effective integration of object boundaries
  and regions for improving the performance of medical image segmentation by
  using two cascaded networks, Computer Methods and Programs in Biomedicine 212
  (2021) 106423.
\newblock \href {https://doi.org/10.1016/j.cmpb.2021.106423}
  {\path{doi:10.1016/j.cmpb.2021.106423}}.

\bibitem{chen2016dcan}
H.~Chen, X.~Qi, L.~Yu, P.-A. Heng, Dcan: deep contour-aware networks for
  accurate gland segmentation, in: Proceedings of the IEEE conference on
  Computer Vision and Pattern Recognition, 2016, pp. 2487--2496.

\bibitem{8363791}
C.~{Tan}, L.~{Zhao}, Z.~{Yan}, K.~{Li}, D.~{Metaxas}, Y.~{Zhan}, Deep
  multi-task and task-specific feature learning network for robust shape
  preserved organ segmentation, in: 2018 IEEE 15th International Symposium on
  Biomedical Imaging (ISBI 2018), 2018, pp. 1221--1224.
\newblock \href {https://doi.org/10.1109/ISBI.2018.8363791}
  {\path{doi:10.1109/ISBI.2018.8363791}}.

\bibitem{8906008}
R.~{Zhao}, W.~{Chen}, G.~{Cao}, Edge-boosted u-net for 2d medical image
  segmentation, IEEE Access 7 (2019) 171214--171222.
\newblock \href {https://doi.org/10.1109/ACCESS.2019.2953727}
  {\path{doi:10.1109/ACCESS.2019.2953727}}.

\bibitem{9313420}
B.~{Shi}, H.~{Zhang}, R.~{Yan}, W.~{Jing}, J.~{Zang}, F.~{Zhang}, Lr-net: A
  multi-task model using relationship-based contour information to enhance the
  semantic segmentation of cancer regions, in: 2020 IEEE International
  Conference on Bioinformatics and Biomedicine (BIBM), 2020, pp. 2371--2378.
\newblock \href {https://doi.org/10.1109/BIBM49941.2020.9313420}
  {\path{doi:10.1109/BIBM49941.2020.9313420}}.

\bibitem{MEI2020101988}
L.~Mei, X.~Guo, X.~Huang, Y.~Weng, S.~Liu, C.~Lei,
  \href{https://www.sciencedirect.com/science/article/pii/S1746809420301440}{Dense
  contour-imbalance aware framework for colon gland instance segmentation},
  Biomedical Signal Processing and Control 60 (2020) 101988.
\newblock \href {https://doi.org/https://doi.org/10.1016/j.bspc.2020.101988}
  {\path{doi:https://doi.org/10.1016/j.bspc.2020.101988}}.
\newline\urlprefix\url{https://www.sciencedirect.com/science/article/pii/S1746809420301440}

\bibitem{Amyar2020MultitaskDL}
A.~Amyar, R.~Modzelewski, H.~Li, S.~Ruan, Multi-task deep learning based ct
  imaging analysis for covid-19 pneumonia: Classification and segmentation,
  Computers in Biology and Medicine 126 (2020) 104037 -- 104037.

\bibitem{Du2019CrossInfoNetMI}
K.~Du, X.~Lin, Y.~Sun, X.~Ma, Crossinfonet: Multi-task information sharing
  based hand pose estimation, 2019 IEEE/CVF Conference on Computer Vision and
  Pattern Recognition (CVPR) (2019) 9888--9897.

\bibitem{cciccek20163d}
{\"O}.~{\c{C}}i{\c{c}}ek, A.~Abdulkadir, S.~S. Lienkamp, T.~Brox,
  O.~Ronneberger, 3d u-net: learning dense volumetric segmentation from sparse
  annotation, in: International conference on medical image computing and
  computer-assisted intervention, Springer, 2016, pp. 424--432.

\bibitem{zhou2019unet++}
Z.~Zhou, M.~M.~R. Siddiquee, N.~Tajbakhsh, J.~Liang, Unet++: Redesigning skip
  connections to exploit multiscale features in image segmentation, IEEE
  transactions on medical imaging 39~(6) (2019) 1856--1867.

\bibitem{schlemper2019attention}
J.~Schlemper, O.~Oktay, M.~Schaap, M.~Heinrich, B.~Kainz, B.~Glocker,
  D.~Rueckert, Attention gated networks: Learning to leverage salient regions
  in medical images, Medical image analysis 53 (2019) 197--207.

\bibitem{Roth2016}
H.~Roth, A.~Farag, E.~B. Turkbey, L.~Lu, J.~Liu, R.~M. Summers, Data from
  pancreas-ct (2016).
\newblock \href {https://doi.org/10.7937/K9/TCIA.2016.TNB1KQBU}
  {\path{doi:10.7937/K9/TCIA.2016.TNB1KQBU}}.

\bibitem{landmanbvc}
B.~Landman, Z.~Xu, J.~E. Igelsias, M.~Styner, T.~R. Langerak, A.~Klein, Miccai
  multi-atlas labeling beyond the cranial vault - workshop and challenge. 2015,
  accessed August 2020.

\bibitem{heimann2009statistical}
T.~Heimann, H.-P. Meinzer, Statistical shape models for 3d medical image
  segmentation: a review, Medical image analysis 13~(4) (2009) 543--563.

\bibitem{cootes2001active}
T.~F. Cootes, G.~J. Edwards, C.~J. Taylor, Active appearance models, IEEE
  Transactions on pattern analysis and machine intelligence 23~(6) (2001)
  681--685.

\bibitem{xu2016evaluation}
Z.~Xu, C.~P. Lee, M.~P. Heinrich, M.~Modat, D.~Rueckert, S.~Ourselin, R.~G.
  Abramson, B.~A. Landman, Evaluation of six registration methods for the human
  abdomen on clinically acquired ct, IEEE Transactions on Biomedical
  Engineering 63~(8) (2016) 1563--1572.

\bibitem{lombaert2014laplacian}
H.~Lombaert, D.~Zikic, A.~Criminisi, N.~Ayache, Laplacian forests: semantic
  image segmentation by guided bagging, in: International Conference on Medical
  Image Computing and Computer-Assisted Intervention, Springer, 2014, pp.
  496--504.

\bibitem{peng2019method}
Z.~Peng, X.~Fang, P.~Yan, H.~Shan, T.~Liu, X.~Pei, G.~Wang, B.~Liu, M.~Kalra,
  X.~G. Xu, A method of rapid quantification of patient-specific organ dose for
  ct using coupled deep multi-organ segmentation algorithms and gpu-accelerated
  monte carlo dose computing code, arXiv preprint arXiv:1908.00360 (2019).

\bibitem{chen2017rbnet}
Z.~Chen, Z.~Chen, Rbnet: A deep neural network for unified road and road
  boundary detection, in: International Conference on Neural Information
  Processing, Springer, 2017, pp. 677--687.

\bibitem{Dong2017AutomaticBT}
H.~Dong, G.~Yang, F.~Liu, Y.~Mo, Y.~Guo, Automatic brain tumor detection and
  segmentation using u-net based fully convolutional networks, ArXiv
  abs/1705.03820 (2017).

\bibitem{Zhang2020BlockLS}
L.~Zhang, J.~Zhang, P.~Shen, G.~Zhu, P.~Li, X.~Lu, H.~Zhang, S.~A.~A. Shah,
  M.~Bennamoun, Block level skip connections across cascaded v-net for
  multi-organ segmentation, IEEE Transactions on Medical Imaging 39 (2020)
  2782--2793.

\bibitem{Peng2020AMO}
Z.~Peng, X.~Fang, P.~Yan, H.~Shan, T.~Liu, X.~Pei, G.~Wang, B.~Liu, M.~K.
  Kalra, X.~G. Xu, A method of rapid quantification of patient-specific organ
  doses for ct using deep-learning based multi-organ segmentation and
  gpu-accelerated monte carlo dose computing., Medical physics (2020).

\bibitem{liu2020ct}
Y.~Liu, Y.~Lei, Y.~Fu, T.~Wang, X.~Tang, X.~Jiang, W.~J. Curran, T.~Liu,
  P.~Patel, X.~Yang, Ct-based multi-organ segmentation using a 3d
  self-attention u-net network for pancreatic radiotherapy, Medical physics
  47~(9) (2020) 4316--4324.

\bibitem{Lee_2020_CVPR}
H.~J. Lee, J.~U. Kim, S.~Lee, H.~G. Kim, Y.~M. Ro, Structure boundary
  preserving segmentation for medical image with ambiguous boundary, in:
  Proceedings of the IEEE/CVF Conference on Computer Vision and Pattern
  Recognition (CVPR), 2020.

\bibitem{Oda2018BESNetBS}
H.~Oda, H.~Roth, K.~Chiba, J.~Sokolic, T.~Kitasaka, M.~Oda, A.~Hinoki,
  H.~Uchida, J.~A. Schnabel, K.~Mori, Besnet: Boundary-enhanced segmentation of
  cells in histopathological images, in: MICCAI, 2018.

\bibitem{Murugesan2019PsiNetSA}
B.~Murugesan, K.~Sarveswaran, S.~M. Shankaranarayana, K.~Ram, M.~Sivaprakasam,
  Psi-net: Shape and boundary aware joint multi-task deep network for medical
  image segmentation, 2019 41st Annual International Conference of the IEEE
  Engineering in Medicine and Biology Society (EMBC) (2019) 7223--7226.

\bibitem{zhang2019net}
Z.~Zhang, H.~Fu, H.~Dai, J.~Shen, Y.~Pang, L.~Shao, Et-net: A generic
  edge-attention guidance network for medical image segmentation, in:
  International Conference on Medical Image Computing and Computer-Assisted
  Intervention, Springer, 2019, pp. 442--450.

\bibitem{Wang2019AUT}
L.~Wang, H.~Zhen, X.~Fang, S.~Wan, W.~Ding, Y.~Guo, A unified
  two-parallel-branch deep neural network for joint gland contour and
  segmentation learning, Future Gener. Comput. Syst. 100 (2019) 316--324.

\bibitem{milletari2016v}
F.~Milletari, N.~Navab, S.-A. Ahmadi, V-net: Fully convolutional neural
  networks for volumetric medical image segmentation, in: 2016 fourth
  international conference on 3D vision (3DV), IEEE, 2016, pp. 565--571.

\bibitem{Ruder2017AnOO}
S.~Ruder, An overview of multi-task learning in deep neural networks, ArXiv
  abs/1706.05098 (2017).

\bibitem{DBLP:journals/corr/ClevertUH15}
D.~Clevert, T.~Unterthiner, S.~Hochreiter, Fast and accurate deep network
  learning by exponential linear units (elus), in: Y.~Bengio, Y.~LeCun (Eds.),
  4th International Conference on Learning Representations, {ICLR} 2016, San
  Juan, Puerto Rico, May 2-4, 2016, Conference Track Proceedings, 2016.

\bibitem{RothLFSLTS15}
H.~R. Roth, L.~Lu, A.~Farag, H.~Shin, J.~Liu, E.~B. Turkbey, R.~M. Summers,
  Deeporgan: Multi-level deep convolutional networks for automated pancreas
  segmentation, in: Medical Image Computing and Computer-Assisted Intervention
  - {MICCAI} 2015 - 18th International Conference Munich, Germany, October 5-9,
  2015, Proceedings, Part {I}, Vol. 9349 of Lecture Notes in Computer Science,
  Springer, 2015, pp. 556--564.

\bibitem{Clark2013}
K.~Clark, B.~Vendt, K.~Smith, J.~Freymann, J.~Kirby, P.~Koppel, S.~Moore,
  S.~Phillips, D.~Maffitt, M.~Pringle, L.~Tarbox, F.~Prior, The cancer imaging
  archive ({TCIA}): Maintaining and operating a public information repository,
  Journal of Digital Imaging 26~(6) (2013) 1045--1057.
\newblock \href {https://doi.org/10.1007/s10278-013-9622-7}
  {\path{doi:10.1007/s10278-013-9622-7}}.

\bibitem{Gibson2018}
E.~Gibson, F.~Giganti, Y.~Hu, E.~Bonmati, S.~Bandula, K.~Gurusamy, B.~Davidson,
  S.~P. Pereira, M.~J. Clarkson, D.~C. Barratt, Multi-organ abdominal ct
  reference standard segmentations (2018).
\newblock \href {https://doi.org/10.5281/ZENODO.1169361}
  {\path{doi:10.5281/ZENODO.1169361}}.

\bibitem{paszke2017automatic}
A.~Paszke, S.~Gross, S.~Chintala, G.~Chanan, E.~Yang, Z.~DeVito, Z.~Lin,
  A.~Desmaison, L.~Antiga, A.~Lerer, Automatic differentiation in pytorch, in:
  NIPS-W, 2017.

\bibitem{kingma2014adam}
D.~P. Kingma, J.~Ba, Adam: A method for stochastic optimization, arXiv preprint
  arXiv:1412.6980 (2014).

\bibitem{4587417}
P.~Kohli, L.~Ladicky, P.~H.~S. Torr, Robust higher order potentials for
  enforcing label consistency, in: 2008 IEEE Conference on Computer Vision and
  Pattern Recognition, 2008, pp. 1--8.
\newblock \href {https://doi.org/10.1109/CVPR.2008.4587417}
  {\path{doi:10.1109/CVPR.2008.4587417}}.

\bibitem{Cheng2021BoundaryII}
B.~Cheng, R.~B. Girshick, P.~Doll'ar, A.~C. Berg, A.~Kirillov, Boundary iou:
  Improving object-centric image segmentation evaluation, 2021 IEEE/CVF
  Conference on Computer Vision and Pattern Recognition (CVPR) (2021)
  15329--15337.

\end{thebibliography}





\end{document}


\begin{frontmatter}


\author[a]{Samra Irshad\corref{aaa}}
\ead{sam.ershad@yahoo.com}
\author[b]{Douglas P.S. Gomes}
\author[c]{Seong Tae Kim}
\address[a]{Swinburne University of Technology, Hawthorn, Australia}
\address[b]{Victoria University, Melbourne, Australia}
\address[c]{Kyung Hee University, Yongin-si, Gyeonggi-do, South Korea}
\cortext[aaa]{Corresponding Author.}
\title{Supplementary Material}


\end{frontmatter}
\section{Baseline models}
\begin{figure*}[!hbt]
\centering
\includegraphics[width=0.8\textwidth]{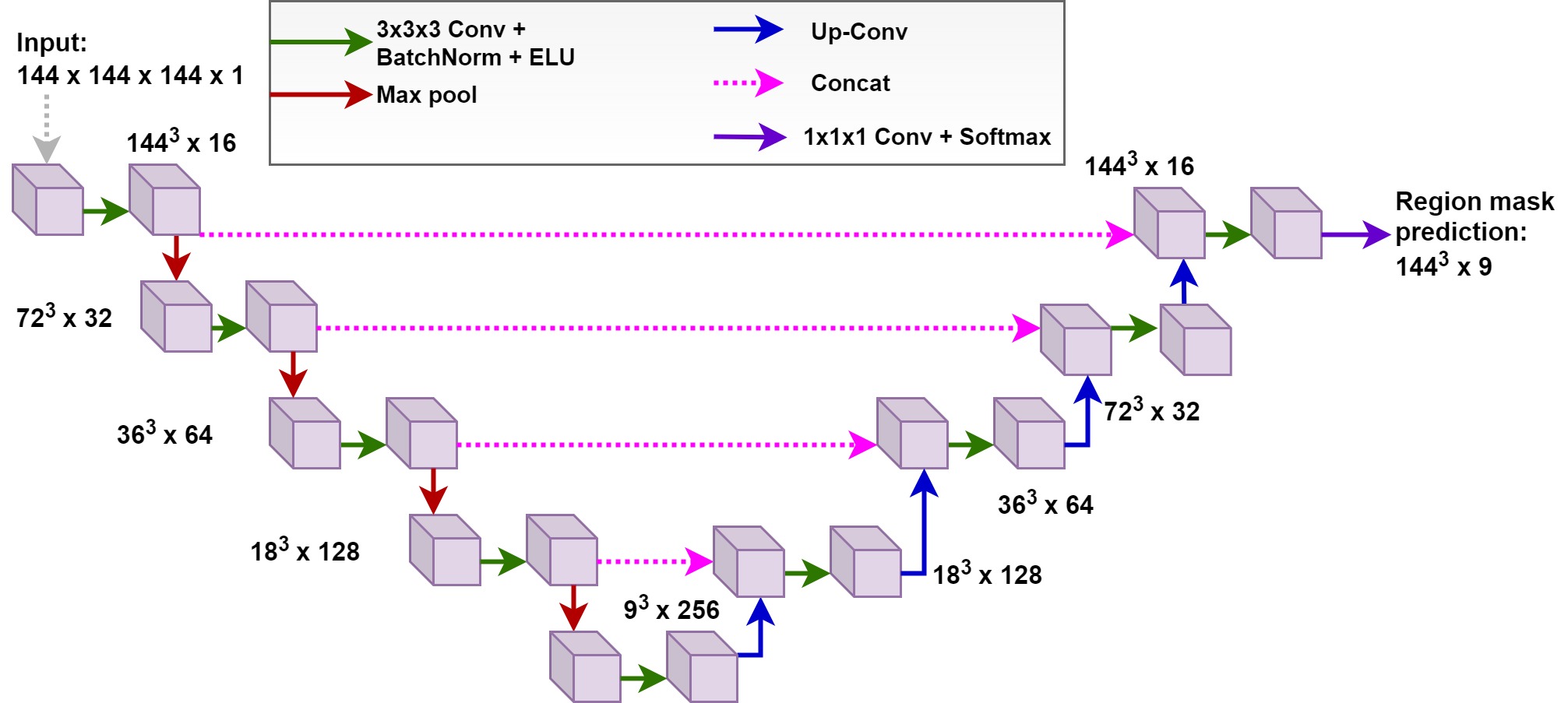} 
\caption{3D UNet model.}
\label{fig:sub1s}
\end{figure*}

\begin{figure*}[!hbt]
\centering
\includegraphics[width=0.6\textwidth]{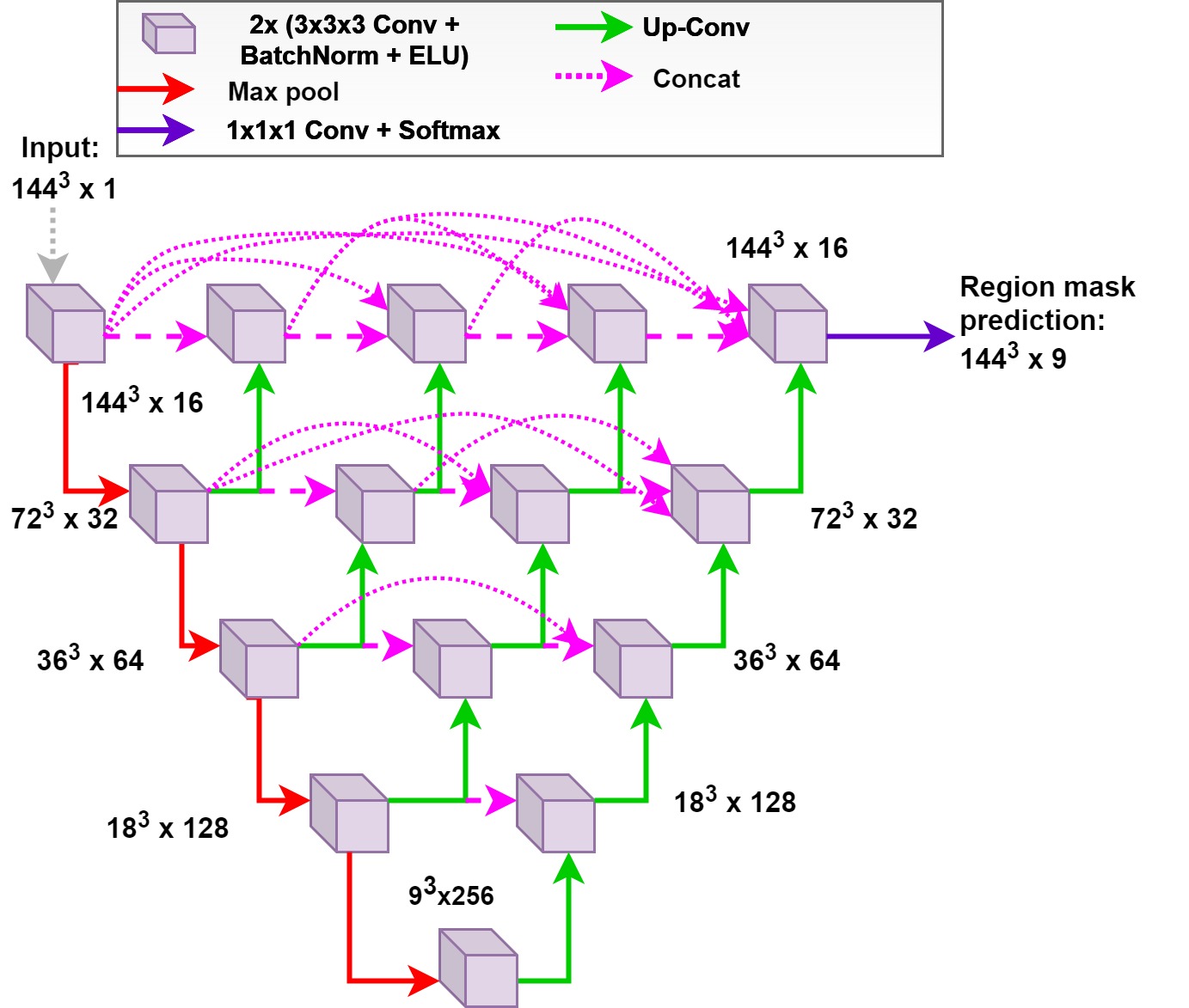} 
\caption{3D UNet$_{++}$.}
\label{fig:sub2s}
\end{figure*}

\begin{figure*}[!hbt]
\centering
\includegraphics[width=0.8\textwidth]{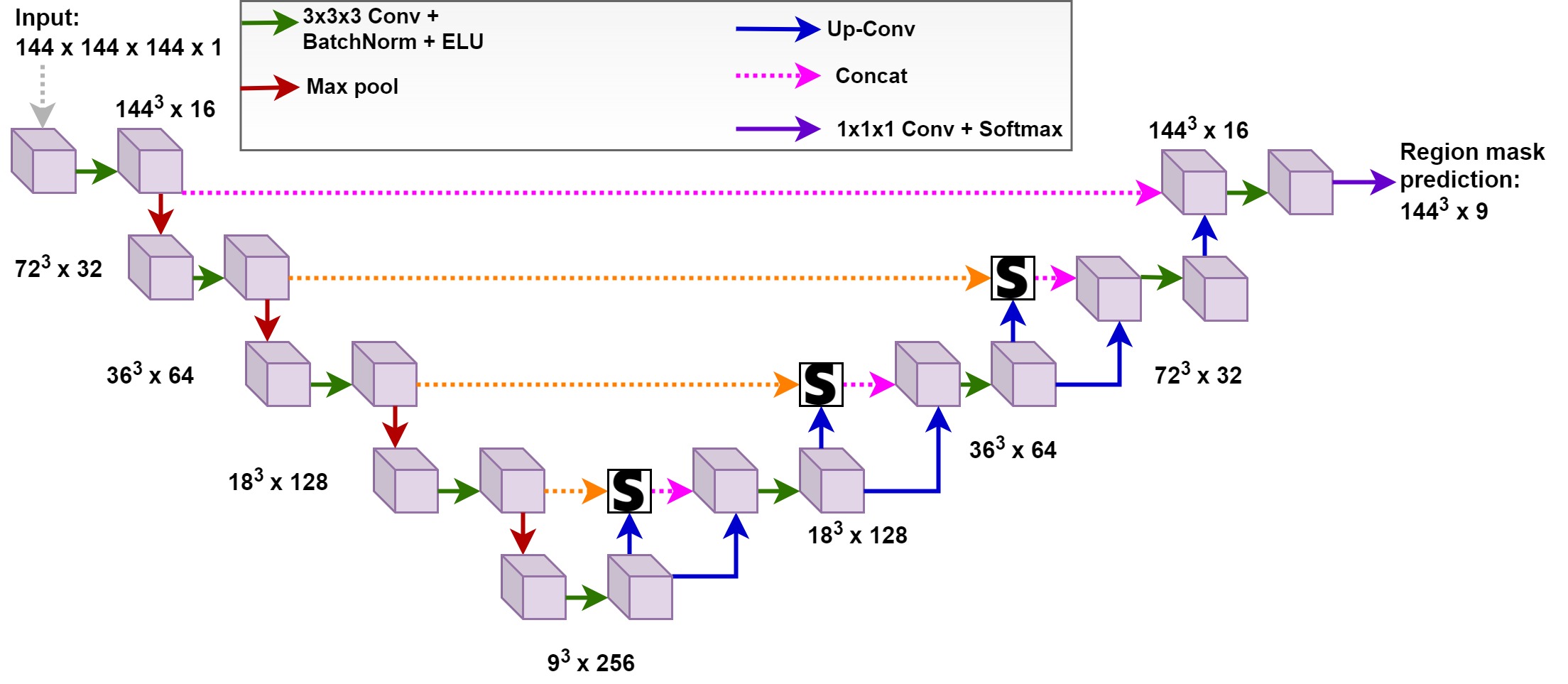} 
\caption{3D Att-UNet.}
\label{fig:sub3s}
\end{figure*}

\begin{figure*}[!hbt]
\centering
\includegraphics[width=0.5\textwidth]{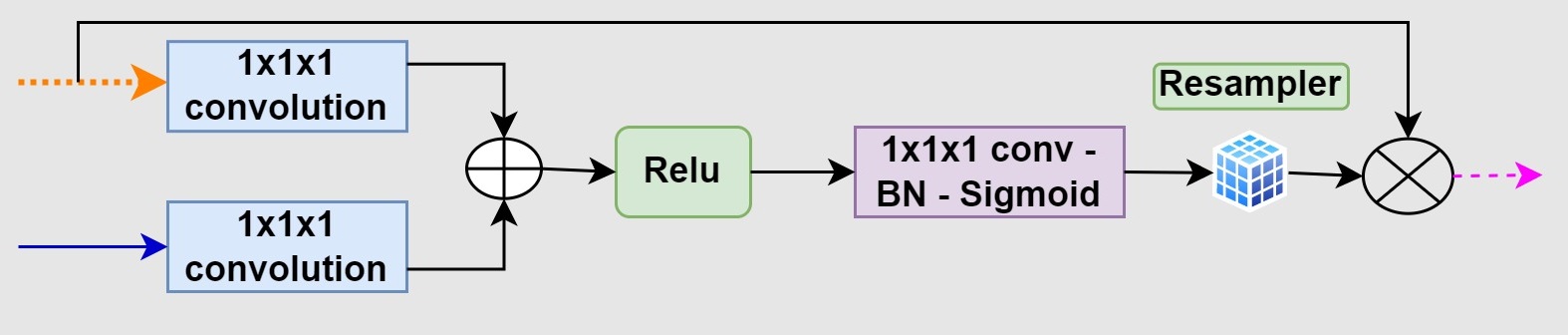} 
\caption{Design of attention gate utilized on 3D Att-UNet.}
\label{fig:sub4s}
\end{figure*}

We illustrate here the architectural design of baseline models, used in our work. \Cref{fig:sub1s,fig:sub2s} shows the baseline 3D UNet model and 3D UNet$_{++}$ models, respectively. \Cref{fig:sub3s,fig:sub4s} presents the baseline 3D Att-UNet model and the design of attention gate used in Att-UNet model.
\section{Implementation details}
The experiments conducted for selecting the model training parameters for Pancreas-CT and BTCV datasets are summarized in \Cref{table:Table 1supp,table:Table 2supp}, respectively. We also show the variation in validation dice scores when values of $\lambda$ are changed for Pancreas-CT and BTCV dataset in \Cref{table:Table 3supp,table:Table 4supp}, respectively.

\begin{table*}[ht] 
\centering
\caption{Effect of using different values of training parameters for Pancreas-CT dataset. CE loss represents Cross-Entropy loss.}\label{table:Table 1supp}
\begin{tabular}{llllllll}
\hline
Model & \multicolumn{6}{c}{\makecell[c]{Training\\ parameters}} & \makecell[c]{Validation\\ dice \\score}\\\hline
&\multicolumn{2}{c}{Optimizer}&\multicolumn{2}{c}{\makecell[c]{Initial\\learning\\rate}}&\multicolumn{2}{c}{Loss}\\\cline{2-7}
&Adam&RMSprop&0.01&0.001&CE Loss&Dice Loss&\\
\multirow{4}{*}{3D UNet} & \makecell[c]{\CheckmarkBold} &\makecell[c]{\xmark}&\makecell[c]{\xmark}& \makecell[c]{\CheckmarkBold}&\makecell[c]{\xmark}&\makecell[c]{\CheckmarkBold}&\makecell[c]{\textbf{0.769}}\\
 & \makecell[c]{\xmark} &\makecell[c]{\CheckmarkBold}&\makecell[c]{\xmark}& \makecell[c]{\CheckmarkBold}&\makecell[c]{\xmark}&\makecell[c]{\CheckmarkBold}&\makecell[c]{0.748}\\
  & \makecell[c]{\CheckmarkBold} &\makecell[c]{\xmark}&\makecell[c]{\CheckmarkBold}&\makecell[c]{\xmark} &\makecell[c]{\xmark}&\makecell[c]{\CheckmarkBold}&\makecell[c]{0.755}\\
    & \makecell[c]{\CheckmarkBold} &\makecell[c]{\xmark}&\makecell[c]{\xmark}&\makecell[c]{\CheckmarkBold} &\makecell[c]{\CheckmarkBold}&\makecell[c]{\xmark}&\makecell[c]{0.728}\\\hline
   \multirow{4}{*}{3D UNet$_{++}$} & \makecell[c]{\CheckmarkBold} &\makecell[c]{\xmark}&\makecell[c]{\xmark}& \makecell[c]{\CheckmarkBold}&\makecell[c]{\xmark}&\makecell[c]{\CheckmarkBold}&\makecell[c]{\textbf{0.755}}\\
 & \makecell[c]{\xmark} &\makecell[c]{\CheckmarkBold}&\makecell[c]{\xmark}& \makecell[c]{\CheckmarkBold}&\makecell[c]{\xmark}&\makecell[c]{\CheckmarkBold}&\makecell[c]{0.751}\\
  & \makecell[c]{\CheckmarkBold} &\makecell[c]{\xmark}&\makecell[c]{\CheckmarkBold}&\makecell[c]{\xmark} &\makecell[c]{\xmark}&\makecell[c]{\CheckmarkBold}&\makecell[c]{0.732}\\
    & \makecell[c]{\CheckmarkBold} &\makecell[c]{\xmark}&\makecell[c]{\xmark}&\makecell[c]{\CheckmarkBold} &\makecell[c]{\CheckmarkBold}&\makecell[c]{\xmark}&\makecell[c]{0.744}\\\hline
    \multirow{4}{*}{3D Att-UNet} & \makecell[c]{\CheckmarkBold} &\makecell[c]{\xmark}&\makecell[c]{\xmark}& \makecell[c]{\CheckmarkBold}&\makecell[c]{\xmark}&\makecell[c]{\CheckmarkBold}&\makecell[c]{\textbf{0.769}}\\
 & \makecell[c]{\xmark} &\makecell[c]{\CheckmarkBold}&\makecell[c]{\xmark}& \makecell[c]{\CheckmarkBold}&\makecell[c]{\xmark}&\makecell[c]{\CheckmarkBold}&\makecell[c]{0.754}\\
  & \makecell[c]{\CheckmarkBold} &\makecell[c]{\xmark}&\makecell[c]{\CheckmarkBold}&\makecell[c]{\xmark} &\makecell[c]{\xmark}&\makecell[c]{\CheckmarkBold}&\makecell[c]{0.757}\\
    & \makecell[c]{\CheckmarkBold} &\makecell[c]{\xmark}&\makecell[c]{\xmark}&\makecell[c]{\CheckmarkBold} &\makecell[c]{\CheckmarkBold}&\makecell[c]{\xmark}&\makecell[c]{0.676}\\\hline
\end{tabular}
\end{table*}
\begin{table}[ht] 
\centering
\caption{Effect of using different values of training parameters for BTCV dataset. CE loss represents Cross-Entropy loss.}\label{table:Table 2supp}
\begin{tabular}{llllllll}
\hline
Model & \multicolumn{6}{c}{\makecell[c]{Training\\ parameters}} & \makecell[c]{Validation\\ dice \\score}\\\hline
&\multicolumn{2}{c}{Optimizer}&\multicolumn{2}{c}{\makecell[c]{Initial\\learning\\rate}}&\multicolumn{2}{c}{Loss}\\\cline{2-7}
&Adam&RMSprop&0.01&0.001&CE Loss&Dice Loss&\\
\multirow{4}{*}{3D UNet} & \makecell[c]{\CheckmarkBold} &\makecell[c]{\xmark}&\makecell[c]{\xmark}& \makecell[c]{\CheckmarkBold}&\makecell[c]{\xmark}&\makecell[c]{\CheckmarkBold}&\makecell[c]{\textbf{0.693}}\\
 & \makecell[c]{\xmark} &\makecell[c]{\CheckmarkBold}&\makecell[c]{\xmark}& \makecell[c]{\CheckmarkBold}&\makecell[c]{\xmark}&\makecell[c]{\CheckmarkBold}&\makecell[c]{0.683}\\
  & \makecell[c]{\CheckmarkBold} &\makecell[c]{\xmark}&\makecell[c]{\CheckmarkBold}&\makecell[c]{\xmark} &\makecell[c]{\xmark}&\makecell[c]{\CheckmarkBold}&\makecell[c]{0.687}\\
    & \makecell[c]{\CheckmarkBold} &\makecell[c]{\xmark}&\makecell[c]{\xmark}&\makecell[c]{\CheckmarkBold} &\makecell[c]{\CheckmarkBold}&\makecell[c]{\xmark}&\makecell[c]{0.586}\\\hline
   \multirow{4}{*}{3D UNet$_{++}$} & \makecell[c]{\CheckmarkBold} &\makecell[c]{\xmark}&\makecell[c]{\xmark}& \makecell[c]{\CheckmarkBold}&\makecell[c]{\xmark}&\makecell[c]{\CheckmarkBold}&\makecell[c]{\textbf{0.681}}\\
 & \makecell[c]{\xmark} &\makecell[c]{\CheckmarkBold}&\makecell[c]{\xmark}& \makecell[c]{\CheckmarkBold}&\makecell[c]{\xmark}&\makecell[c]{\CheckmarkBold}&\makecell[c]{0.661}\\
  & \makecell[c]{\CheckmarkBold} &\makecell[c]{\xmark}&\makecell[c]{\CheckmarkBold}&\makecell[c]{\xmark} &\makecell[c]{\xmark}&\makecell[c]{\CheckmarkBold}&\makecell[c]{0.647}\\
    & \makecell[c]{\CheckmarkBold} &\makecell[c]{\xmark}&\makecell[c]{\xmark}&\makecell[c]{\CheckmarkBold} &\makecell[c]{\CheckmarkBold}&\makecell[c]{\xmark}&\makecell[c]{0.646}\\\hline
    \multirow{4}{*}{3D Att-UNet} & \makecell[c]{\CheckmarkBold} &\makecell[c]{\xmark}&\makecell[c]{\xmark}& \makecell[c]{\CheckmarkBold}&\makecell[c]{\xmark}&\makecell[c]{\CheckmarkBold}&\makecell[c]{\textbf{0.703}}\\
 & \makecell[c]{\xmark} &\makecell[c]{\CheckmarkBold}&\makecell[c]{\xmark}& \makecell[c]{\CheckmarkBold}&\makecell[c]{\xmark}&\makecell[c]{\CheckmarkBold}&\makecell[c]{0.620}\\
  & \makecell[c]{\CheckmarkBold} &\makecell[c]{\xmark}&\makecell[c]{\CheckmarkBold}&\makecell[c]{\xmark} &\makecell[c]{\xmark}&\makecell[c]{\CheckmarkBold}&\makecell[c]{0.626}\\
    & \makecell[c]{\CheckmarkBold} &\makecell[c]{\xmark}&\makecell[c]{\xmark}&\makecell[c]{\CheckmarkBold} &\makecell[c]{\CheckmarkBold}&\makecell[c]{\xmark}&\makecell[c]{0.636}\\\hline
\end{tabular}
\end{table}
\begin{table}[ht] 
\centering
\caption{Value of $\lambda$ used to balance the boundary loss for Pancreas-CT dataset. The first column represents the model's name; the second column shows the $\lambda$ value. The last column shows the standard deviation from the mean of validation dice scores generated using different $\lambda$ values in the search grid.}\label{table:Table 3supp}
\begin{tabular}{lll}
\hline
Method& $\lambda$&Std. Dev.\\\hline
3D UNet-MTL-TSOL & 2&0.009\\\hline
3D UNet-MTL-TSD & 1&0.008\\\hline
3D UNet$_{++}$-MTL-TSOL & 1.5&0.004\\\hline
3D UNet$_{++}$-MTL-TSD & 0.5&0.004\\\hline
3D Att-UNet-MTL-TSOL & 0.5&0.011\\\hline
3D Att-UNet-MTL-TSD & 1&0.005\\\hline
\end{tabular}
\end{table}
\vspace{-30cm}
\begin{table}[ht] 
\centering
\caption{Value of $\lambda$ used to balance the boundary loss for BTCV dataset. The first column represents the model's name; the second column shows the $\lambda$ value. The last column shows the standard deviation from the mean of validation dice scores generated using different $\lambda$ values in the search grid.}\label{table:Table 4supp}
\begin{tabular}{lll}
\hline
Method& $\lambda$&Std. Dev.\\\hline
3D UNet-MTL-TSOL & 1.5&0.012\\\hline
3D UNet-MTL-TSD & 1.5&0.011\\\hline
3D UNet$_{++}$-MTL-TSOL & 0.5&0.016\\\hline
3D UNet$_{++}$-MTL-TSD & 1.5&0.0004\\\hline
3D Att-UNet-MTL-TSOL & 2&0.017\\\hline
3D Att-UNet-MTL-TSD & 1&0.015\\\hline
\end{tabular}
\end{table}